\documentclass{aa}
\usepackage[varg]{txfonts}
\usepackage{graphicx}
\usepackage{morefloats}
\usepackage{natbib}
\bibpunct{(}{)}{;}{a}{}{,}

\begin{document}

\title{Simulations of an inhomogeneous stellar wind interacting with a pulsar wind in a binary system}

\author{X. Paredes-Fortuny\inst{1}
\and V. Bosch-Ramon\inst{1}
\and M. Perucho\inst{2}
\and M. Rib\'o\inst{1,}\thanks{Serra H\'unter Fellow.}}

\institute{Departament d'Astronomia i Meteorologia, Institut de Ci\`ences del Cosmos,
Universitat de Barcelona, IEEC-UB, Mart\'{\i} i Franqu\`es 1,
E-08028 Barcelona, Spain, e-mail: xparedes@am.ub.es
\and Departament d'Astronomia i Astrof\'{\i}sica, Universitat de Val\`encia, 
Av.\ Vicent Andr\'es Estell\'es s/n, 46100 Burjassot (Val\`encia), Spain
}
\date{Received - / Accepted -}

\abstract
{Binary systems containing a massive star and a non-accreting pulsar
present strong interaction between the stellar and the 
pulsar winds. The properties of this interaction, which largely determine the non-thermal radiation in these systems, 
strongly depend on the structure of the stellar wind, which can be clumpy or strongly anisotropic, as in Be stars.}
{We study numerically the influence of inhomogeneities in the stellar wind on the structure of the two-wind
interaction region.}
{We carried out for the first time axisymmetric, relativistic hydrodynamical simulations, with Lorentz factors of $\sim 6$
and accounting for the impact of instabilities, to study the impact in the two-wind interaction structure of an over-dense region of the stellar wind. 
We also followed the evolution of this over-dense region or clump as it faces the impact of the pulsar wind.}
{For typical system parameters, and adopting a stellar wind inhomogeneity with a density contrast $\chi\gtrsim 10$, 
clumps with radii of 
a few percent of the binary size can significantly perturb the two-wind interaction region, 
pushing the two-wind interface to $\lesssim 40$\% of the initial distance to the pulsar.
After it is shocked, the inhomogeneity quickly expands and is disrupted
when it reaches the smallest distance to the pulsar. 
It eventually fragments, being advected away from the binary system.
The whole interaction region is quite unstable, and the shocked pulsar wind can strongly change under small 
perturbations.}
{We confirm the sensitive nature of the two-wind interaction structure to perturbations, in
particular when the stellar wind is inhomogeneous. For realistic over-dense regions of the stellar wind,
the interaction region may shrink by a factor of a few, with the shocked flow presenting a
complex spatial and temporal pattern. This can lead to strong variations in the non-thermal radiation.}


\maketitle

\section{Introduction}
In binaries consisting of a massive star and a young non-accreting pulsar, the relativistic wind of the pulsar interacts with the non-relativistic wind of the stellar companion. This can result in efficient particle acceleration and in the production of non-thermal radiation, from radio to gamma rays. At present, PSR~B1259$-$63 has been confirmed as a member of this class of objects, and there are several other 
high-mass binaries hosting a compact object whose nature is still unknown and may belong to this class as well \citep[e.g.,][]{Dubus2013,Paredes2013}. 

The properties of the non-thermal radiation in 
high-mass binaries hosting a non-accreting pulsar are determined to a large extent by the dynamics of the two-wind interaction structure. To study different aspects of this structure and its evolution along the orbit, heavy and detailed numerical simulations have been performed. \cite{Romero2007} carried out non-relativistic, smoothed particle hydrodynamic (SPH) simulations in three dimensions (3D) to study the orbital evolution of a hypothetic two-wind interaction region in LS~I~$+$61~303. 
\cite{Bogovalov2008} studied numerically the collision of the pulsar and the stellar wind in PSR~B1259$-$63, treating the flows as laminar fluids by relativistic hydrodynamical simulations in two dimensions (2D).
For the same system, \cite{Okazaki2011} and \cite{Takata2012} performed 3D SPH non-relativistic simulations to study the tidal and wind interactions between the pulsar and the decretion disc of the Be star, and \cite{Bogovalov2012} studied the collision of a magnetized anisotropic pulsar wind and the stellar
wind adopting the same geometry and using a similar treatment as adopted in \cite{Bogovalov2008}. 
\cite{LambertsAAA} performed non-relativistic hydrodynamical simulations in planar geometry of the two-wind interaction structure for several orbits on large scales. \cite{LambertsBBB} and \cite{Lamberts2013} performed 2D relativistic hydrodynamical (RHD) simulations in planar geometry of the interaction between the pulsar and the stellar wind on the scales of the binary system. Finally, \cite{Bosch-Ramon2012} performed 2D RHD simulations in planar geometry aimed at studying the interaction between the stellar and pulsar winds on scales at which the orbital motion is important. 

Some relevant conclusions of the numerical work done up to now are the importance of instabilities, in particular the
Rayleigh-Taylor (RT) and Kelvin-Helmholtz (KH)\footnote{RT: instability developed in the contact surface between two fluids of different densities, with the lightest fluid exerting a force on the densest one. KH: instability developed in the contact surface between two fluids of different parallel velocities (see \citealt{Chandrasekhar1961}).} instabilities, in the spatial and temporal properties of the two-wind interaction structure on both small and large scales, the occurrence of shocked-flow re-acceleration, the impact of orbital motion through lateral, strong shocks, the effective mixing of the shocked winds downstream of the flow, the strong effects of the pulsar wind ram pressure on Be discs, and the relatively minor role of pulsar wind anisotropies and magnetic fields.  However, all the simulations of binaries hosting a pulsar considered smooth winds, while the stellar wind is expected to be inhomogeneous,
in particular for earlier spectral types \citep{Lucy1970}. The wind inhomogeneities (clumps hereafter) are expected to play an important role in the structure of the  two-wind interaction region, and therefore in the non-thermal emission. The origin of these clumps can be explained by different mechanisms depending on their scale, either the stellar surface at the wind formation region, or a circumstellar disc in the Be system. For instance, the impact of a fragment of a Be star disc on the two-wind interaction region has been proposed to cause the GeV flare observed in the gamma-ray binary PSR~B1259$-$63 \citep{Abdo2011,Chernyakova2014}. In addition, some sort of wind clumping could explain short flares on scales of seconds to hours found in the X-ray light curves of LS 5039 and LS I +61 303 \citep[e.g.,][]{Bosch-Ramon2005,Paredes2007,Smith2009,Li2011}.
The clump-pulsar wind interaction problem has previously been studied analytically \citep[see][]{Bosch-Ramon2013}, but numerical calculations of the interaction between an inhomogeneous stellar wind and a relativistic pulsar wind have not been conducted yet.

In this work, we present 2D axisymmetric RHD simulations of the interaction between a wind clump and a relativistic pulsar wind.
Orbital motion has not been accounted for because of the axisymmetric nature of the simulations; given the small spatial and temporal scales considered, this is a reasonable approximation. 
We simulated clumps with different sizes and densities to study the global variations induced in the two-wind interaction structure, as well as on the evolution of the clumps under the impact of the pulsar wind.

We note that this is the first time that relativistic simulations are carried out in a more realistic set-up, with axisymmetry along the two-star line, without suppressing the development of instabilities. The numerical results are compared with the analytical results found in \cite{Bosch-Ramon2013}. The paper is structured as follows: in Sect.~\ref{frame} we describe the physical framework: the two-wind interaction region, the origin of the clumps in the winds of massive stars, and the analytical equations; in Sect.~\ref{num}, we describe the numerical simulations: the numerical set-up and the obtained results; finally, a discussion and the conclusions can be found in Sects.~\ref{disc} and ~\ref{conc}. 

\section{Physical framework}
\label{frame}

\begin{figure}
\centering
\resizebox{0.8\hsize}{!}{\includegraphics{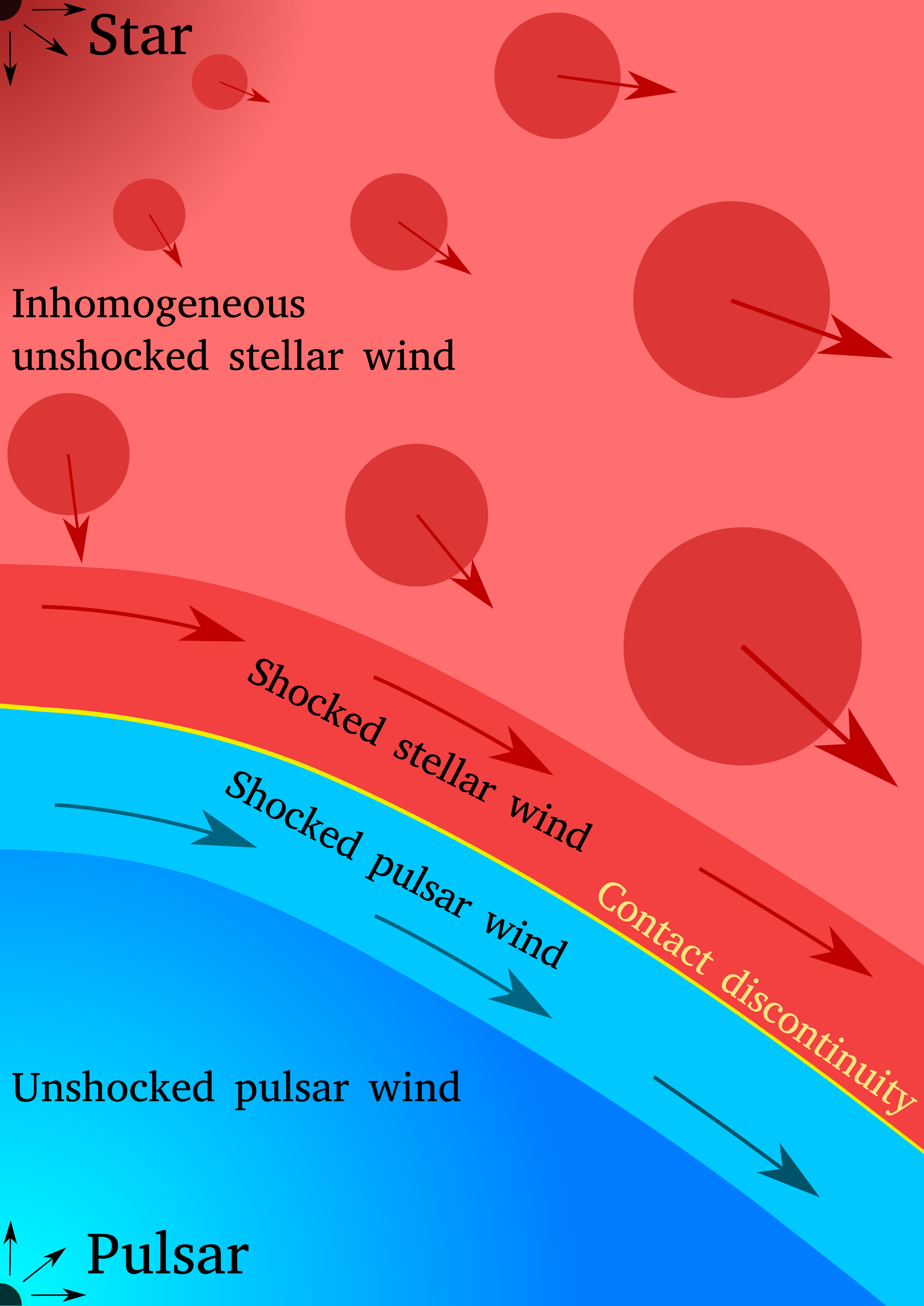}}
\caption{Cartoon representing a density map of the physical scenario: an inhomogeneous stellar wind interacts with the 
relativistic pulsar wind. The shocked stellar wind and the shocked pulsar wind are separated by a contact 
discontinuity. The distances are not to scale. 
The simulations presented in this paper
only considered a stellar wind with one clump centered at the axis
joinning the two stars.
}
\label{sce}
\end{figure}

The collision between the stellar wind and the wind from the young pulsar creates a shock structure bow-shaped towards the wind with the lower momentum flux. The contact discontinuity is located where the ram pressure of the two winds balances. As discussed in \cite{Bosch-Ramon2013}, the presence of clumps can significantly distort the overall interaction structure, although their dynamical impact is determined by their size and mass \citep[see, e.g.,][for a similar non-relativistic scenario]{Pittard2007}. Winds formed by small and light clumps will behave as uniform winds, while massive clumps will tend to come less frequently and be more damaging for the global stability of the interaction region (see \citealt{Perucho2012} for the effect of an inhomogeneous wind on a high-mass microquasar jet). A sketch of the physical scenario is shown in Fig.~\ref{sce}. 

The instability of the line-driving mechanism in the inner wind is thought to be an important source of clumps in the wind structure of hot stars of spectral types O and B \citep{Lucy1970}. \cite{Runacres2002} suggested that the outer evolution of the inhomogeneous wind can be approximated as a pure gas dynamical problem, and the stellar wind clumps initiated close to the star can survive up to long distances ($d\sim 1000~{\rm R_\odot}$). \cite{Moffat2008} suggested that all hot stellar winds are likely to be inhomogeneous because of radiative instabilities, with a multi-scale distribution of masses and sizes. Therefore, considering the outer wind evolution as a gas dynamical problem, the clumps could expand with the wind flow and large clumps could reach the interaction region between the stellar and the pulsar wind. However, the relatively small initial sizes ($\sim 0.01\,R_*$), and limitations on the clump growth (which can be parameterized as $R_{\rm c}=R_{\rm{c}_0}\,(d/R_*)^{\alpha}$, with $\alpha \le 1$ and $R_{\rm{c}_0}$ being the initial clump radius) could prevent the formation of clumps as large as, $R_{\rm c}\sim R_*$, for instance, which is required for the strongest variations of the  two-wind interaction structure \citep[see][and references there in]{Bosch-Ramon2013}. Typical massive star radii are $R_*\sim 10\,R_\odot$.

Large clumps different from those related to radiative instabilities may be found in the stellar wind. For instance, early-type stars known as Be stars present a decretion disc formed by material ejected from the stellar equator by rapid rotation \citep{Hanuschik1996}. The truncation of this disc, either caused by tidal forces or by direct pulsar wind ram pressure \citep[see][]{Okazaki2011}, could explain the presence of large clumps in the stellar wind formed by chunks of disc. This possibility has been considered for instance to explain the GeV flare detected from the gamma-ray binary PSR~B1259$-$63 by \cite{Chernyakova2014} (see Sect.~\ref{psr}). Other types of large-scale structures in the stellar wind, of size of the order of the stellar radius have also been inferred from the observed discrete absorption components (DACs) in the UV \citep[e.g.,][]{Kaper1997}. 
The appearance of DACs can be interpreted as co-rotating dense structures produced at the stellar surface and extending several tens of stellar radii
\citep[e.g.,][]{Lobel2008},
and can be attributed to the dynamical effects of rotation, magnetic fields, or non-radial pulsations \citep{Cranmer1996}.
The arrival of any of those structures at the two-wind interaction region probably modify the latter in space and time.

\cite{Bosch-Ramon2013}, developed an analytical model for the two-wind interaction dynamics accounting for the lifetime of clumps under the pulsar-wind impact. It was concluded that for a clump radius $R_{\rm c} \ll \chi^{-1/2}\Delta R$, being $\Delta R$ the thickness of the two-wind interaction region and $\chi = \rho_{\rm c}/\rho_{\rm w}$ the density contrast between the clump density and the wind density, the clump is destroyed and deflected by the shocked wind medium before crossing the two-wind interaction region. Otherwise, for  $R_{\rm c}$ approaching $\chi^{-1/2}\Delta R$ or larger, the clump will penetrate into the unshocked pulsar wind. When $R$ and $R^{\prime}$ are the initial and final distances between the pulsar and the contact discontinuity, following \cite{Bosch-Ramon2013}, this approximate relation applies

\begin{equation}
R^{\prime} \sim R - \chi^{1/2}R_{\rm c}\,,
\label{equ}
\end{equation}
which for $R_{\rm c}\sim R^{\prime}$ implies
\begin{equation}
R_{\rm c}\sim R^{\prime} \sim \frac{R}{\left( 1+\chi^{1/2}\right)}\quad \text{(where $R^{\prime}>0$ )}\,.
\label{equ2}
\end{equation}

The simulations presented here are intended to investigate in detail the dynamical consequences of the presence of clumps in the stellar wind. Below we compare analytical estimates with numerical results.

\section{Numerical simulations}
\label{num}

\subsection{Numerical set-up}

The simulations were performed using a finite-difference code based on a high-resolution shock-capturing scheme that solves the equations of relativistic hydrodynamics in two dimensions in a conservation form \citep{Marti1997}. The code is parallelized using open message passing (OMP) \citep{Perucho2005}. The simulations were run in a workstation with two Intel(R) Xeon(R) CPU E5-2643 processors (3.30 GHz, $4\times2$ cores, with two threads for each core) and four modules of 4096 MB of memory (DDR3 at 1600 MHz).

We adopted an ideal relativistic gas without a magnetic field, one particle species, and a polytropic index of $\gamma =1.444$, intermediate between a non-relativistic (stellar wind) and a relativistic (pulsar wind) index. The physical size of the domain is $r \in \lbrack 0, 30~a\rbrack$ and $z \in \lbrack 0, 50~a \rbrack$, where $a = 8\times10^{10}$~cm. The adopted resolution is $150\times250$ cells. The pulsar is located at $(r,z) = (0,5~a)$, and the star at $(r,z) = (0,60~a)$, outside the simulated grid. This yields a two-star separation of $d = 4.4\times10^{12}~{\rm cm}$. The typical orbital separation distances in gamma-ray binaries are $\sim 10^{13}~{\rm cm}$ \citep[see, e.g.,][and references therein]{Dubus2013}.

The initial conditions of the simulation were computed in spherical coordinates assuming adiabatic gas radial propagation and solving the Bernoulli equation for the pulsar and stellar winds. The initial two-wind separation point is
derived as the approximate location where the on-axis wind ram pressures become equal ($z = 22~a$).

The chosen physical parameters of the pulsar wind are
the total luminosity $L_{\rm sd} = 10^{37}~{\rm erg~s^{-1}}$, 
the Lorentz factor $\Gamma= 5$,
and the specific internal energy $\epsilon_{\rm pw} = 9.0\times10^{19}~{\rm erg~g^{-1}}$.
As a result of resolution limitations, the adopted Lorentz factor is smaller than expected in pulsar winds, $\Gamma\sim 10^4$--10$^6$ \citep[see][and references therein]{Khangulyan2012,Aharonian2012}, but high enough to capture important relativistic effects \citep[see the discussion in][]{Bosch-Ramon2012}.
For the stellar wind, the physical parameters are
a mass-loss rate $\dot{M}=10^{-7}~{\rm M_\odot~yr^{-1}}$, 
a radial velocity $v_{\rm sw}= 3000~{\rm km~s^{-1}}$,
and a specific internal energy $\epsilon_{\rm sw} = 1.8\times10^{15}~{\rm erg~g^{-1}}$.
The derived pulsar wind and stellar wind densities are $\rho_{\rm pw} = 1.99\times10^{-19}~{\rm g~cm^{-3}}$ and $\rho_{\rm sw} = 2.68\times10^{-13}~{\rm g~cm^{-3}}$, respectively. 
All the previous quantities, except for $L_{\rm sd}$ and $\dot{M}$, are given at a distance $r = a$ with respect to the pulsar and star centres. The values of $L_{\rm sd}$, $\dot{M}$ and  $v_{\rm sw}$ were chosen as representative values of gamma-ray binaries hosting a pulsar because $L_{\rm sd}$ is to be high enough to power the gamma-ray emission, and the stellar wind properties correspond to those of an OB star. All these parameters are summarized in Tables~\ref{table_sim} and \ref{table_wind}.

The previous physical values lead to 
a pulsar-to-stellar wind thrust ratio of
\begin{equation}
\eta = \frac{F_{\rm pw} S_{\rm pw}}{F_{\rm sw} S_{\rm sw}} = \frac{\left( \rho_{\rm pw} \Gamma^2 v_{\rm pw}^2 h_{\rm pw}+p_{\rm pw} \right)S_{\rm pw}}{\dot{M}v_{\rm sw}+p_{\rm sw} S_{\rm sw}} \approx \frac{L_{\rm sd}}{\dot{M} v_{\rm sw} {c}} \sim 0.2{\rm,}
\label{eta}
\end{equation}
where $F_{\rm pw}$ ($F_{\rm sw}$) is the momentum flux of the pulsar wind (stellar wind), $S_{\rm pw}$ ($S_{\rm sw}$) the spherical surface at a distance $r = a$ with respect to the pulsar (star), $h_{\rm pw}$ the specific enthalpy of the pulsar wind given by $h_{\rm pw}=1+\frac{\epsilon_{\rm pw}}{c^2}+\frac{p_{\rm pw}}{\rho_{\rm pw} {c^2}}$, and $p_{\rm pw/sw}$ the pressure of the pulsar/stellar wind given by $p_{\rm pw/sw} = \left( \gamma-1\right)\rho_{\rm pw/sw}\epsilon_{\rm pw/sw}$.

The pulsar wind is introduced by defining an injector with the mentioned properties and radius $3~a$ $(2.4\times10^{11}~{\rm cm};~ 15~{\rm cells)}$ as a boundary condition, and the stellar wind is injected according to its characterization at the upper boundary of the grid. 
The lower and right boundaries of the grid are set to outflow, while the left boundary of the grid is set to reflection.

The pulsar wind is injected with a Lorentz factor of 5, but because of adiabatic propagation, it reaches Lorentz factors of $\sim 6$ before termination.

\begin{table}
\caption{Simulation parameters.}
\label{table_sim}
\centering
\begin{tabular}{c c}
\hline\hline
Parameter & Value \\
\hline
$\gamma$  & $1.444$ \\
$\rho_0$  & $22.5\times10^{-22}~{\rm g~cm^{-3}}$ \\
$a$       & $8\times10^{10}~\rm{cm}$ \\
$l_r$     & $30~a~(2.4\times10^{12}~\rm{cm})$ \\
$l_z$     & $50~a~(4.0\times10^{12}~\rm{cm})$ \\
$n_r$     & $150$ \\       
$n_z$     & $250$ \\       
\hline
\end{tabular}
\tablefoot{Polytropic index $\gamma$, density unit $\rho_0$, distance unit $a$, 
physical $r$ grid size $l_r$, physical $z$ grid size $l_z$, 
number of cells in the $r$ axis $n_r$, and number of cells in the $z$ axis $n_z$.}
\end{table}

\begin{table*}
\caption{Pulsar and stellar parameters in code and CGS units.}
\label{table_wind}
\centering
\begin{tabular}{c c c}
\hline\hline
Parameter    & Pulsar wind                                              & Stellar wind \\
\hline
$\rho$       & $88.45~\rho_{0} \ (1.99\times10^{-19}~{\rm g~cm^{-3}})$       & $1.19\times10^8~\rho_{0}\ (2.68\times10^{-13}~{\rm g~cm^{-3}})$ \\
$\epsilon$   & $0.1~{c^2} \ (9\times10^{19}~{\rm erg~g^{-1}})$             & $2.0\times10^{-6}~{c^2}\ (1.8\times10^{15}~{\rm erg~g^{-1}})$ \\
$v$          & $0.9798~{c}$ ($2.94\times 10^{10}$~cm~s$^{-1}$)                                         & $0.01~{c}$ ($3\times 10^8$~cm~s$^{-1}$) \\
$(r_0,z_0)$  & $(0,5~a);\ (0,4\times10^{11}~{\rm cm})$             & $(0,60~a);\ (0,4.8\times10^{12}~{\rm cm})$ \\
\hline
\end{tabular}
\tablefoot{
Density $\rho$, specific interal energy $\epsilon$, and wind velocity $v$
at a distance $r = a$ with respect to the pulsar and star centres located at $(r_0,z_0)$.
}
\end{table*}

The stellar wind clumps are introduced at $(r,z)=(0,33~a)$ after the steady state is reached.
The clumps are characterized by their radius $R_{\rm c}$ and their density contrast 
$\chi$ with respect to the average stellar wind value at their location.
We present here the results for four different cases:
$\chi = 10$ and $R_{\rm c} =1~a$, $2.5~a$, $5~a$ and $\chi = 30$ and $R_{\rm c} =1~a$, 
corresponding to different degrees of wind inhomogeneity, from modest $\chi=10$ and $R_{\rm c}=1~a$ to rather high $\chi=10$ (30) and $R_{\rm c}=5$ (1) $a$.

The axisymmetric nature of the simulations, in particular the reflection boundary conditions, the presence of a coordinate singularity, plus the concentration of fluxes, all at $r=0$, lead to the generation of perturbations that quickly grow, posing difficulties to achieve a steady state. This effect makes the simulation also quite sensitive to the initial conditions, and the initial two-wind contact discontinuity has to be carefully chosen so as not to enhance the axis perturbations further. In this context, the modest resolution of most of the simulations and the pulsar wind Lorentz factor adopted were chosen to stabilize the solution of the calculations through a larger numerical dissipation and moderate wind density contrast. Given the fast growth of any perturbation (numerical or physical) in the simulated problem (as explained below), higher resolutions will enhance (as also shown below) the instability of the simulated structures through the penetration of small fragments of stellar wind material into the pulsar wind region. On the other hand, higher Lorentz factors would have enhanced the wind density contrast and thereby the instability growth rate. Finally, the grid size was also limited to prevent the instabilities from fully disrupting the two-wind interaction structure within the simulated region. With our resolution, larger grid sizes would have given enough time to the instabilities to develop and grow towards the downstream edge of the grid. The resulting structures would be similar to those found in the higher resolution simulations also presented in this work, and would also lead to the collapse of the two-wind interaction region. Albeit small, the grid size does not significantly affect the hydrodynamics within as the flow is supersonic at the outflow boundaries. We remark that because the clump is the strongest dynamical factor, its impact can be studied neglecting the role of numerical perturbations on the axis after a (quasi-)steady state has been reached.

\subsection{Results}

\begin{figure}
\resizebox{\hsize}{!}{\includegraphics{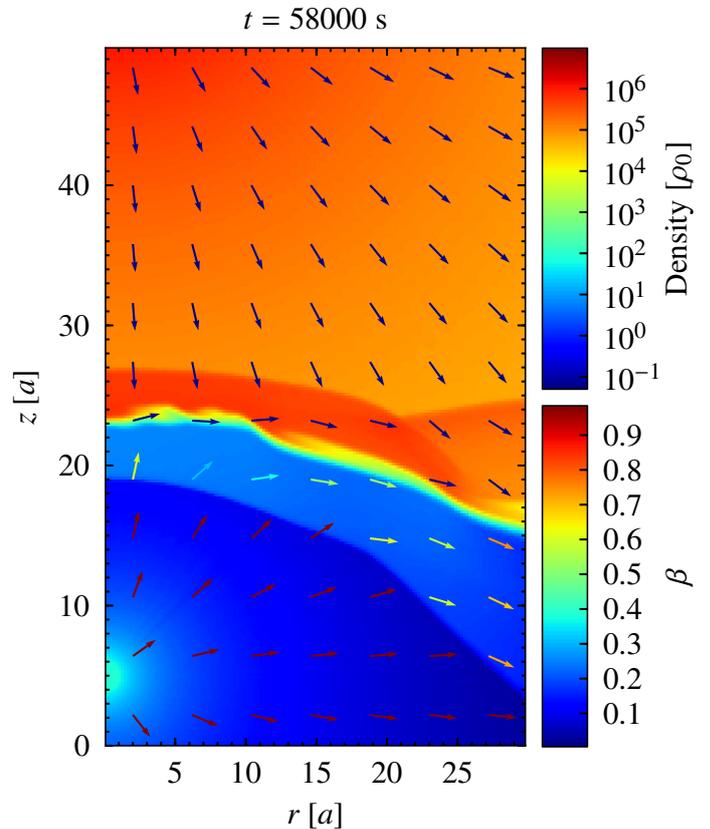}}
\caption{Density distribution by colour
at time $t = 58000$ s. 
The coloured arrows represent the three-velocity at different locations ($\beta = v/c$).
The axes units are $a = 8\times10^{10}$~cm.
The density units are $\rho_0 = 22.5\times10^{-22}~{\rm g~cm^{-3}}$. 
The pulsar and the star are located at $(r, z) = (0,5~a)$ and $(r, z) = (0,60~a)$, respectively.
The same units are used in all the figures.
}
\label{steady}
\end{figure}

Figure \ref{steady} shows the density map of the simulated flow in the (quasi-)steady state at $t = 58000$~s. The shock structure is bow-shaped towards the pulsar, with the contact discontinuity at $\sim 18~a$ from the pulsar. The thickness of the shocked two-wind region is $\sim 8~a$ on the simulation axis. After being shocked, winds are accelerated side-wards because of the strong pressure gradient in the downstream region, as pointed out in \cite{Bogovalov2008}. 
This is seen in the velocity maps presented in this section, for which the two-wind interaction structure has not yet been affected by the clump arrival. Because of this flow re-acceleration, the ratio of the momentum-flux to pressure shows that the flow becomes supersonic at the boundaries.

These axisymmetric simulations are quite sensitive to the initial conditions, and in particular, reaching steady state, is highly dependent on the initial set-up of the simulation. 
It is still an open question, to be answered with 3D simulations, to which extent this sensitivity to the initial set-up is caused by the geometry of the grid. Nevertheless, it is expected that instabilities quickly grow if perturbations are present, as shown by planar coordinate relativistic simulations \citep{Bosch-Ramon2012,Lamberts2013}. In the present simulations, the boundary conditions at the axis induce perturbations there that grow under the RT instability. These perturbations are then amplified by the KH instability as the flow propagates along the contact discontinuity. For grid sizes larger than $30~a$ in the $r$ direction, the structures resulting from instability growth can reach deep into the pulsar wind, eventually disrupting the two-wind interaction region and filling the pulsar vicinity with shocked material. For small grids, on the other hand, these structures are advected out of the grid before it happens. The RT and KH instabilities grow faster for faster (and lighter) pulsar winds, which for a fixed $\eta$ implies a stronger wind density contrast and thus a more unstable contact discontinuity. A Lorentz factor of $\Gamma\sim$ 5--6 was adopted as a compromise: a rather relativistic flow with tenable instability-growth rates. 

Despite numerical perturbations introducing a degree of deformation in the global structure of the two-wind collision region, as shown by the wave at $r\sim 20-30~a$ in the two-wind interaction region in Fig.~\ref{steady}, the opening angle and width of the shocked wind zone are similar to those found in \cite{Bogovalov2008,Bogovalov2012}. Those simulations were similar to ours given their axisymmetry, although the simulated flows were laminar. 

\begin{figure*}
\centering
\includegraphics[width=6.4cm]{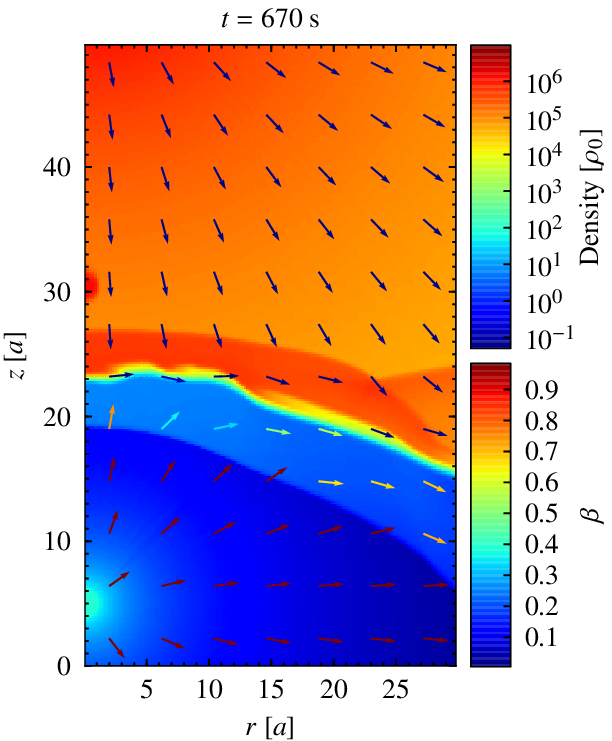}
\includegraphics[width=6.4cm]{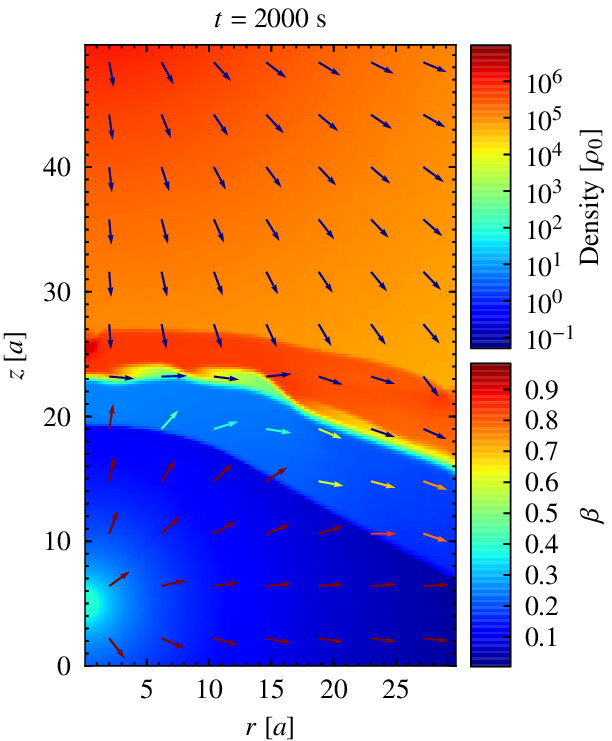}\\\vspace{0.2cm}
\includegraphics[width=6.4cm]{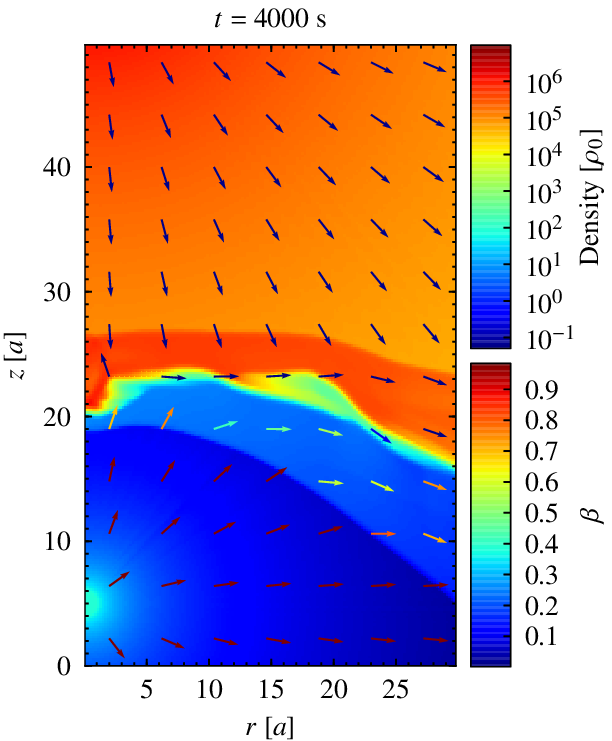}
\includegraphics[width=6.4cm]{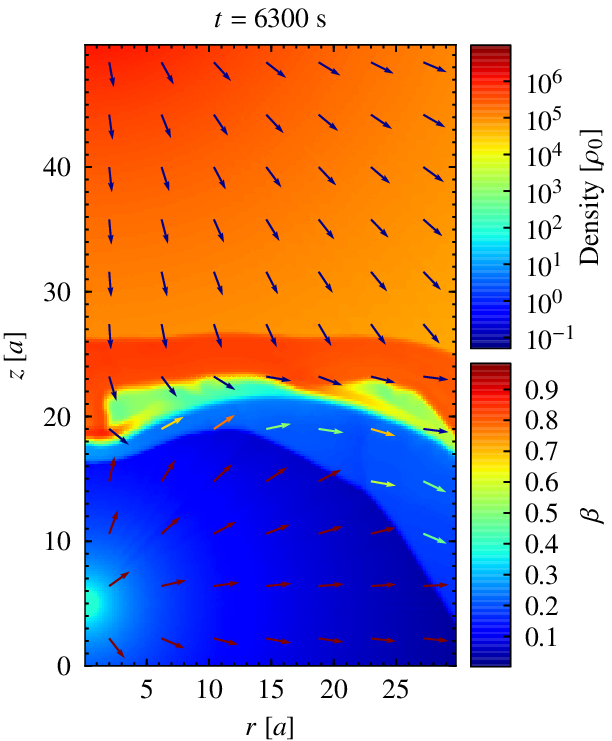}\\\vspace{0.2cm}
\includegraphics[width=6.4cm]{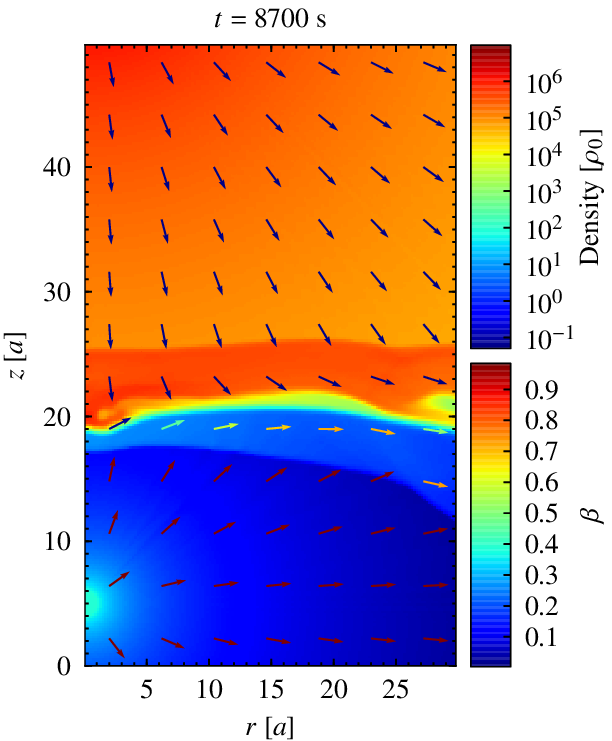}
\includegraphics[width=6.4cm]{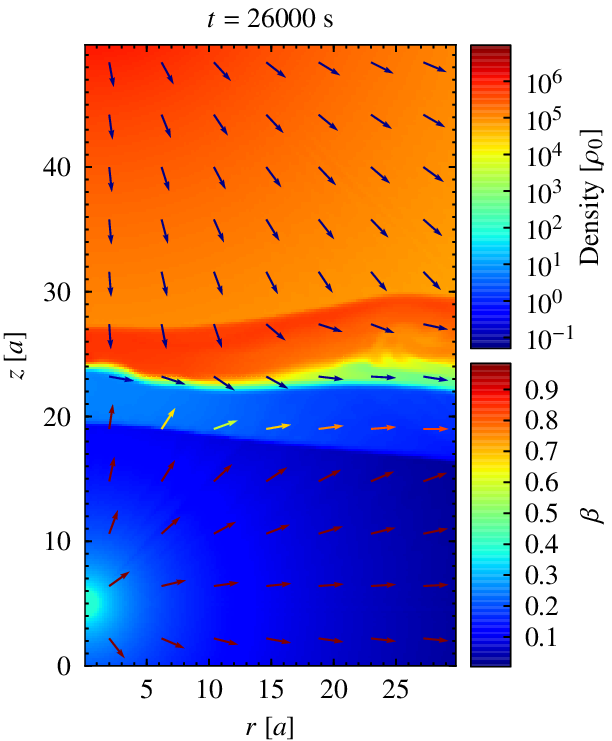}
\caption{
Density distribution by colour for the case with clump parameters
$\chi = 10$ and $R_{\rm c} =1~a$
for the times shown at the top of each plot.
The remaining plot properties are the same as those of Fig.~\ref{steady}.
}
\label{f10r1}
\end{figure*}

\begin{figure*}
\centering
\includegraphics[width=6.4cm]{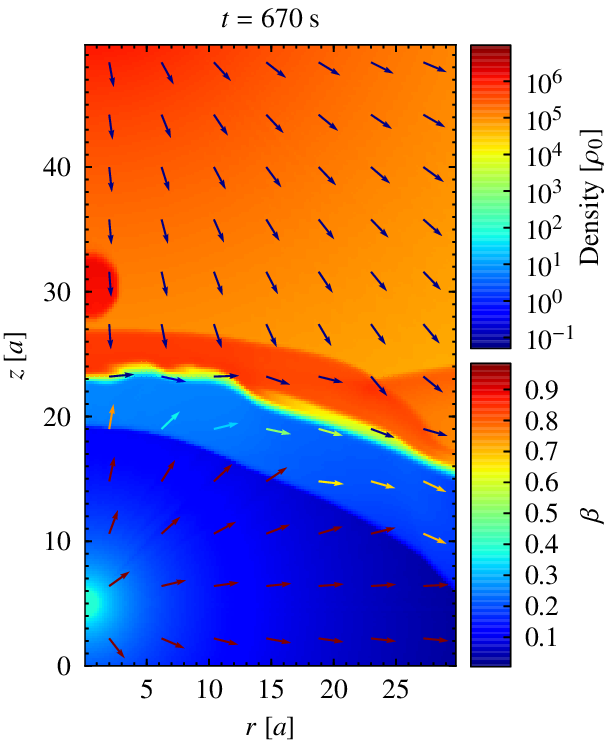}
\includegraphics[width=6.4cm]{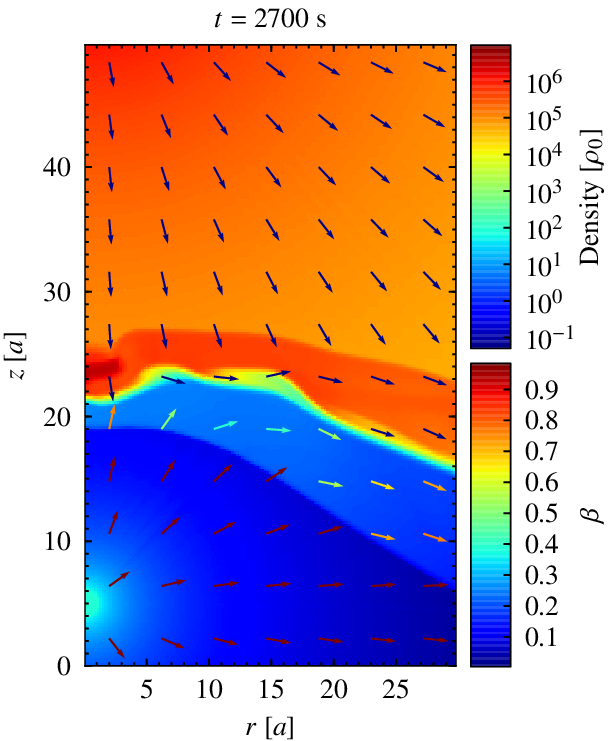}\\\vspace{0.2cm}
\includegraphics[width=6.4cm]{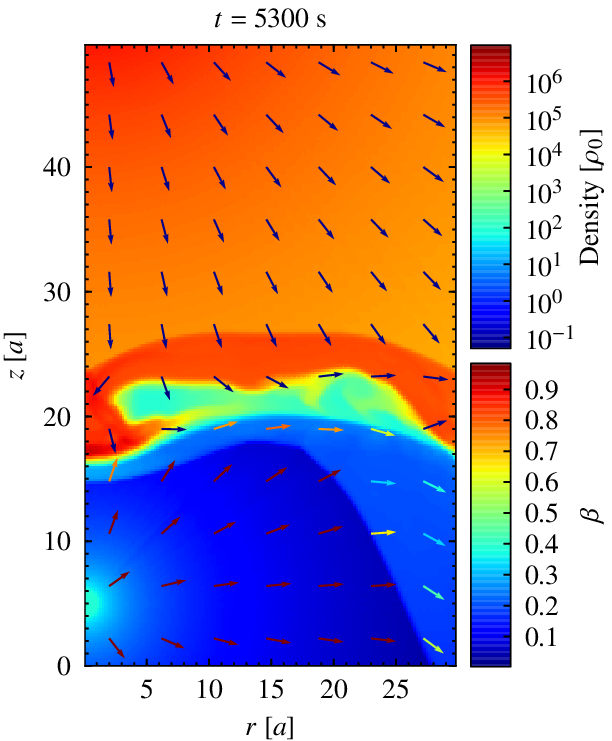}
\includegraphics[width=6.4cm]{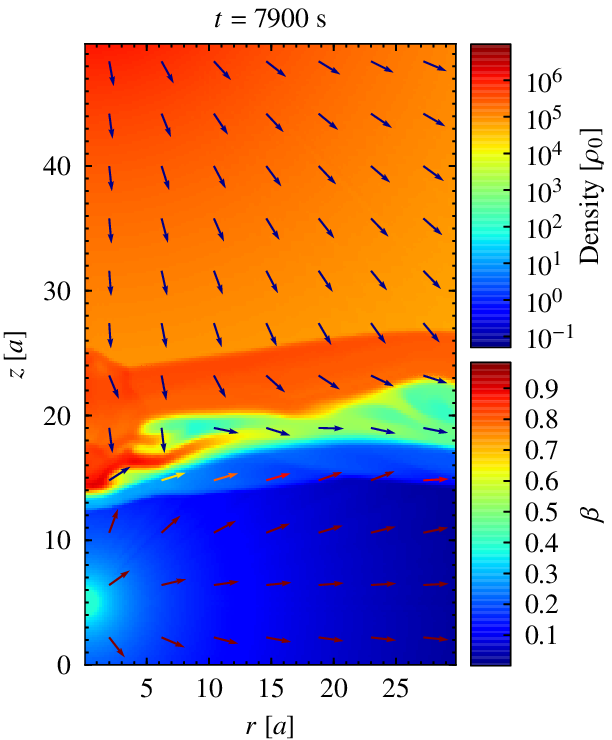}\\\vspace{0.2cm}
\includegraphics[width=6.4cm]{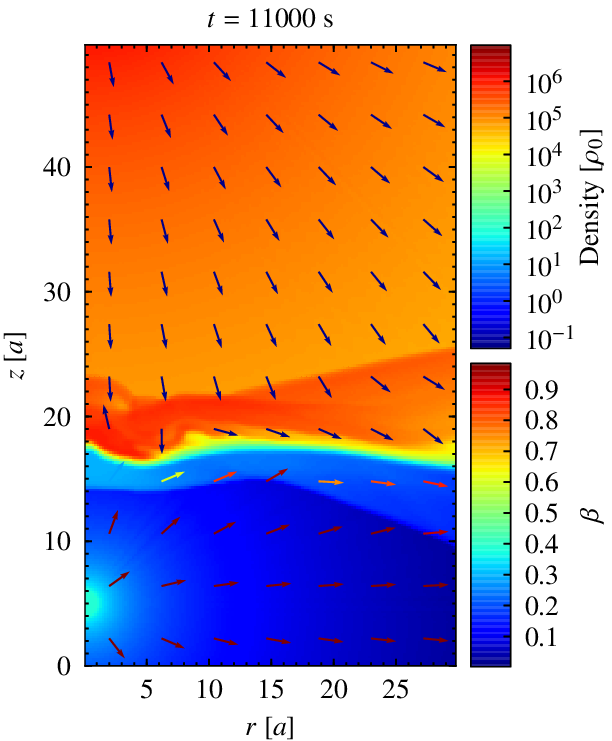}
\includegraphics[width=6.4cm]{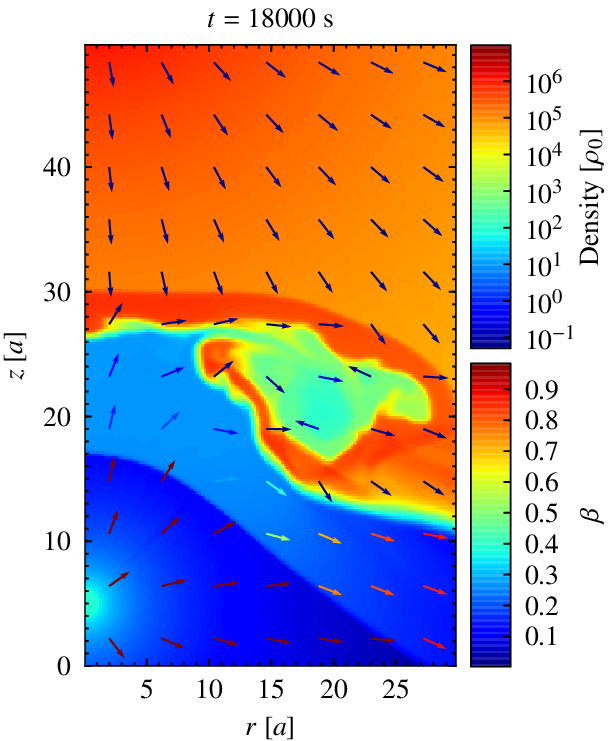}
\caption{
Density distribution by colour for the case with clump parameters
$\chi = 10$ and $R_{\rm c} =2.5~a$
for the times shown at the top of each plot.
The remaining plot properties are the same as those of Fig.~\ref{steady}.
}
\label{f10r2p5}
\end{figure*}

\begin{figure*}
\centering
\includegraphics[width=6.4cm]{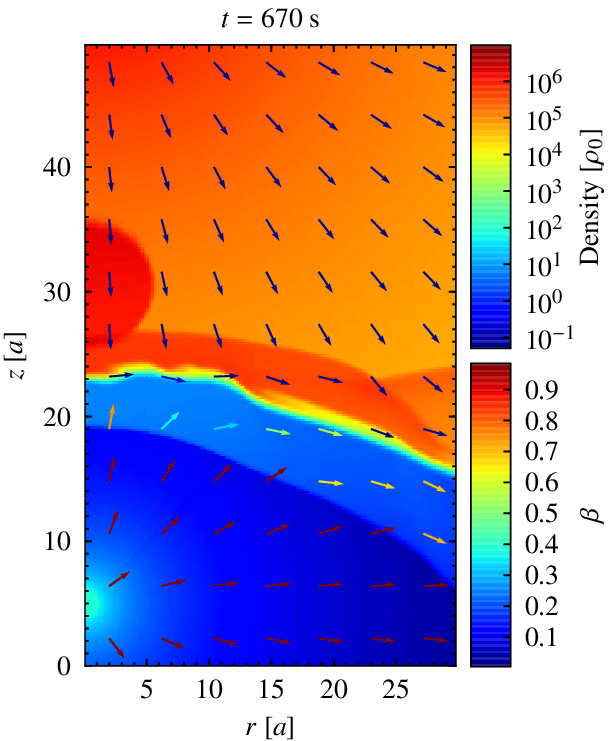}
\includegraphics[width=6.4cm]{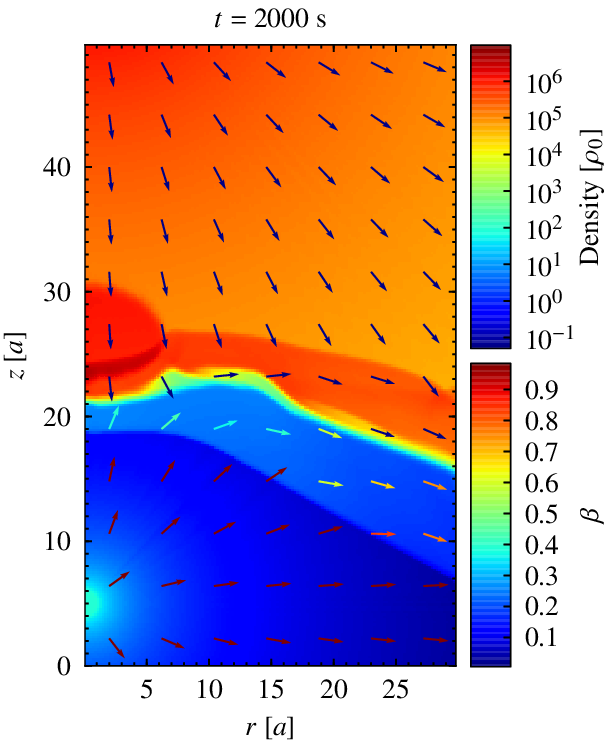}\\\vspace{0.2cm}
\includegraphics[width=6.4cm]{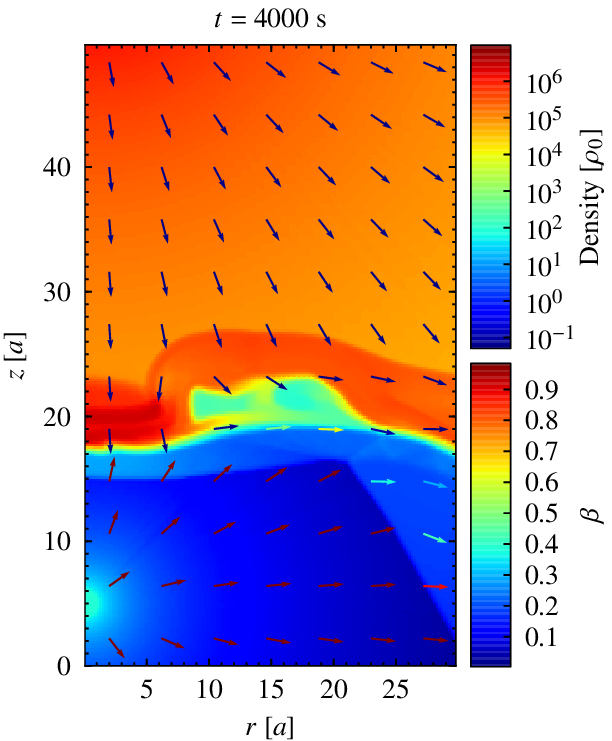}
\includegraphics[width=6.4cm]{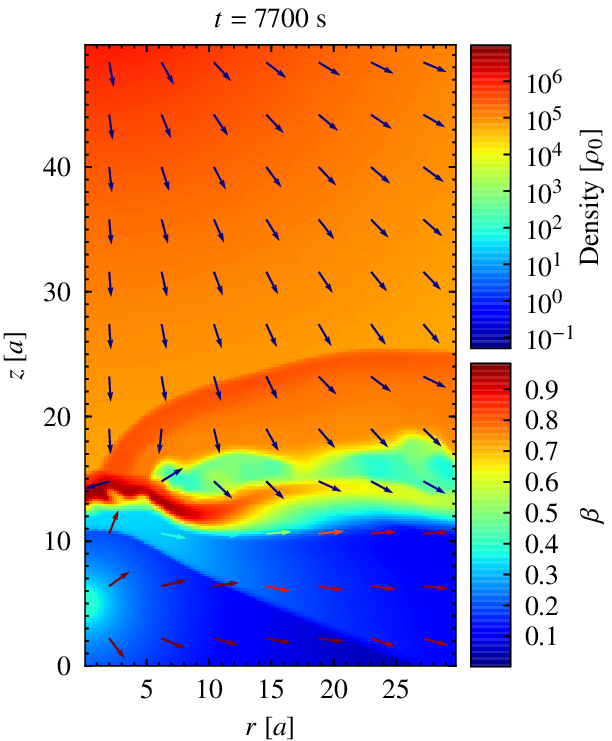}\\\vspace{0.2cm}
\includegraphics[width=6.4cm]{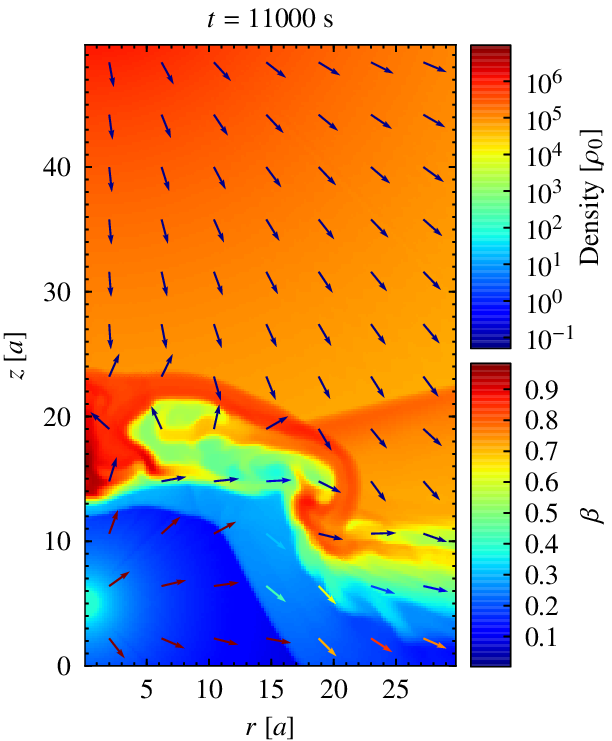}
\includegraphics[width=6.4cm]{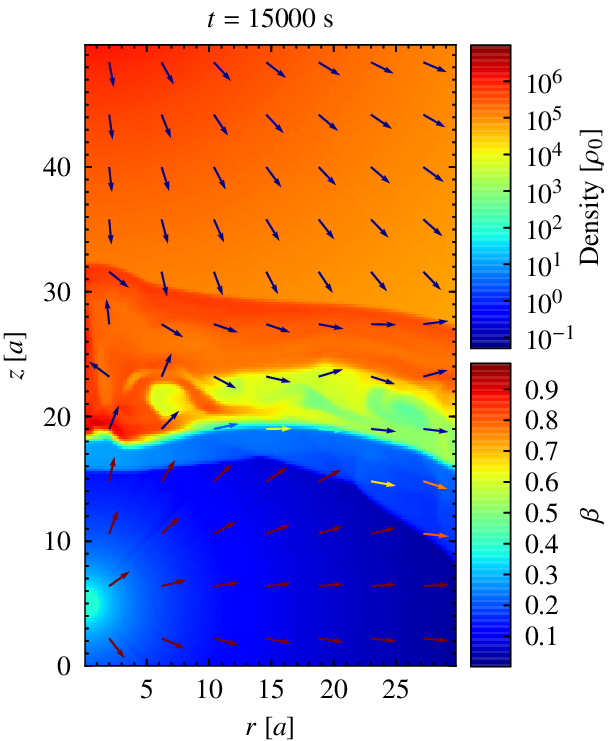}
\caption{
Density distribution by colour for the case with clump parameters
$\chi = 10$ and $R_{\rm c} =5~a$
for the times shown at the top of each plot.
The remaining plot properties are the same as those of Fig.~\ref{steady}.
}
\label{f10r5}
\end{figure*}

\begin{figure*}
\centering
\includegraphics[width=6.4cm]{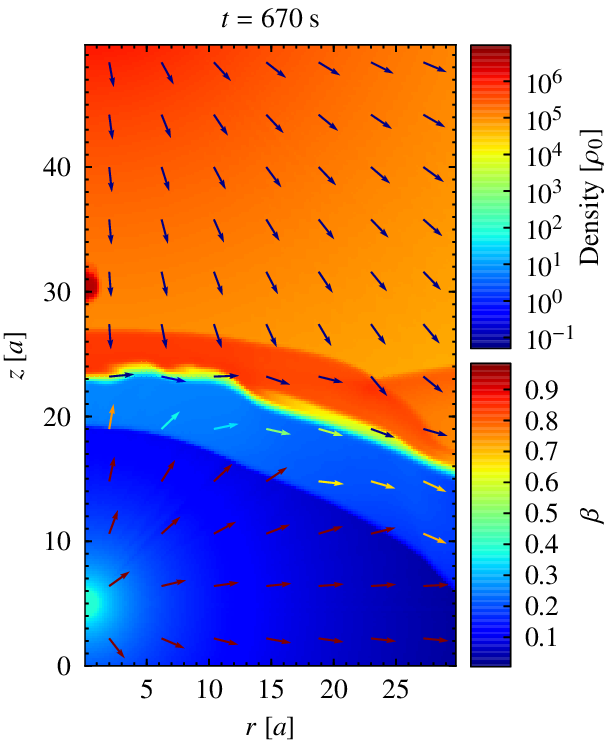}
\includegraphics[width=6.4cm]{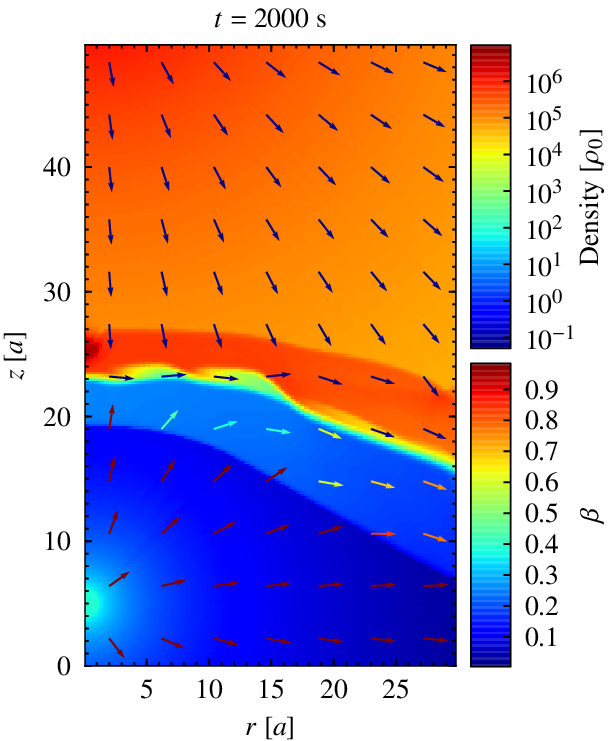}\\\vspace{0.2cm}
\includegraphics[width=6.4cm]{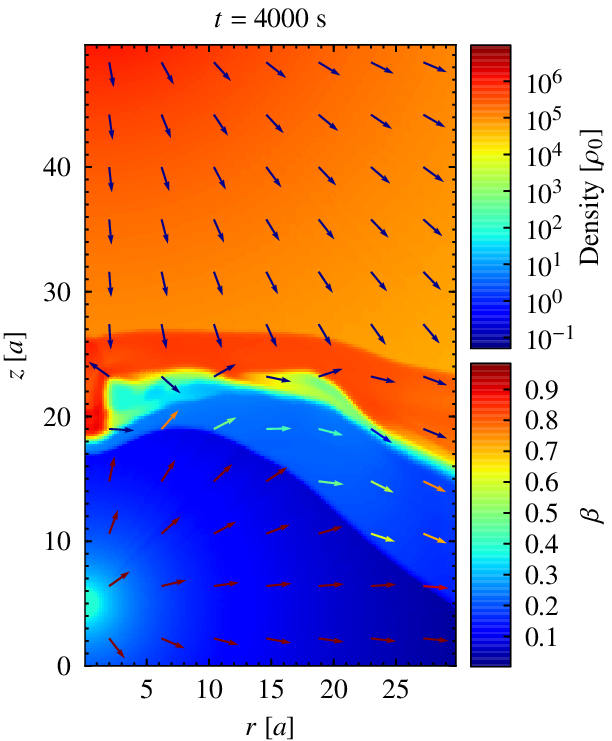}
\includegraphics[width=6.4cm]{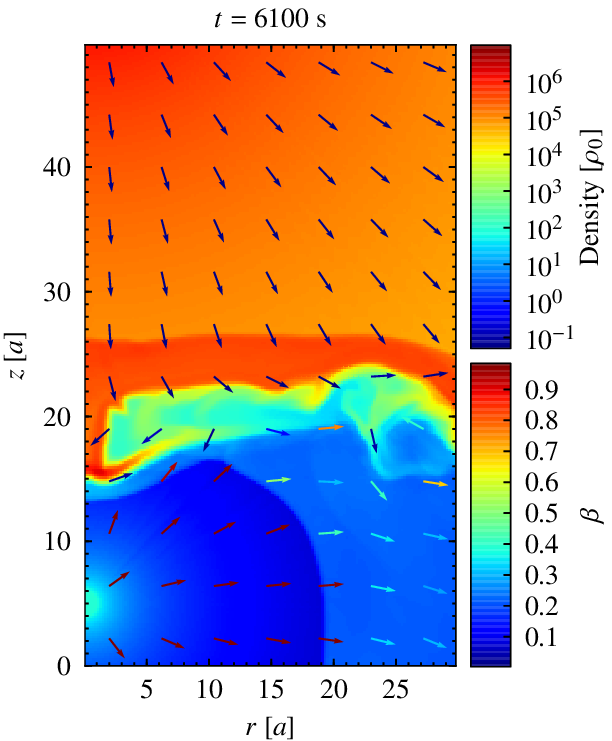}\\\vspace{0.2cm}
\includegraphics[width=6.4cm]{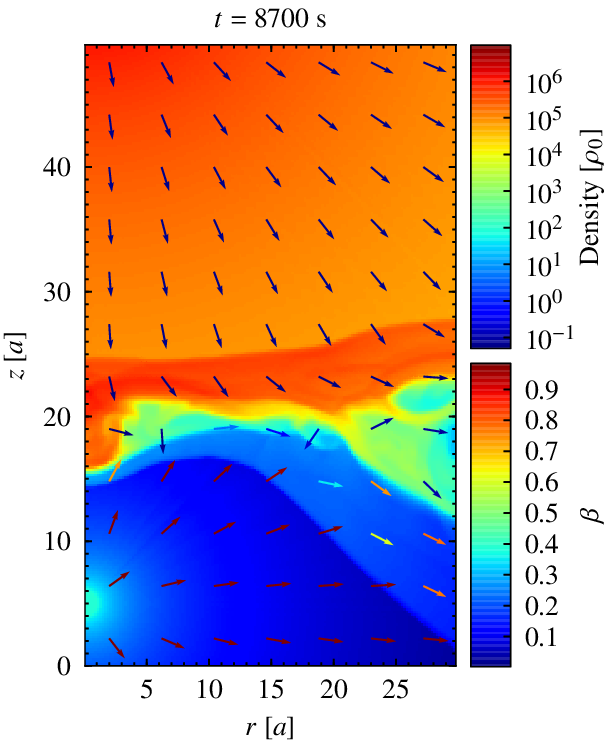}
\includegraphics[width=6.4cm]{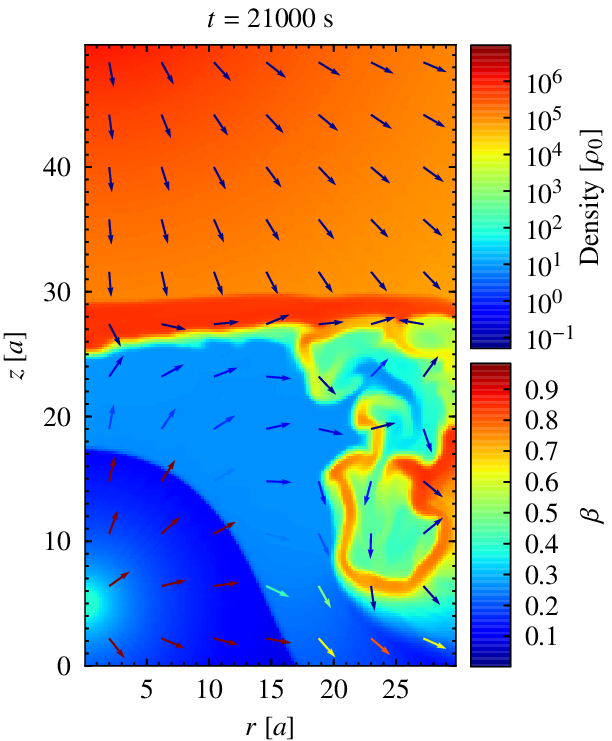}
\caption{
Density distribution by colour for the case with clump parameters
$\chi = 30$ and $R_{\rm c} =1~a$
for the times shown at the top of each plot.
The remaining plot properties are the same as those of Fig.~\ref{steady}.
}
\label{f30r1}
\end{figure*}

Figures \ref{f10r1}, \ref{f10r2p5}, \ref{f10r5}, and \ref{f30r1} show the density map evolution for the clumps characterized by 
$\chi = 10$ and $R_{\rm c} =1~a$, $2.5~a$, $5~a$, and $\chi = 30$ and $R_{\rm c} =1~a$, respectively. 
The times of the map snapshots were chosen such that the images are illustrative of the structure evolution. 
The bigger clumps, characterized by $\chi = 10$ and $R_{\rm c} =2.5~a$ and $R_{\rm c} =5~a$, 
and the smallest but densest one, with $\chi = 30$ and $R_{\rm c} =1~a$,
strongly perturb the two-wind interaction region, pushing the contact discontinuity to less than a half its initial distance (see Fig.~\ref{steady})
to the pulsar (see Figs.~\ref{f10r2p5}, \ref{f10r5}, and \ref{f30r1}, respectively).


\begin{figure*}
\centering
\includegraphics[height=4.4cm, trim= 0.0cm 0.9cm 0.0cm 0.0cm,clip]{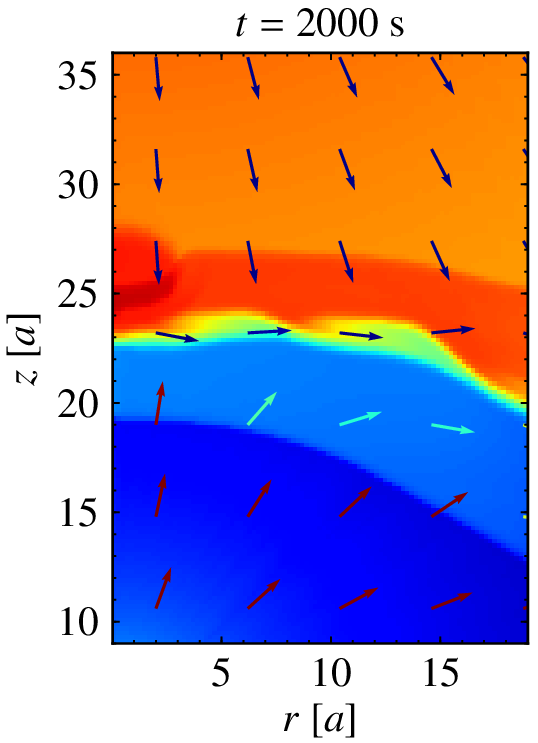}
\includegraphics[height=4.4cm, trim= 1.0cm 0.9cm 0.0cm 0.0cm,clip]{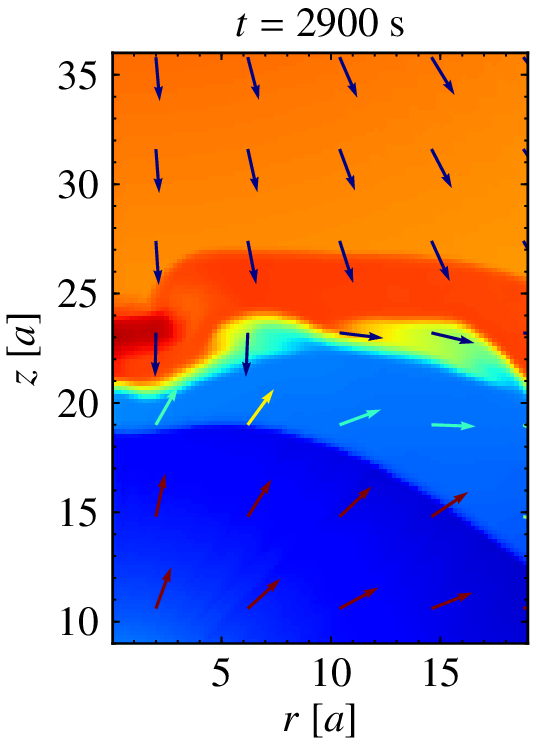}
\includegraphics[height=4.4cm, trim= 1.0cm 0.9cm 0.0cm 0.0cm,clip]{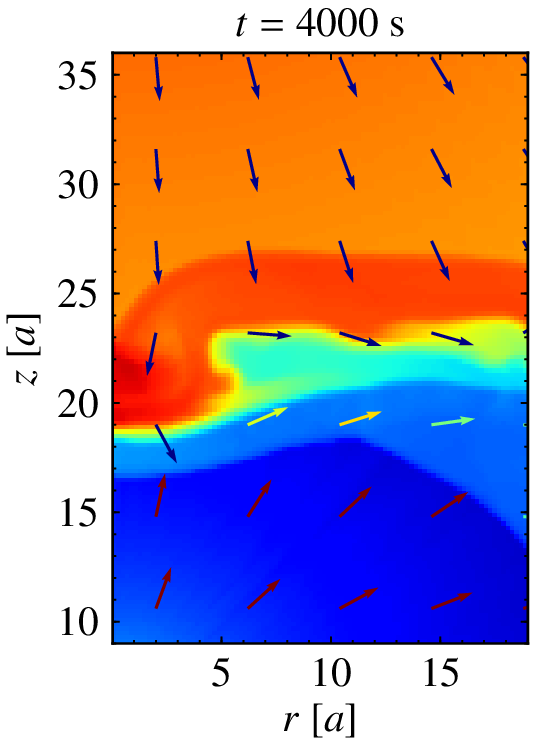}
\includegraphics[height=4.4cm, trim= 1.0cm 0.9cm 0.0cm 0.0cm,clip]{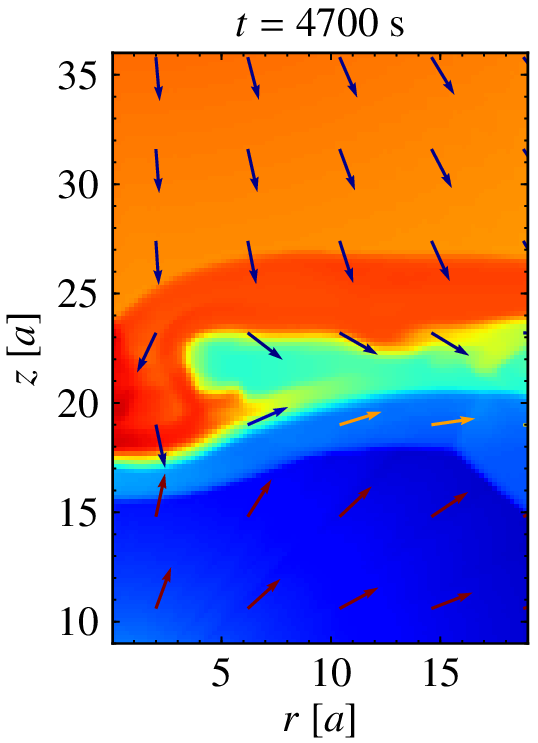}
\includegraphics[height=4.4cm, trim= 1.0cm 0.9cm 0.0cm 0.0cm,clip]{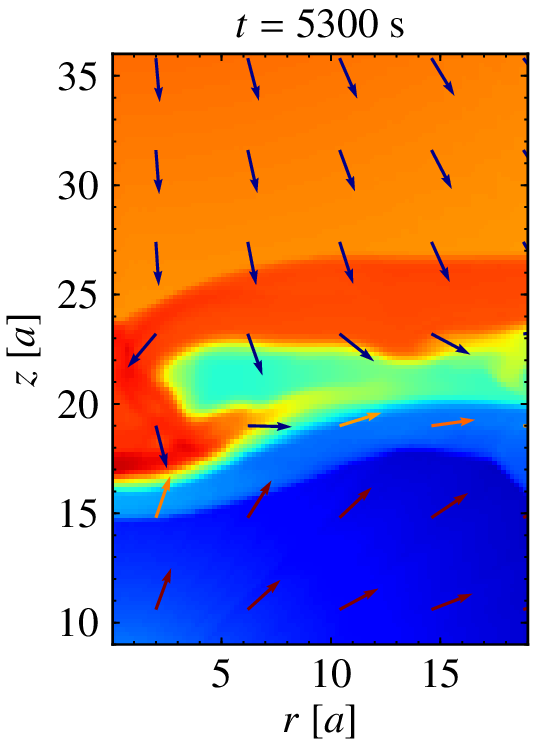}\\
\includegraphics[height=5.0cm, trim= 0.0cm 0.0cm 0.0cm 0.0cm,clip]{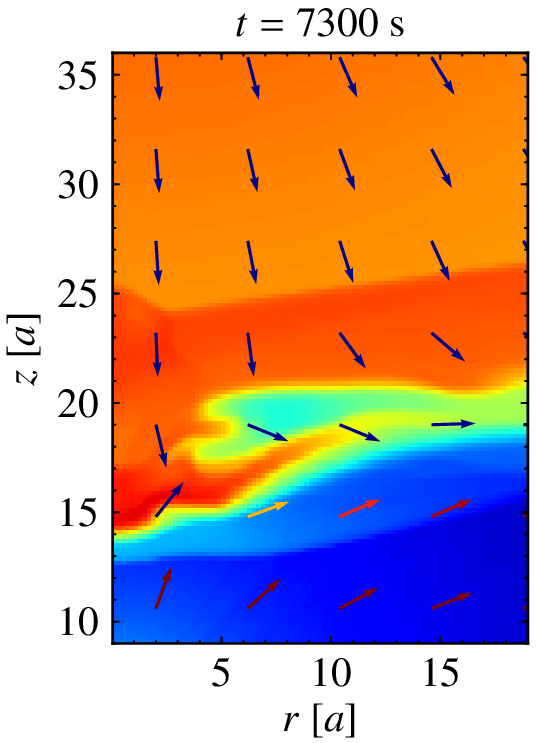}
\includegraphics[height=5.0cm, trim= 1.0cm 0.0cm 0.0cm 0.0cm,clip]{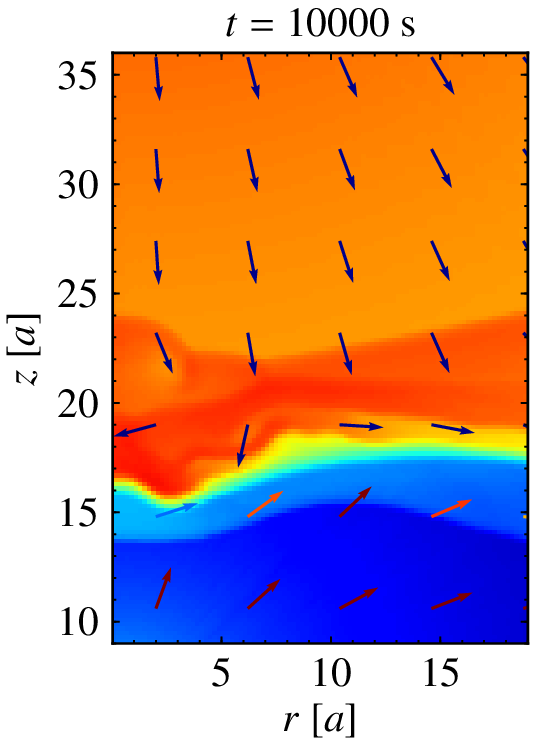}
\includegraphics[height=5.0cm, trim= 1.0cm 0.0cm 0.0cm 0.0cm,clip]{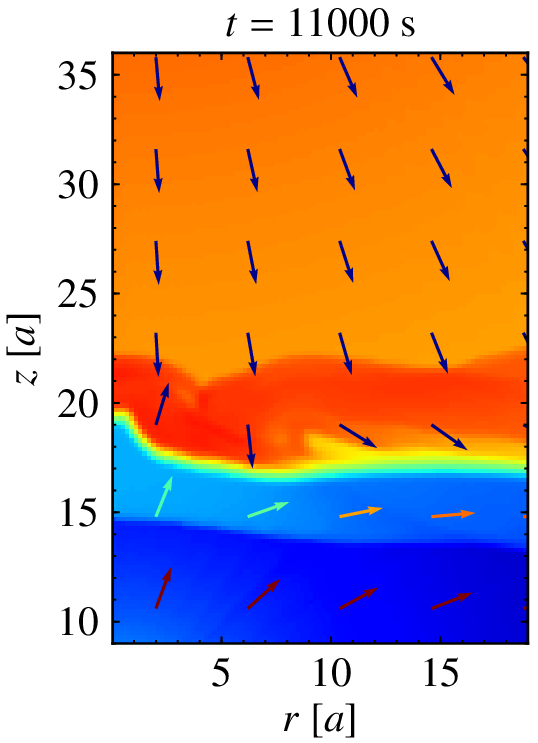}
\includegraphics[height=5.0cm, trim= 1.0cm 0.0cm 0.0cm 0.0cm,clip]{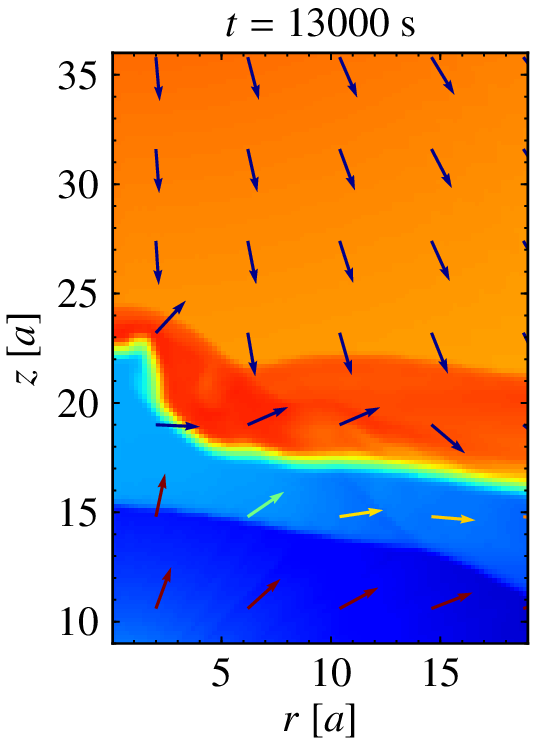}
\includegraphics[height=5.0cm, trim= 1.0cm 0.0cm 0.0cm 0.0cm,clip]{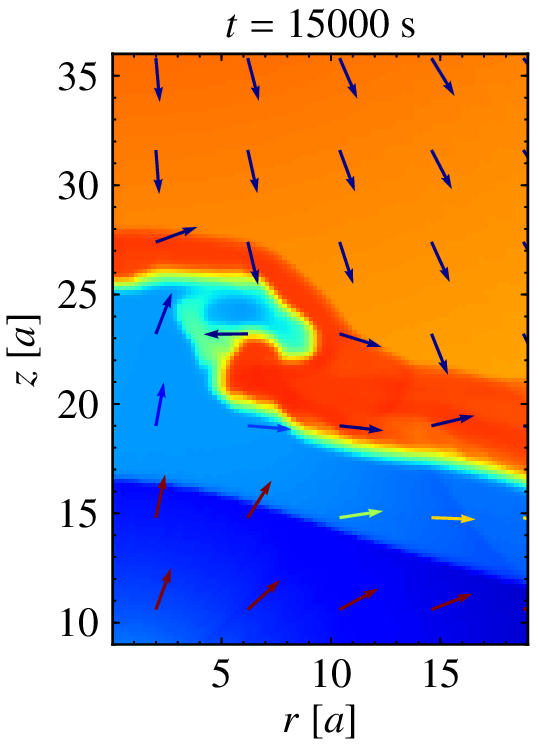}
\caption{
Zoom-in of the density distribution by colour for the case with clump parameters
$\chi = 10$ and $R_{\rm c} =2.5~a$
for the times shown at the top of each plot.
The remaining plot properties are the same as those of Fig.~\ref{f10r2p5}.
The pulsar and the star are located at $(r, z) = (0,5~a)$ and $(r, z) = (0,60~a)$, respectively.} 
\label{f10r2p5_zoom}
\end{figure*}

\begin{figure*}
\centering
\includegraphics[width=6.4cm]{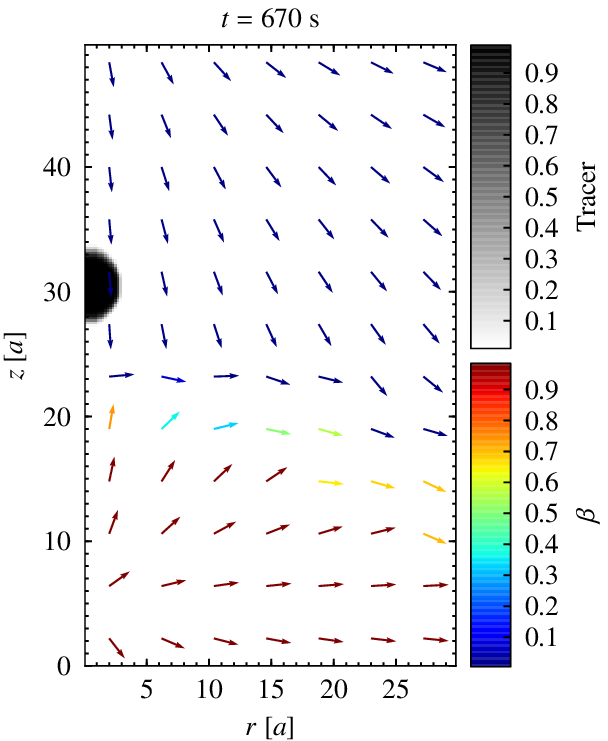}
\includegraphics[width=6.4cm]{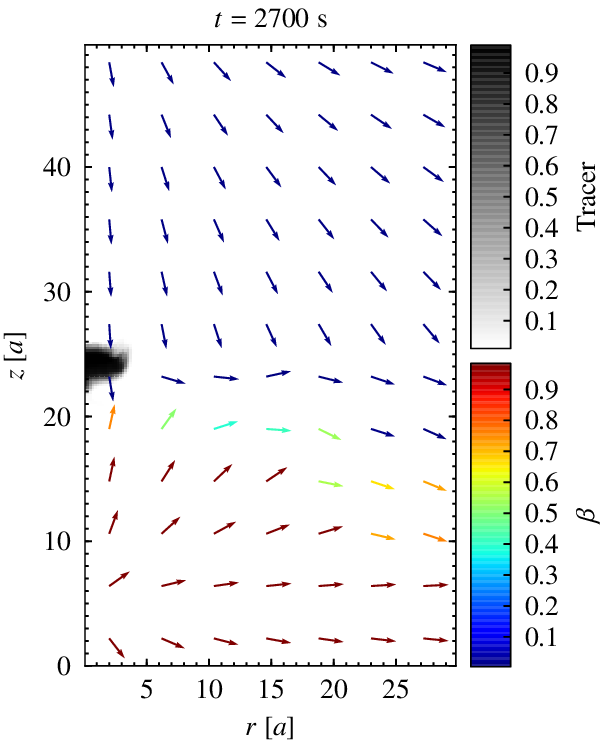}\\\vspace{0.2cm}
\includegraphics[width=6.4cm]{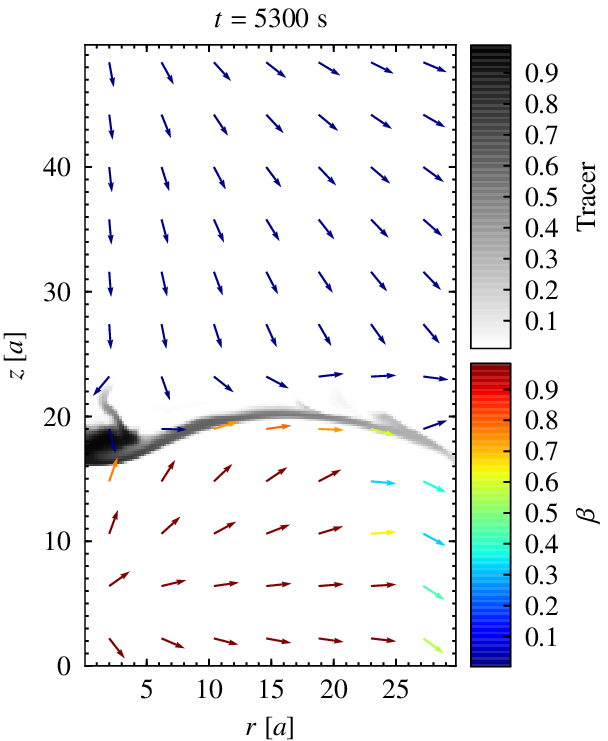}
\includegraphics[width=6.4cm]{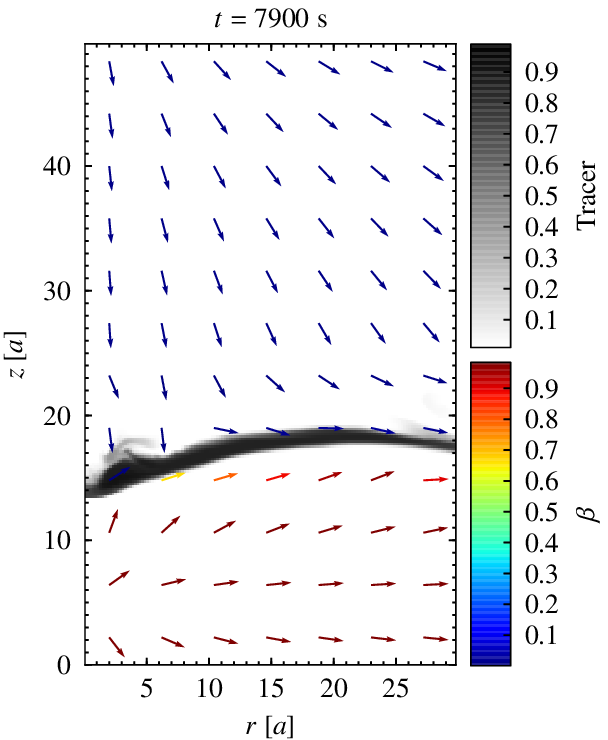}\\\vspace{0.2cm}
\includegraphics[width=6.4cm]{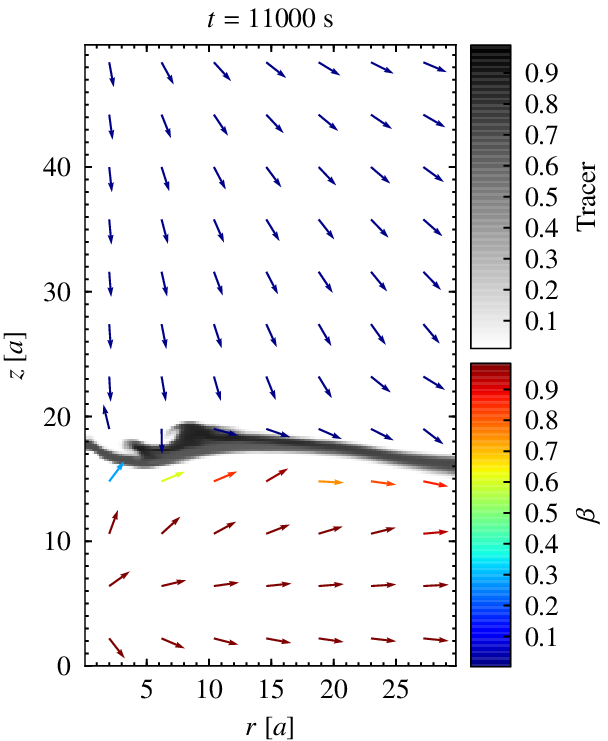}
\includegraphics[width=6.4cm]{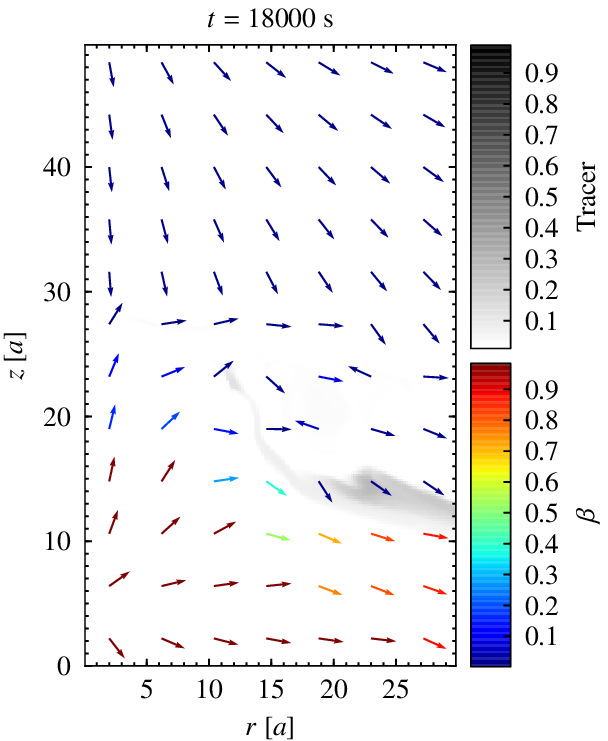}
\caption{Tracer distribution by colour for the case with clump parameters
$\chi = 10$ and $R_{\rm c} =2.5~a$
for the times shown at the top of each plot.
The tracer value ranges from 0 (pulsar and stellar wind) to 1 (clump).
The remaining plot properties are the same as those of Fig.~\ref{steady}.}
\label{f10r2p5_tracer}
\end{figure*}

\begin{figure*}
\centering
\includegraphics[width=6.5cm]{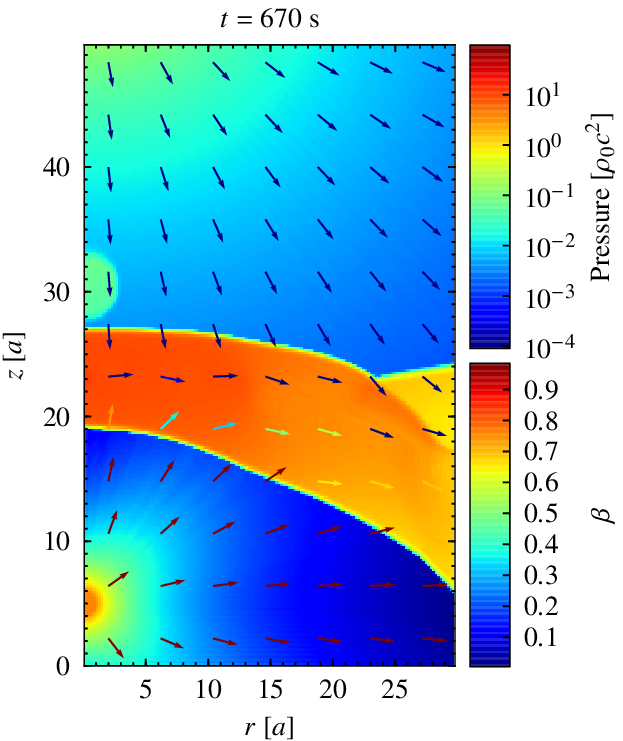}
\includegraphics[width=6.5cm]{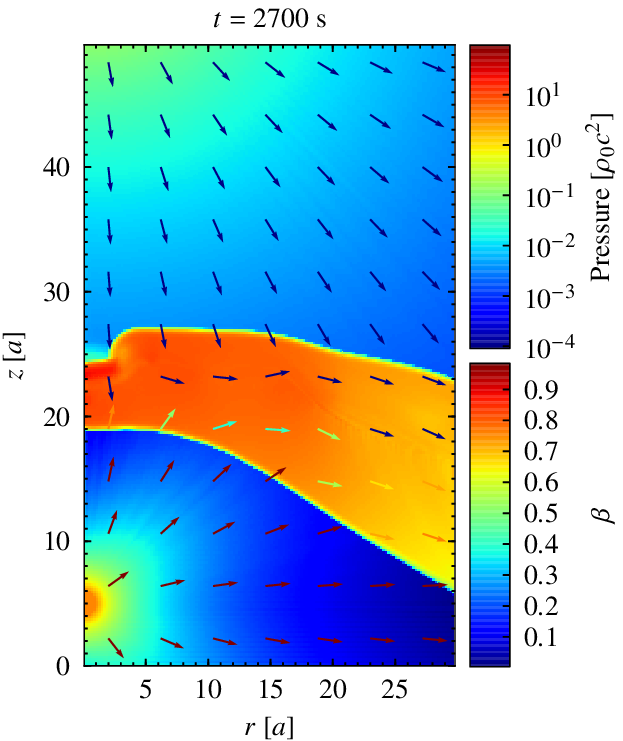}\\\vspace{0.2cm}
\includegraphics[width=6.5cm]{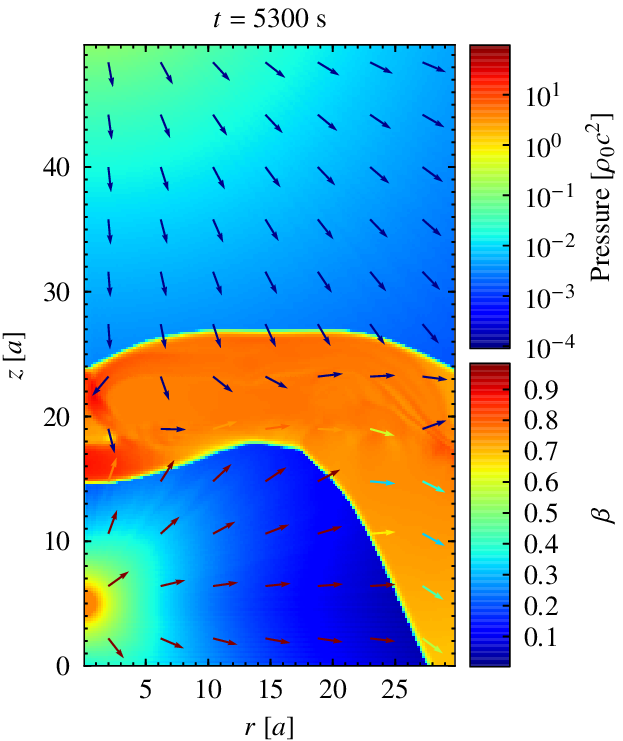}
\includegraphics[width=6.5cm]{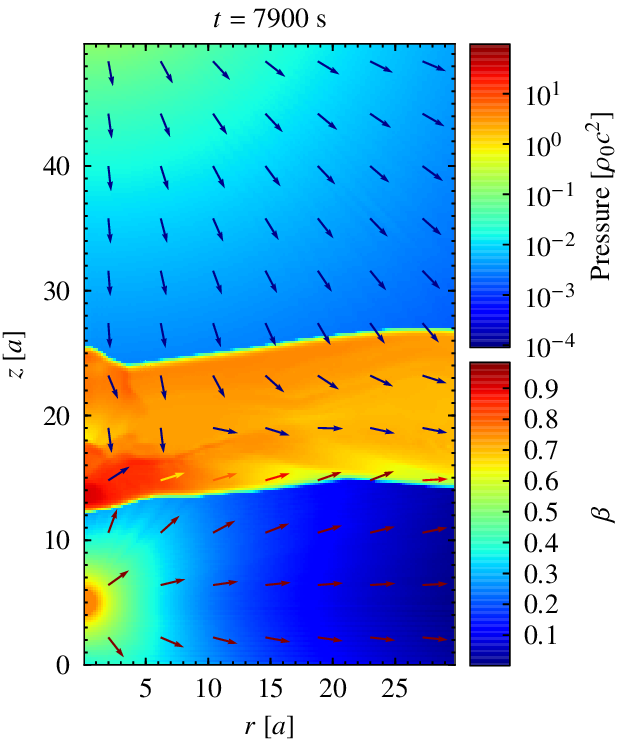}\\\vspace{0.2cm}
\includegraphics[width=6.5cm]{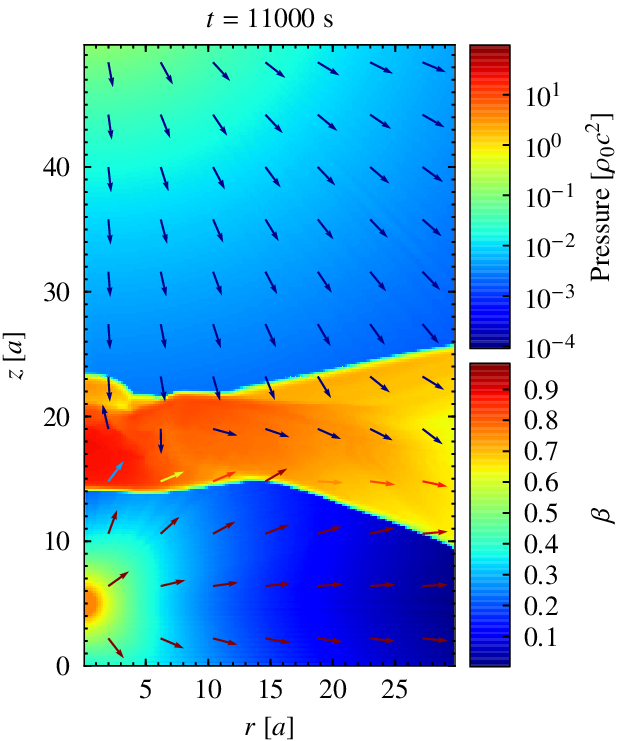}
\includegraphics[width=6.5cm]{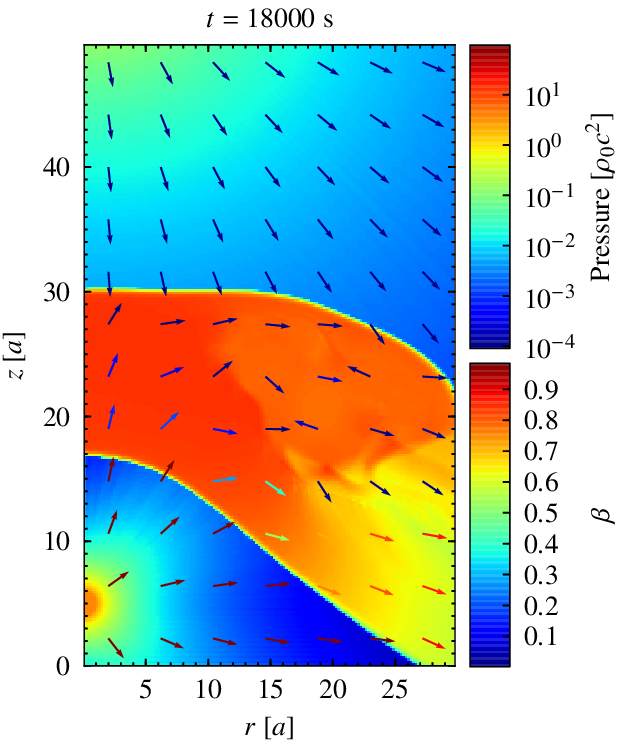}
\caption{Pressure distribution in units of $\rho_0{c^2}$ by colour for the case with clump parameters
$\chi = 10$ and $R_{\rm c} =2.5~a$
for the times shown at the top of each plot.
The remaining plot properties are the same as those of Fig.~\ref{steady}.}
\label{f10r2p5_pressure}
\end{figure*}

\begin{figure*}
\centering
\includegraphics[width=6.4cm]{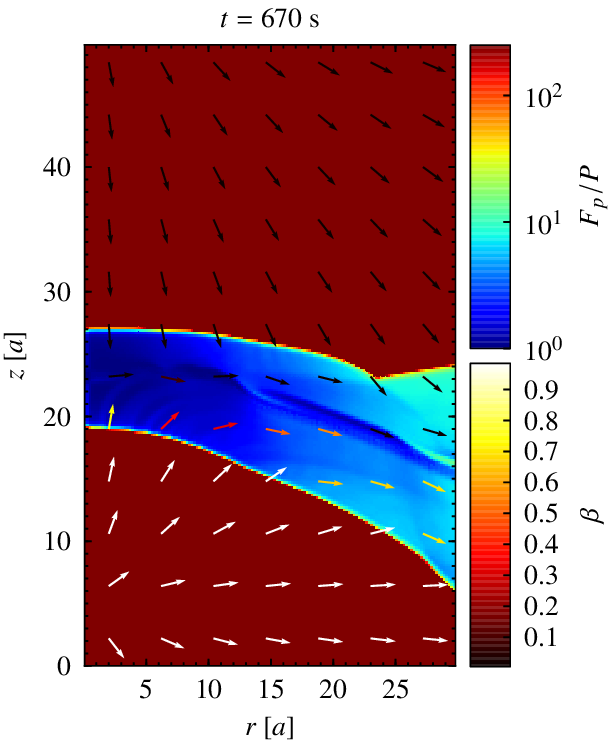}
\includegraphics[width=6.4cm]{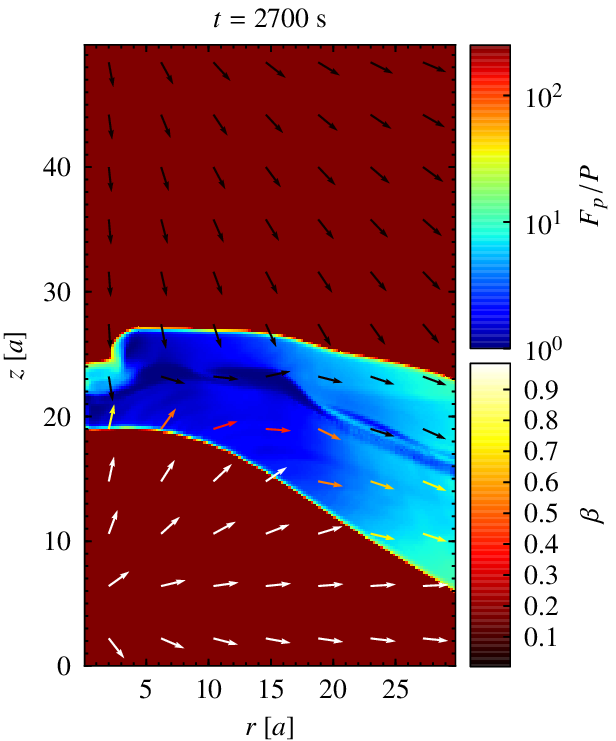}\\\vspace{0.2cm}
\includegraphics[width=6.4cm]{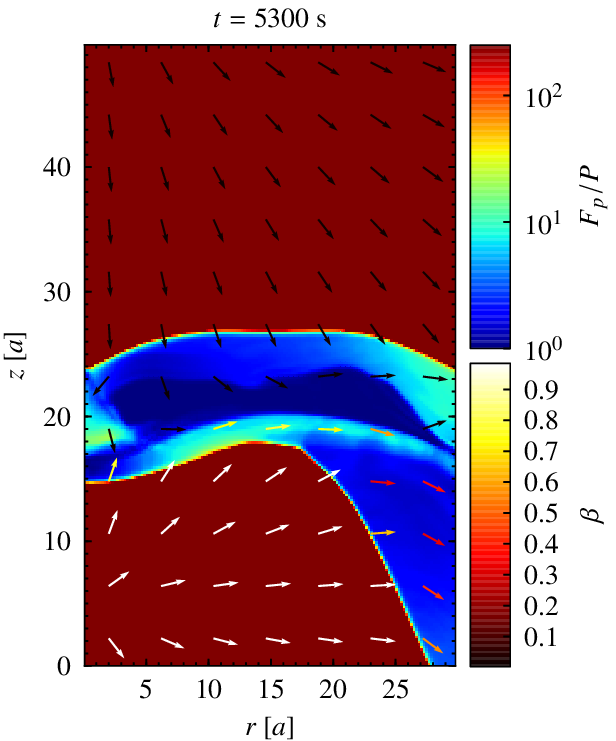}
\includegraphics[width=6.4cm]{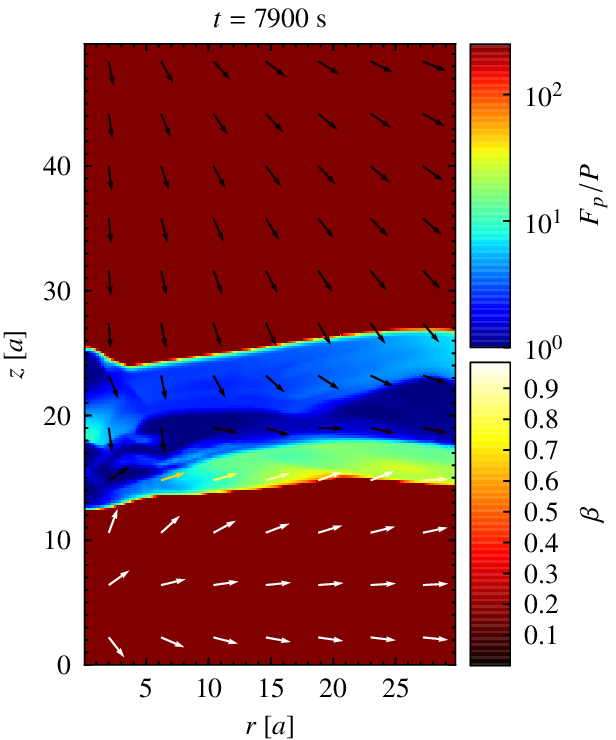}\\\vspace{0.2cm}
\includegraphics[width=6.4cm]{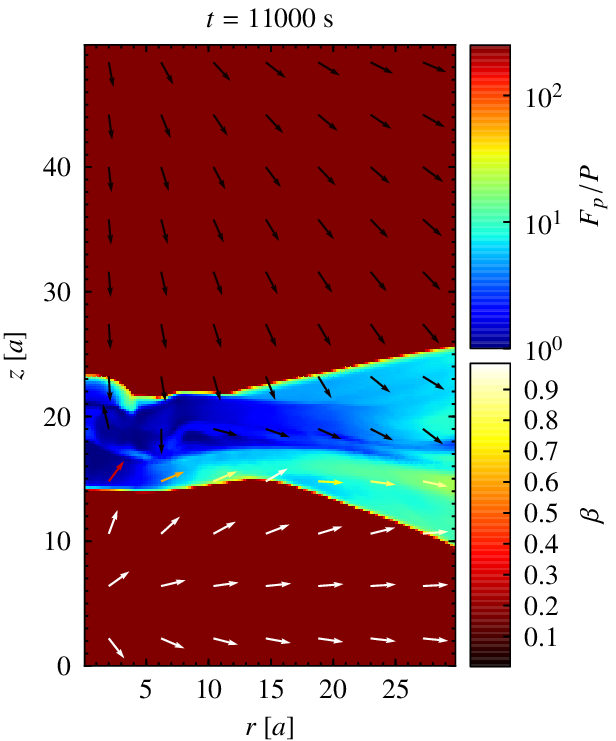}
\includegraphics[width=6.4cm]{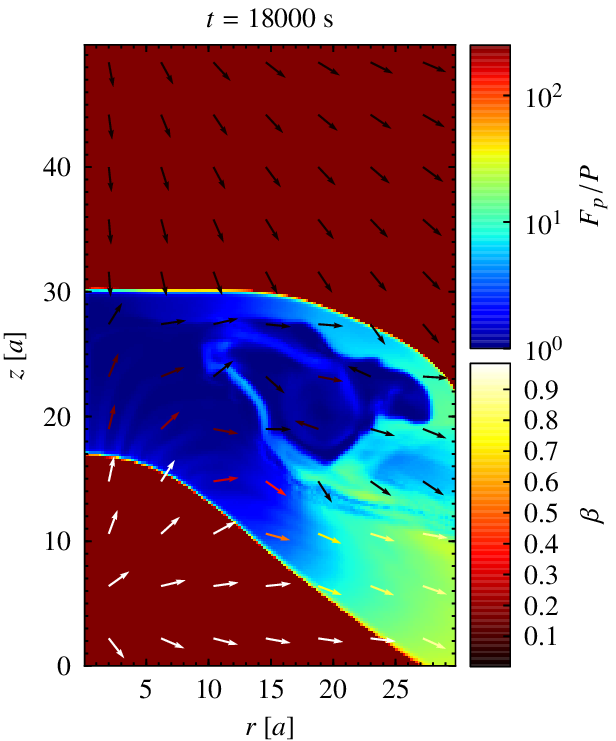}
\caption{Momentum flux over pressure distribution by colour for the case with clump parameters
$\chi = 10$ and $R_{\rm c} =2.5~a$
for the times shown at the top of each plot.
The momentum flux is given by $F_p = \rho~\Gamma^2 v^2 (1+\epsilon+P/{\rho})+P$.
The remaining plot properties are the same as those of Fig.~\ref{steady}.}
\label{f10r2p5_sonic}
\end{figure*}

\begin{figure*}
\centering
\includegraphics[width=6.4cm]{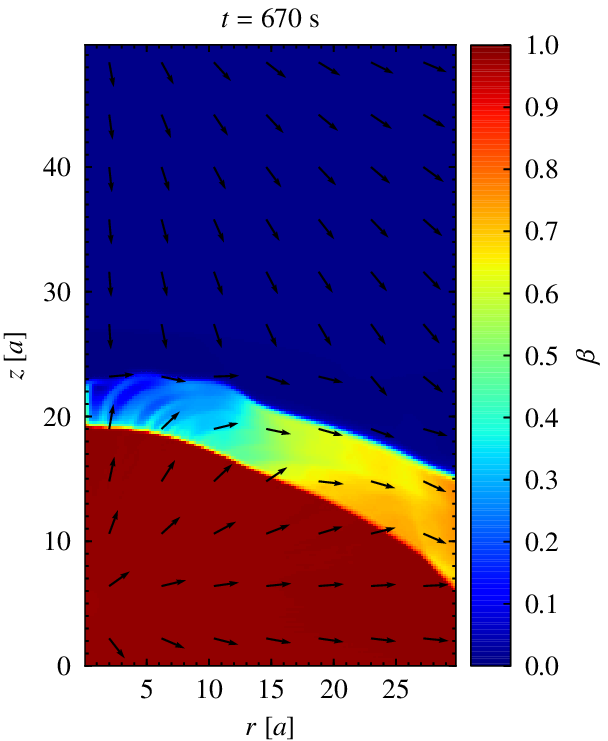}
\includegraphics[width=6.4cm]{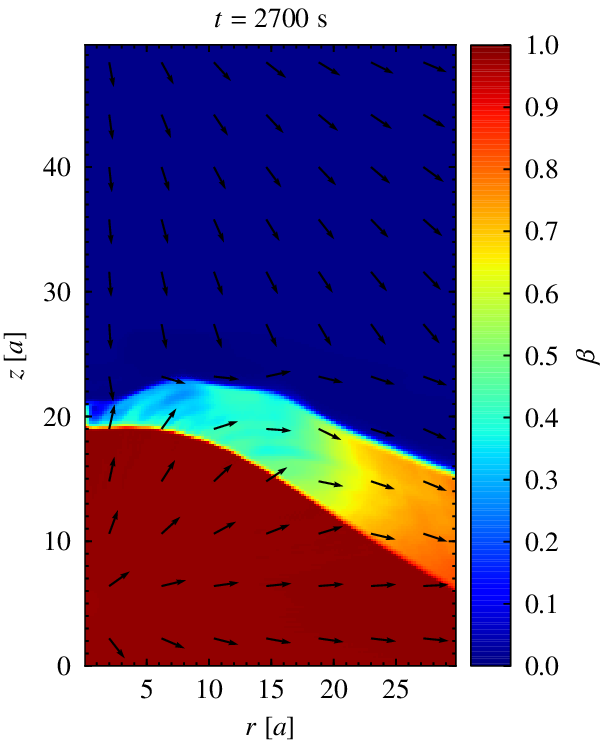}\\\vspace{0.2cm}
\includegraphics[width=6.4cm]{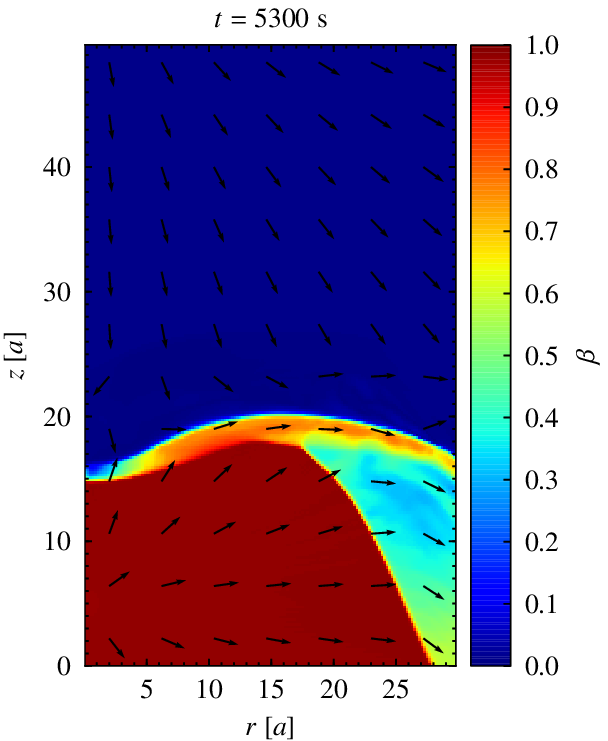}
\includegraphics[width=6.4cm]{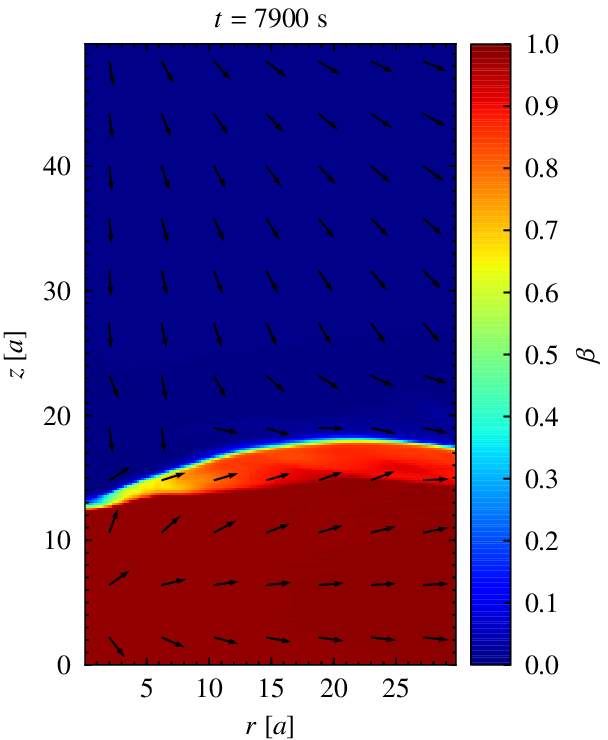}\\\vspace{0.2cm}
\includegraphics[width=6.4cm]{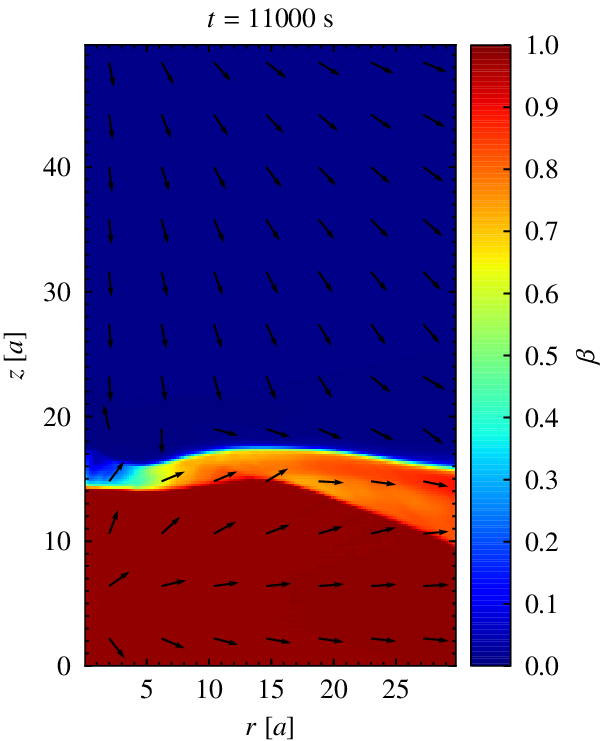}
\includegraphics[width=6.4cm]{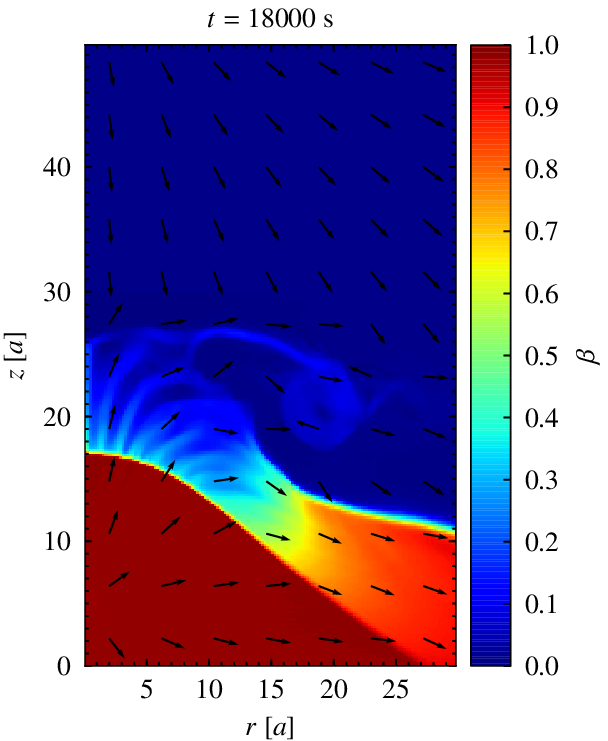}
\caption{$\beta$ distribution by colour for the case with clump parameters
$\chi = 10$ and $R_{\rm c} =2.5~a$
for the times shown at the top of each plot.
The remaining plot properties are the same as those of Fig.~\ref{steady}.}
\label{f10r2p5_W}
\end{figure*}

We now focus on a clump with $\chi = 10$ and $R_{\rm c} =2.5~a$, as it is illustrative of a strong impact by a clump of intermediate size. For this case, we also show Figs.~\ref{f10r2p5_zoom}, \ref{f10r2p5_tracer}, \ref{f10r2p5_pressure}, \ref{f10r2p5_sonic}, and \ref{f10r2p5_W}, which display the evolution of the clump in the density map, the tracer map (1 for the clump material and 0 otherwise), the pressure map, the map of the momentum flux over the pressure (greater than a few for a super-sonic flow), and the map of $\beta$. These figures show how the two-wind interaction region is pushed by the clump, until the contact discontinuity reaches a minimum distance to the pulsar at $R^{\prime}_{\rm num}\approx 8.5~a$  ($R^{\prime}_{\rm num}\approx 7.5~a$ for the termination shock). After that, the clump starts to be pushed backwards by the pulsar wind, is shocked, and decelerates, eventually disrupting, with its fragments driven away from the simulation axis by the shocked flow. All this is clearly seen in the density maps (Figs.~\ref{f10r2p5} and \ref{f10r2p5_zoom}), and in the tracer (Fig.~\ref{f10r2p5_tracer}). In addition, Figs.~\ref{f10r2p5_pressure}, \ref{f10r2p5_sonic} and \ref{f10r2p5_W} provide information on the presence of shocks, apparent as sudden increases in pressure or drops in the ratio of momentum-flux to pressure, and relevant for particle acceleration; flow re-acceleration, important for non-radiative cooling, flow relativistic motion, and flow-to-sound-speed relation; and flow speed and direction downstream of the shocks, determining Doppler effects on the flow emission. Equivalent images to Figs.~\ref{f10r2p5_zoom}, \ref{f10r2p5_tracer}, ~\ref{f10r2p5_pressure}, \ref{f10r2p5_sonic}, and \ref{f10r2p5_W} for the other three simulated clump cases are available in the on-line material. In all the simulations, a $t=0$ is assigned to the steady state plus a clump at the initial position.
%
%
%
%

Concerning stability, under the perturbation induced by the clump with $\chi = 10$ and $R_{\rm c} =5~a$, the system recovers steady state after the clump has been advected. In contrast, for the cases with $\chi = 10$ and $R_{\rm c} =2.5~a$,  
and $\chi = 30$ and $R_{\rm c} =1~a$, the instabilities eventually lead to the collapse of the two-wind interaction region, filling the region close to the pulsar with shocked pulsar wind. The clump with $R_{\rm c} =1~a$ and $\chi = 10$ only slightly perturbs the two-wind interaction region and pushes the contact discontinuity to $\sim 2/3$ of its initial distance (see Fig.~\ref{steady}) to the pulsar (see Fig.~\ref{f10r1}). Afterwards, the system recovers the (quasi-)steady state. We recall that although the steady-state solution is only metastable in the context of our simulations, the main features of the clump phase should be reliable because the clump represents a dominant perturbation over any other, physical or numerical, before it is assimilated by the two-wind flow.

Other simulations, not shown here, were carried out for clumps with $R_{\rm c}=$ 0.5--5 and $\chi=$10--100. As seen before, the impact on the two-wind interacting region is larger the denser the clump or the larger is its radius, so different combinations of these parameters yield results in line with those shown. 

\begin{figure}
\resizebox{\hsize}{!}{\includegraphics{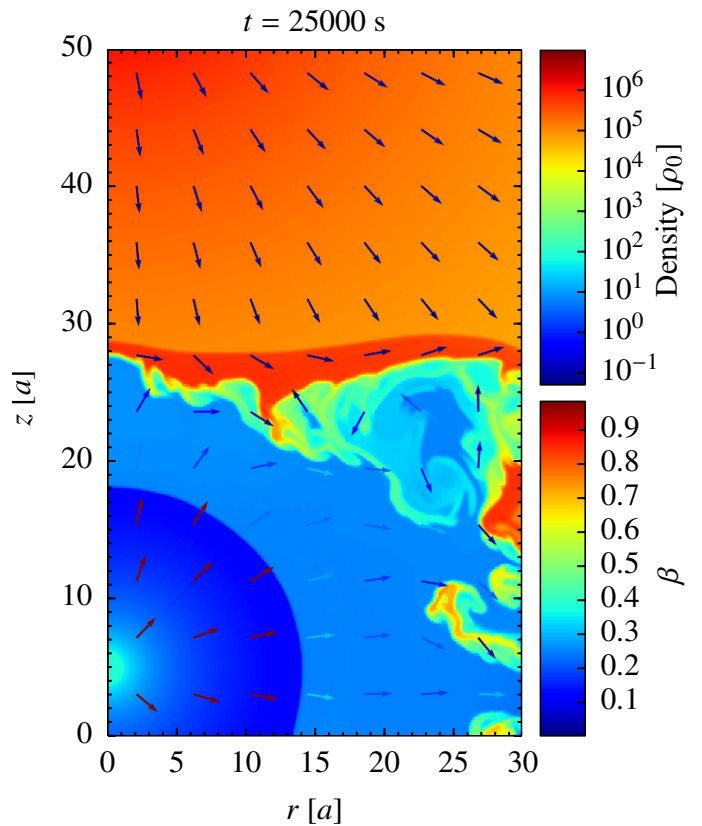}}
\caption{Density distribution by colour
at time $t = 25000$ s for the simulation with a resolution 
of $300\times500$ cells. The remaining plot properties are the same as those of Fig.~\ref{steady}.
}
\label{steady_x2}
\end{figure}

\begin{figure*}
\centering
\includegraphics[width=6.4cm]{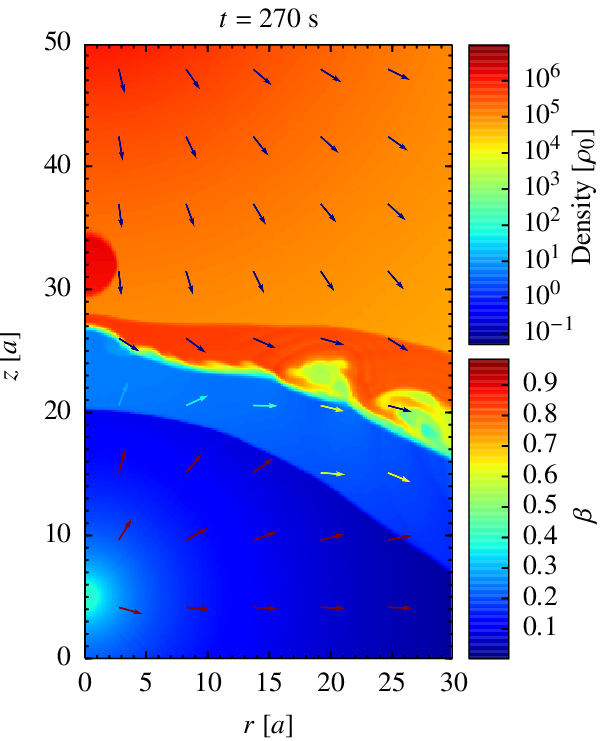}
\includegraphics[width=6.4cm]{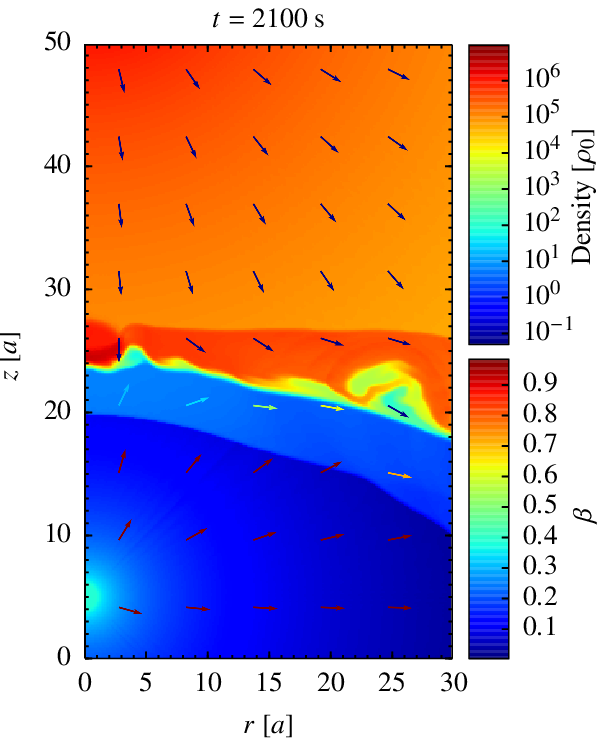}\\\vspace{0.2cm}
\includegraphics[width=6.4cm]{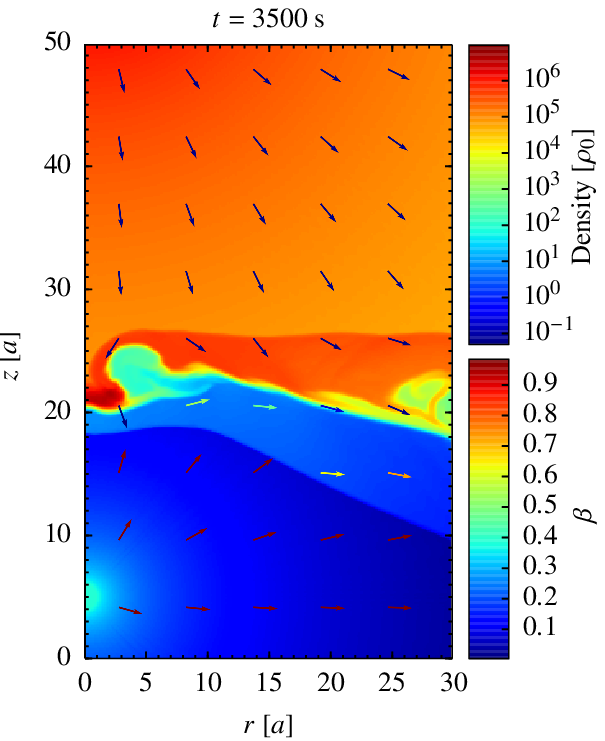}
\includegraphics[width=6.4cm]{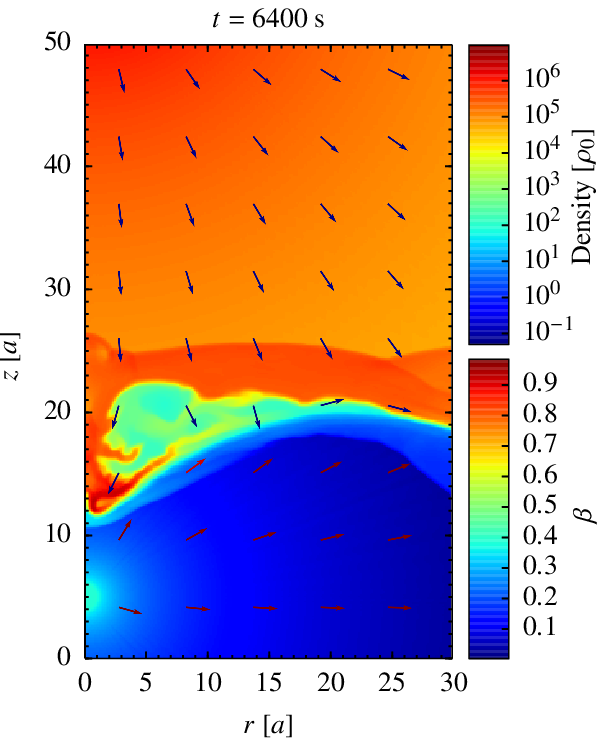}\\\vspace{0.2cm}
\includegraphics[width=6.4cm]{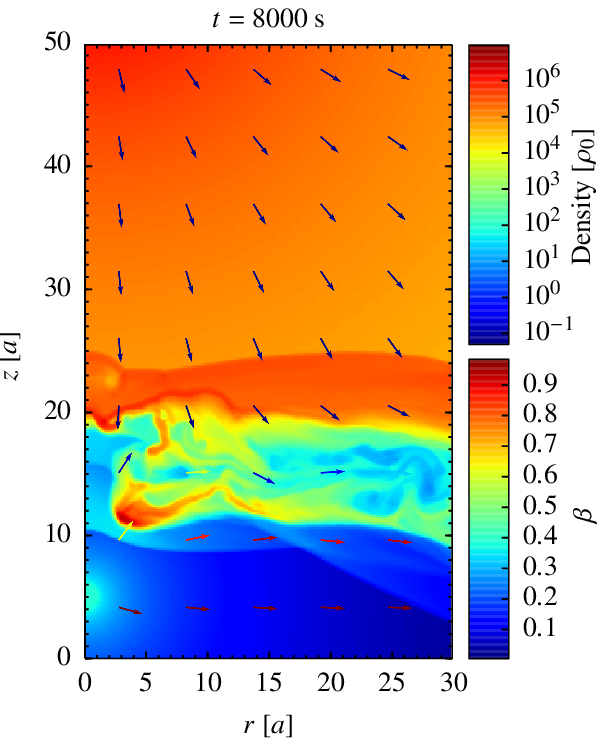}
\includegraphics[width=6.4cm]{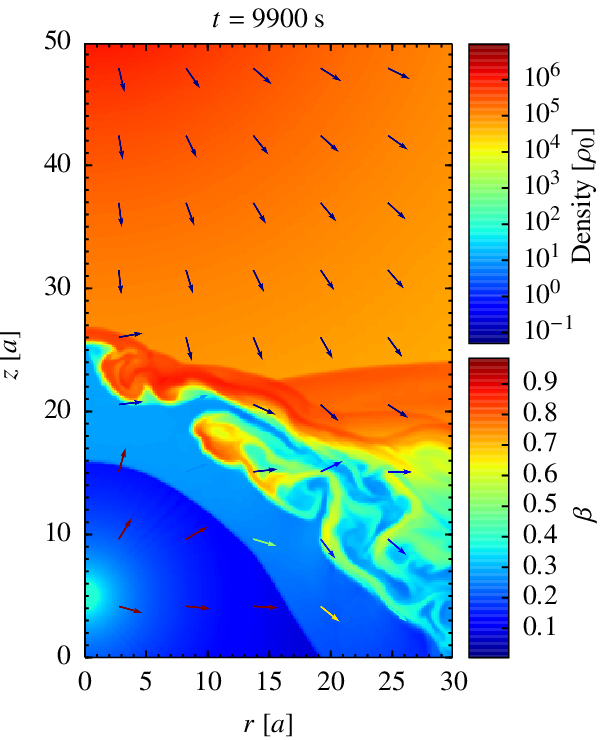}
\caption{
Density distribution by colour for the case with clump parameters
$\chi = 10$ and $R_{\rm c} =2.5~a$ and a resolution of
$225\times375$ cells
for the times shown at the top of each plot.
The remaining plot properties are the same as those of Fig.~\ref{steady}.
}
\label{f10r2p5_x1p5}
\end{figure*}

To show the impact of increasing resolution, we present the results of two simulations with the same set-up but, a resolution 2 and 1.5 times (i.e. $300\times500$ and $225\times375$ cells, respectively) higher than the resolution adopted for most of the simulations in the paper. For the highest resolution case, we were unable to reach steady state (without clump) because the higher resolution allowed the rapid development of instabilities already within the grid, leading to the collapse of the two-wind interaction region. A temporal state of this simulation taken shortly before the collapse of the interaction region is shown in Fig.~\ref{steady_x2}. For the intermediate-resolution case, we reached (quasi-)steady state, but compared with the lower resolution simulations, it presented a more unstable pattern in the two-wind interaction region away from the simulation axis and a more compressed two-wind interaction region close to the axis singularity. A sequence of images showing the evolution of a clump characterized by $\chi = 10$ and $R_{\rm c} =2.5~a$ for the intermediate resolution simulation is shown in Fig.\ref{f10r2p5_x1p5}. The higher resolution allowed the development of denser small-scale structures that pushed the pulsar wind termination shock closer to the neutron star. However, the general behaviour of the higher and lower resolution clump simulations is similar, which suggests that the main features resulting from the lowest resolution simulations, albeit smoother, are reliable.

\section{Discussion}
\label{disc}

The interaction of a relativistic pulsar wind with the non-relativistic wind of a massive star was simulated through relativistic axisymmetric hydrodynamic calculations. The two-wind interaction structure reaches a (quasi-)steady state, although the solution is metastable and quite sensitive to the initial set-up parameters, such as the grid size or the wind density contrast. Therefore, even though the collision between the two winds forms a structure similar to those previously found in axisymmetric relativistic simulations, it shows some irregularity (Fig.~\ref{steady}). In the dynamical problem solved here, the axisymmetric geometry may introduce numerical perturbations to the physical solution that grow affected by the RT instability close to the axis, and by the KH instability farther away. However, similar irregularities have been also found in relativistic 2D simulations in planar geometry that have been attributed to the development of KH instabilities \citep{Bosch-Ramon2012,Lamberts2013}, which implies that the irregularities found in this work are not only related to the axisymmetric geometry adopted.

The growth of instabilities strongly affects the shocked pulsar-wind region in its (quasi-)steady state. Given the very high sound speed, this region can change within time intervals significantly shorter than the dynamical time of the simulation, $\sim d/v_{\rm sw}$, dominated by the slow stellar wind. Nevertheless, we find that the interaction region globally follows the expected geometry both for the contact discontinuity and for the width of the two-wind collision region \citep{Bogovalov2008,Bogovalov2012}. 

The arrival of clumps can have a very strong impact on the whole interaction structure, overcoming the effect of any possible numerical perturbation related to the geometry of the calculations. The clumps trigger violent RT and KH instabilities because the structure is prone to suffer them, and thus the interaction between the clump and the two-wind collision region is highly non-linear, leading not only to a {\it secular} modification of the global geometry, but also to quick changes of the shocked pulsar-wind region. This is apparent in all the  map time sequences (in particular in the zoomed-in sequence: Fig.~\ref{f10r2p5_zoom}), as the largest variations affect the shocked pulsar wind.

Below we discuss a few relevant points: the comparison between the analytical and the numerical approximation; the clump effect on the global structure and radiation, with a mention of the GeV flare of PSR~B1259$-$63 as a possible instance of matter clump-perturbation of the two-wind interaction region, and work under development.

\subsection{Numerical results vs analytical estimates}

An analytical estimate of the minimum distance between the two-wind interaction region and the pulsar, after the clump has penetrated the shocked pulsar wind, has been presented in Sect.~\ref{frame}. For comparison, the analytical and numerical values of the minimum distances from the contact discontinuity, and the termination shock, to the pulsar, are shown in Table~\ref{table_clumps} for all the simulated clumps.

The final distance between the pulsar and the contact discontinuity, for a given density contrast and a clump radius, computed using Eq.~(\ref{equ}), agree well with the numerical results except for the clump with radius $R_{\rm c}=5~a$ (see Table~\ref{table_clumps}). The reason is that such a large clump is already outside of the application range of Eq.~(\ref{equ}), which strictly applies only to clumps with $R_{\rm c}\ll R^{\prime}$. For $R_{\rm c} \rightarrow R^\prime$ or bigger, the clump behaves more as an homogeneous wind than as a discrete obstacle, but $\chi$ times denser than the average stellar wind. In this case, the minimum distance from the contact discontinuity to the pulsar can be computed 
from Eq.~(\ref{eta}) and assuming momentum flux equality
as $R^{\prime}=\eta_{\chi}^{1/2}~{\rm d}/(1+\eta_{\chi}^{1/2})\simeq\eta_{\chi}^{1/2}~{\rm d}$, with $\eta_{\chi} = \chi^{-1} \eta$. For $\chi=10$ it gives $R^{\prime}_{\rm an}\sim 7.9~a$, which agrees well with the numerical result for the clump with $R_{\rm c}=5~a$ ($R^{\prime}_{\rm num}\approx 7.5~a$).

\begin{table}
\caption{Clump impact on the size of the two-wind interaction region.}
\label{table_clumps}
\centering
\begin{tabular}{c c c c c}
\hline\hline
$\chi$ & $R_{\rm c}$ (a) & $R_{\rm num}^{\prime}$ (a)& $R_{\rm an}^{\prime}$ (a)& $R_{\rm num}^{\prime~({\rm TS})} (a)$\\
\hline
10 & 1 &   13 & 14.8 &  11 \\
10 & 2.5 &  8.5 &   10 & 7.5 \\
10 & 5 &  7.5 &  2.2 &   6 \\
30 & 1 &   10 & 12.5 &   8 \\
\hline
\end{tabular}
\tablefoot{
Density contrast $\chi$, clump radius $R_{\rm c}$, 
numerical minimum distance with respect to the contact discontinuity $R_{\rm num}^{\prime}$ (analytical: $R_{\rm an}^{\prime}$), and
numerical minimum distance with respect to the termination shock $R_{\rm num}^{\prime~({\rm TS})}$.
}
\end{table}

The time required for the clump with $\chi=10$ and $R_{\rm c}=2.5~a$ to become shocked is $\sim 4000$~s, and for full clump disruption, meaning that when the clump material is advected away already integrated in the shocked flows,
the time required is $\sim 10000$--15000~s (see Fig.~\ref{f10r2p5_zoom}). These times are consistent with the analytical value of the clump lifetime, $\sim \chi^{1/2}R_{\rm c}/v_{\rm sw}\approx 2000$~s \citep{Bosch-Ramon2013}, or 4000~s, if the clump diameter is adopted as the characteristic clump size.

\subsection{Clump effects on the global structure and radiation}

On the scales simulated in this work, stellar winds with a modest inhomogeneity degree ($\chi = 10$ and $R_{\rm c} = 1~a$) present non-negligible variations of the interaction structure, and even smaller/lighter clumps can provide continuous perturbations for the development of instabilities in the contact discontinuity. In addition, medium-sized or denser clumps ($R_{\rm c} = 2.5$, $5~a$ and $\chi=10$; $R_{\rm c} = 1~a$ and $\chi=30$) can lead to strong variations in the size of the two-wind interaction structure. Both small and medium-sized/denser clumps can generate quick and global variations in the shocked pulsar wind. This affects the location of the pulsar wind termination shock (i) and also introduces seeds for small-scale relativistic and transonic turbulence that would grow downstream of this shock (ii).

(i) The consequences of large variations in the termination shock location have previously been  discussed in \cite{Bosch-Ramon2013}. In short, they can induce variations in the cooling, radiative as well as non-radiative, channels, and non-linear radiation processes, through synchrotron self-Compton or internal pair creation, for large reductions of the emitter size. 

(ii) The relativistic flow variations on small spatial and temporal scales downstream of the pulsar wind shock, already apparent despite the modest resolution in Fig.~\ref{f10r2p5_W}, would lead to a complex radiative pattern in time and direction. This complex radiative pattern is caused by Doppler boosting because of the complex orientation of the fluid lines and the relativistic speeds achieved through re-acceleration of the shocked pulsar wind \citep[see][for a study of the impact of Doppler boosting on radiation]{Khangulyan2014}. 

Finally, weak shocks are present in the shocked pulsar wind (e.g., Fig.~\ref{f10r2p5_sonic}), which suggests that further particle acceleration, additional to that occurring in the pulsar wind termination shock, could take place already deep inside the binary system, well before the postshock flow has been affected by the orbital motion \citep[see, e.g.,][for a discussion of the larger-scale evolution of the shocked structure]{Bosch-Ramon2012}.

\subsubsection{The flare in PSR~B1259$-$63}\label{psr}

The flare in PSR~B1259$-$63, observed by Fermi about 30 days after periastron passage \citep{Abdo2011}, has a potential connection with the Be disc through the disruption and fragmentation of the latter. This may have led to the impact of a dense piece of disc on the two-wind interaction structure, largely reducing the size of the pulsar wind termination shock. This reduction could allow efficient Compton scattering by a population of GeV electrons on local X-ray photons, as proposed in \cite{Dubus2013b} \citep[see also][for a similar proposal involving infrared photons]{Khangulyan2012} as a result of the strong enhancement of the X-ray photon density. This idea is worth to be studied in more detail, although it remains unclear why such a modification of the GeV emitter has no clear effects at other wavelengths \citep{Chernyakova2014}.

It is worth estimating what fraction of the mass of a Be stellar disc the simulated clumps would represent. The mass of a typical Be disc can be derived from the disc mass-loss rate, $\sim 10^{-12}$--10$^{-9}\,M_\odot$~yr$^{-1}$, times the disc extension, say $\sim 1$~AU (quantities similar to those adopted in \citealt{Okazaki2011} and \citealt{Takata2012}), with a typical disc radial velocity of $\sim 1$~km~s$^{-1}$ (\citealt{Okazaki2001}). This yields a disc mass of $\sim 10^{22}$--10$^{25}$~g. 
In particular, for PSR~B1259$-$63, \cite{Chernyakova2014} estimated a
disc mass of $2\times 10^{25}$~g, with a disc radius of 0.42~AU, roughly similar to the values just mentioned. 
The masses of the simulated clumps are $\sim 8\times10^{18}$~g, $\sim 10^{20}$~g, $\sim 10^{21}$~g and $\sim 2\times10^{19}$~g for the clumps characterized by $\chi = 10$ and $R_{\rm c} =1~a$, $2.5~a$, $5~a$ and $\chi = 30$ and $R_{\rm c} =1~a$, respectively. For comparison, the mass of the clump with $R_{\rm c}=2.5\,a$ would be $\sim 0.001$--1\% the disc mass.

\subsection{Future work}

To distinguish the importance of numerical artefacts in our results and study a more
realistic case,
a 3D version of the simulations presented here is under way
to more accurately characterize the instabilities that affect the two-wind interaction region. In addition, we are planning to carry out calculations of the radiation outcome expected from the two-wind interaction region, and most importantly, from the clump interaction with this structure, making full use of the dynamical information provided by these simulations.

\section{Conclusions}
\label{conc}

We presented, for the first time, 2D axisymmetric RHD simulations of the interaction between an inhomogeneous stellar wind and a relativistic pulsar wind, focusing on the region inside the binary system. We simulated clumps with different sizes and densities to study different degrees of the stellar wind inhomogeneity. The presence of the clumps results in significant variations of the interaction region, which are expected to strongly affect the non-thermal radiation as well. Therefore, we confirm the sensitive nature of two-wind interaction structure under the presence of the stellar wind inhomogeneities. The shocked flow presents a complex spatial and temporal pattern, with fast changes in the shocked pulsar wind. This can lead to strong short time-scale flux variability in the non-thermal radiation of gamma-ray binaries, which might be observed for instance in gamma rays with the future Cherenkov Telescope Array (CTA; \citealt{Acharya2013, Paredes2013}).

\begin{acknowledgements}
We thank the anonymous referee for his/her constructive and useful comments.
We acknowledge support by the Spanish Ministerio de Econom\'{\i}a y Competitividad (MINECO) under grants  
AYA2010-21782-C03-01,
AYA2010-21322-C03-01,
AYA2010-21097-C03-01,
AYA2013-47447-C3-1-P,
FPA2010-22056-C06-02 and
FPA2013-48381-C6-6-P.
We also acknowledge support by the ``Generalitat Valenciana'' grant ``PROMETEO-2009-103''. This research has been supported by the Marie Curie Career Integration Grant 321520. X.P.-F. also acknowledges financial support from Universitat de Barcelona and Generalitat de Catalunya under grants APIF and FI, respectively. V.B-R. also acknowledges financial support from MINECO and European Social Funds through a Ram\'on y Cajal fellowship.
\end{acknowledgements}

\appendix
\section{Physical quantity maps for different clump parameters}

\begin{figure*}
\centering
\includegraphics[height=4.4cm,   trim= 0.0cm 0.9cm 0.0cm 0.0cm,clip]{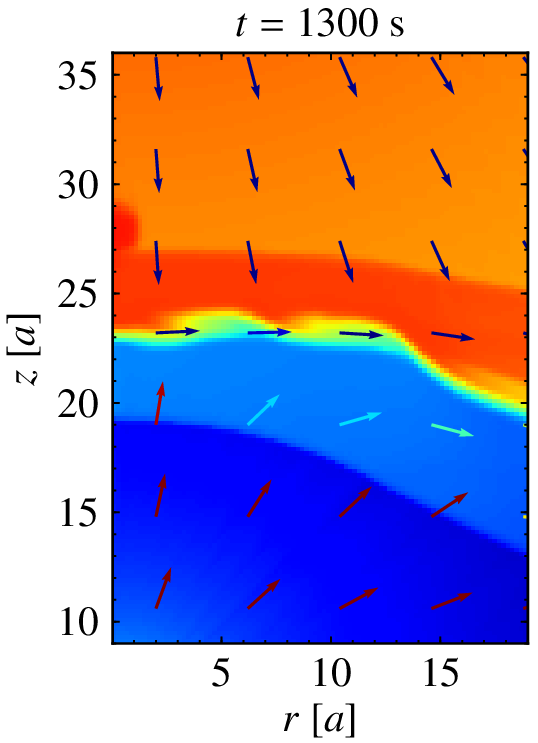}
\includegraphics[height=4.4cm,   trim= 1.0cm 0.9cm 0.0cm 0.0cm,clip]{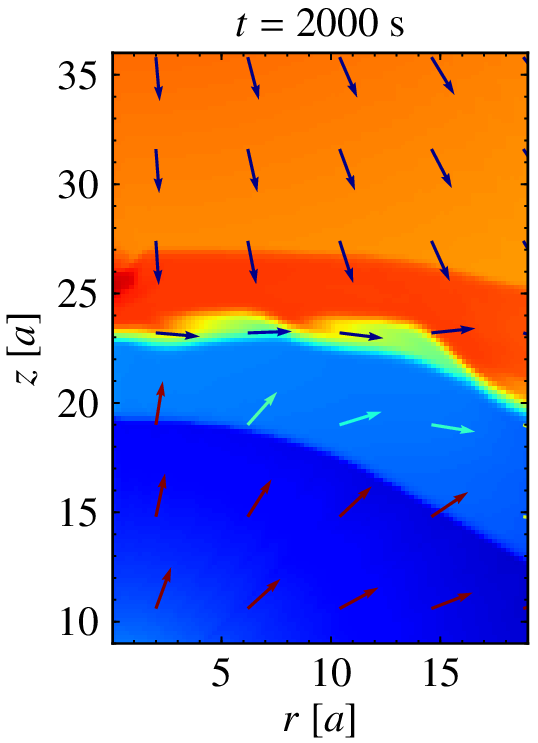}
\includegraphics[height=4.4cm,   trim= 1.0cm 0.9cm 0.0cm 0.0cm,clip]{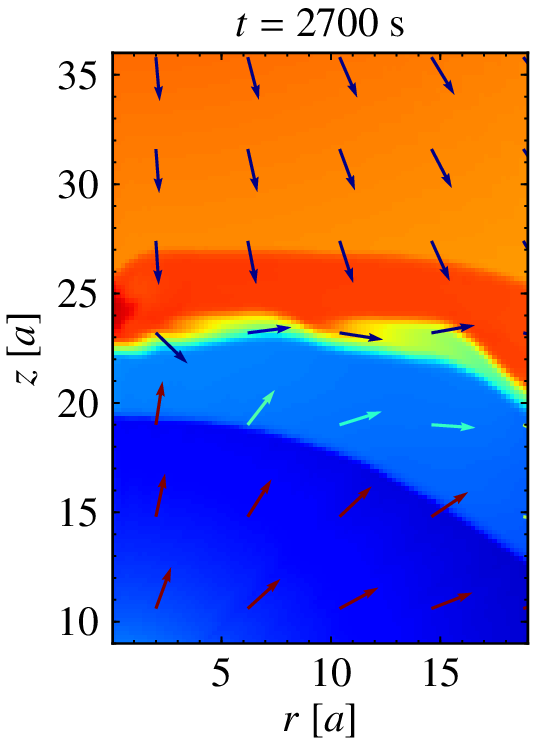}
\includegraphics[height=4.4cm,   trim= 1.0cm 0.9cm 0.0cm 0.0cm,clip]{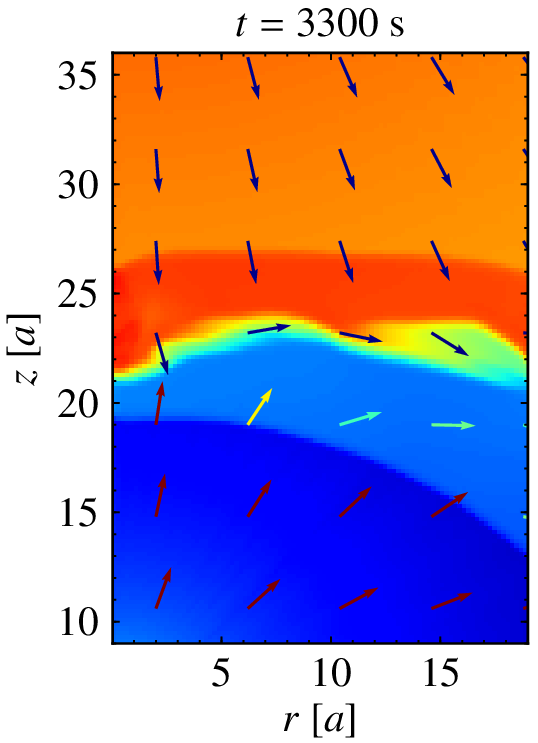}
\includegraphics[height=4.4cm,   trim= 1.0cm 0.9cm 0.0cm 0.0cm,clip]{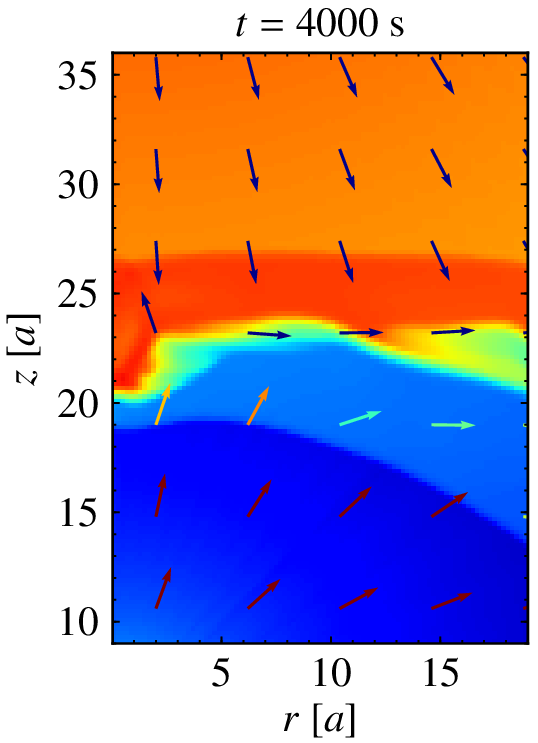}
\includegraphics[height=5.00cm,  trim= 0.0cm 0.0cm 0.0cm 0.0cm,clip]{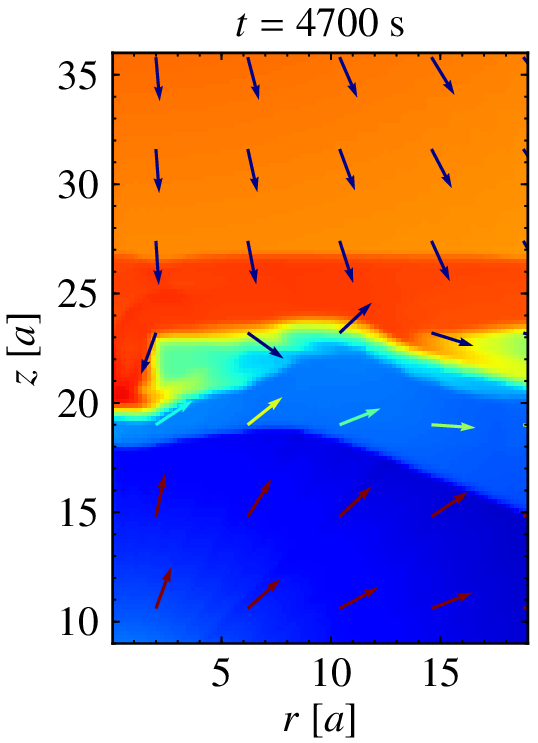}
\includegraphics[height=5.00cm,  trim= 1.0cm 0.0cm 0.0cm 0.0cm,clip]{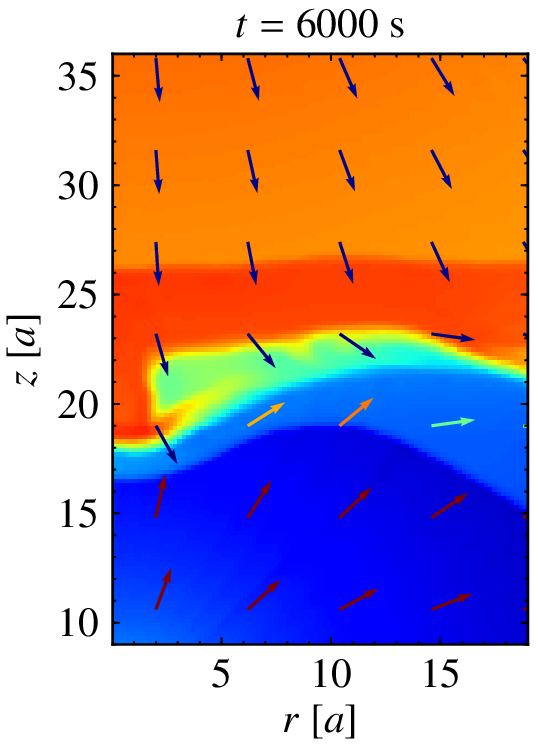}
\includegraphics[height=5.00cm,  trim= 1.0cm 0.0cm 0.0cm 0.0cm,clip]{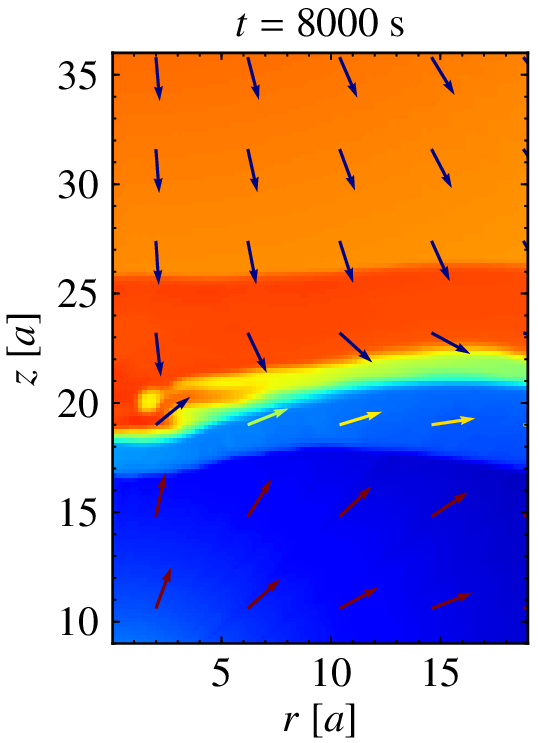}
\includegraphics[height=5.00cm,  trim= 1.0cm 0.0cm 0.0cm 0.0cm,clip]{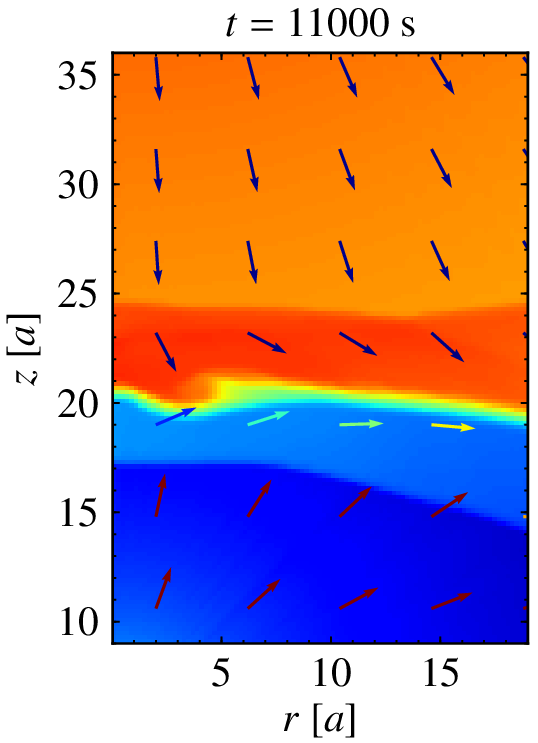}
\includegraphics[height=5.00cm,  trim= 1.0cm 0.0cm 0.0cm 0.0cm,clip]{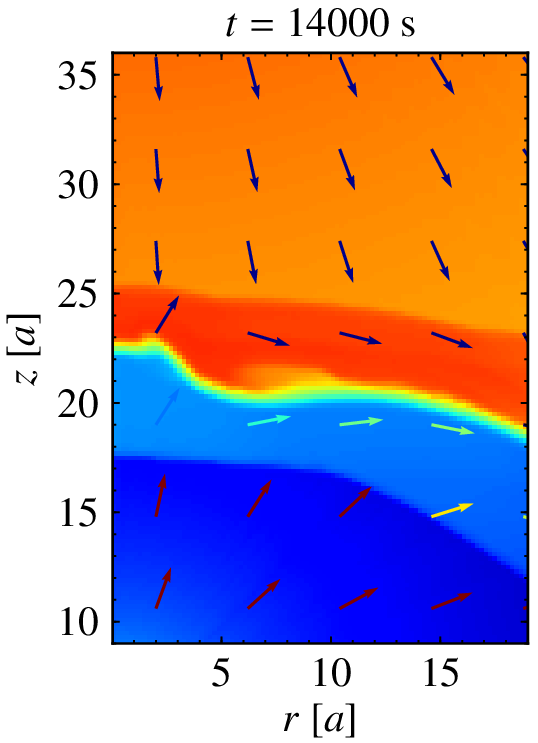}
\caption{
Zoom-in of the density distribution by colour for the case with clump parameters
$\chi = 10$ and $R_{\rm c} =1~a$
for the times shown at the top of each plot.
The coloured arrows represent the three-velocity at different locations ($\beta = v/c$).
The axes units are $a = 8\times10^{10}$~cm.
The pulsar and the star are located at $(r, z) = (0,5~a)$ and $(r, z) = (0,60~a)$, respectively.} 
\label{f10r1_zoom}
\end{figure*}

\begin{figure*}
\centering
\includegraphics[width=6.4cm]{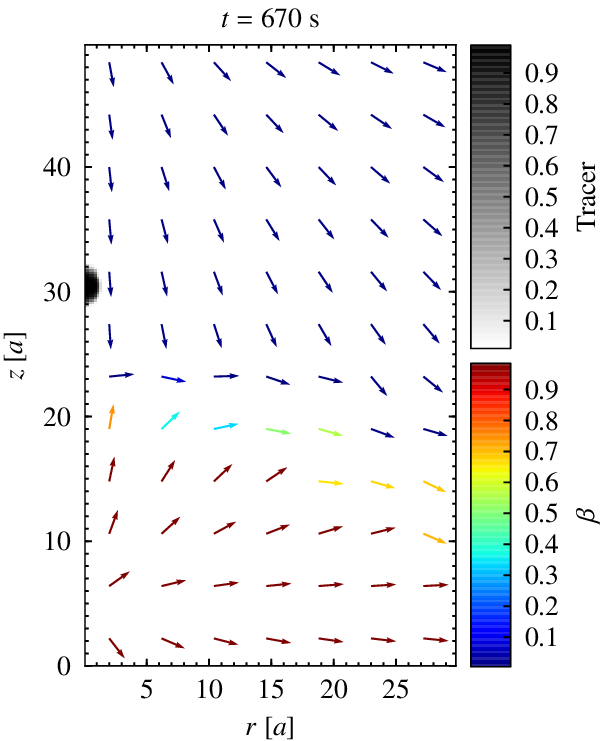}
\includegraphics[width=6.4cm]{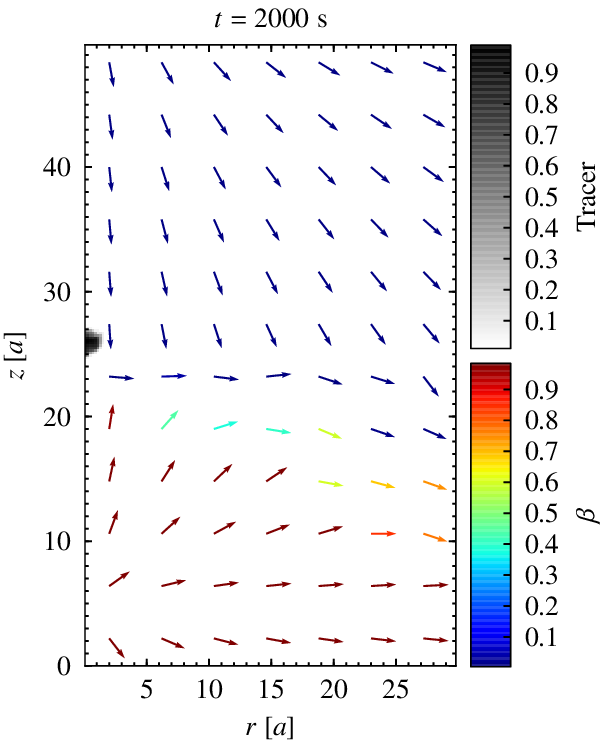}\\\vspace{0.2cm}
\includegraphics[width=6.4cm]{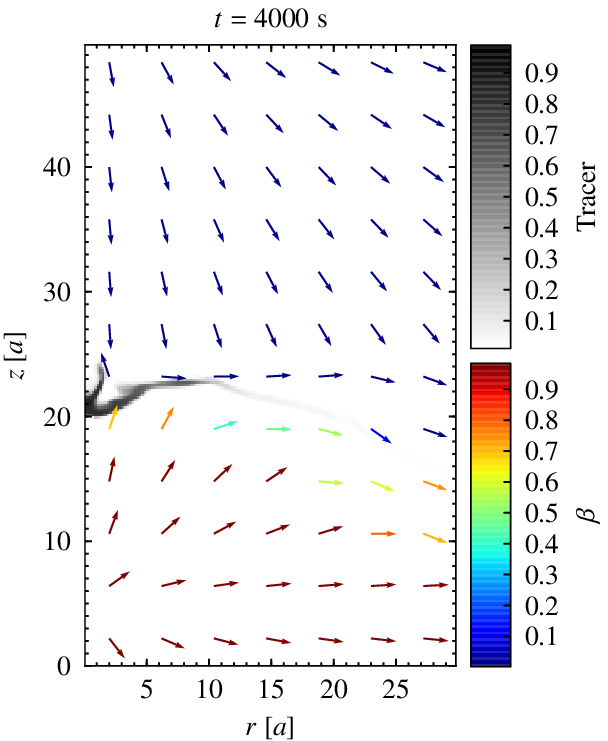}
\includegraphics[width=6.4cm]{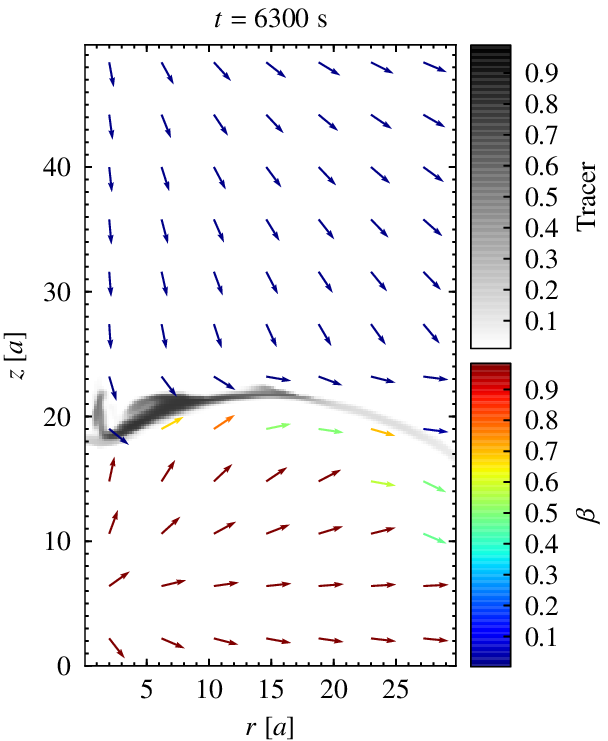}\\\vspace{0.2cm}
\includegraphics[width=6.4cm]{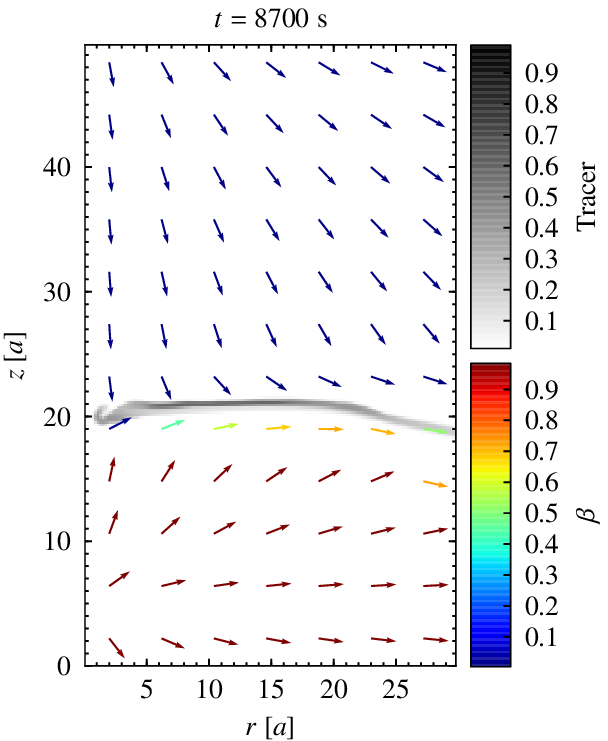}
\includegraphics[width=6.4cm]{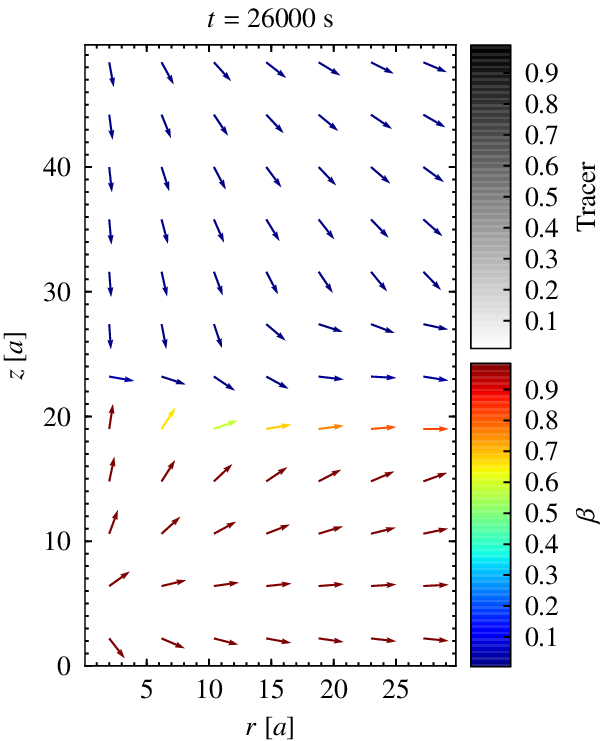}
\caption{Tracer distribution by colour for the case with clump parameters
$\chi = 10$ and $R_{\rm c} =1~a$
for the times shown at the top of each plot.
The tracer value ranges from 0 (pulsar and stellar wind) to 1 (clump).
The remaining plot properties are the same as those of Fig.~\ref{f10r1_zoom}.}
\label{f10r1_tracer}
\end{figure*}

\begin{figure*}
\centering
\includegraphics[width=6.5cm]{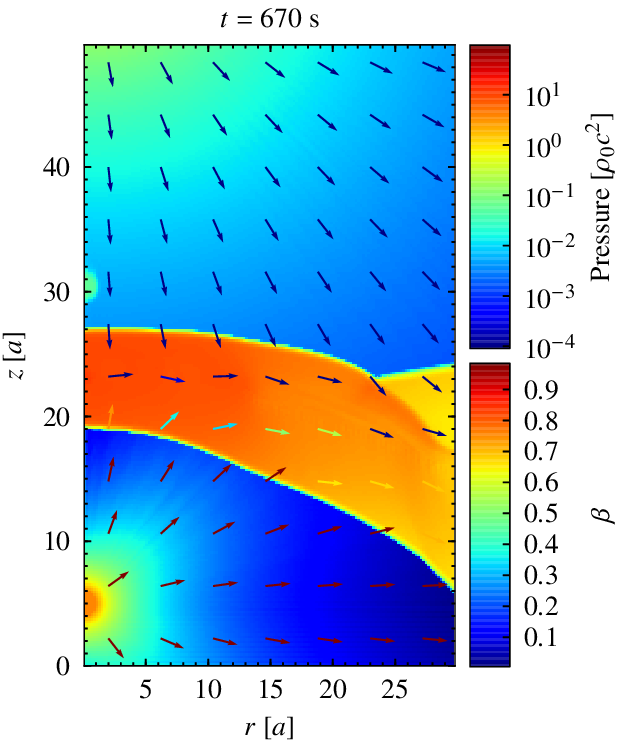}
\includegraphics[width=6.5cm]{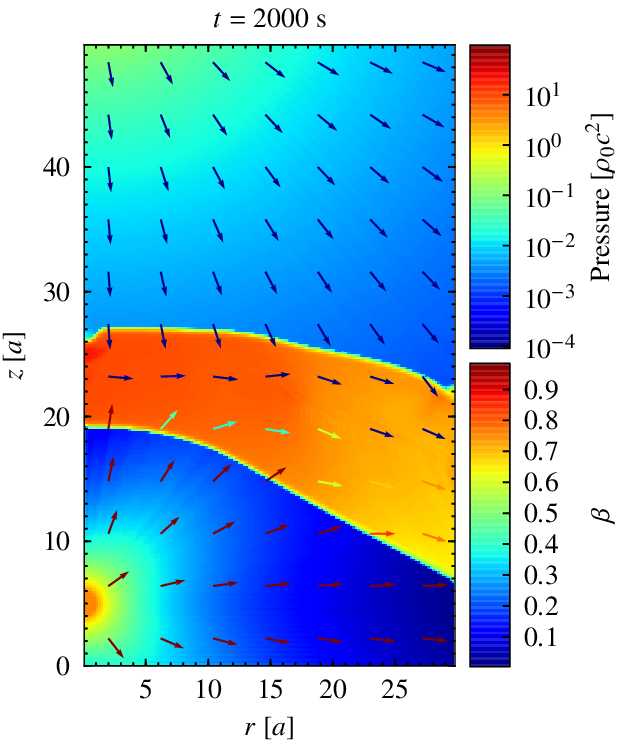}\\\vspace{0.2cm}
\includegraphics[width=6.5cm]{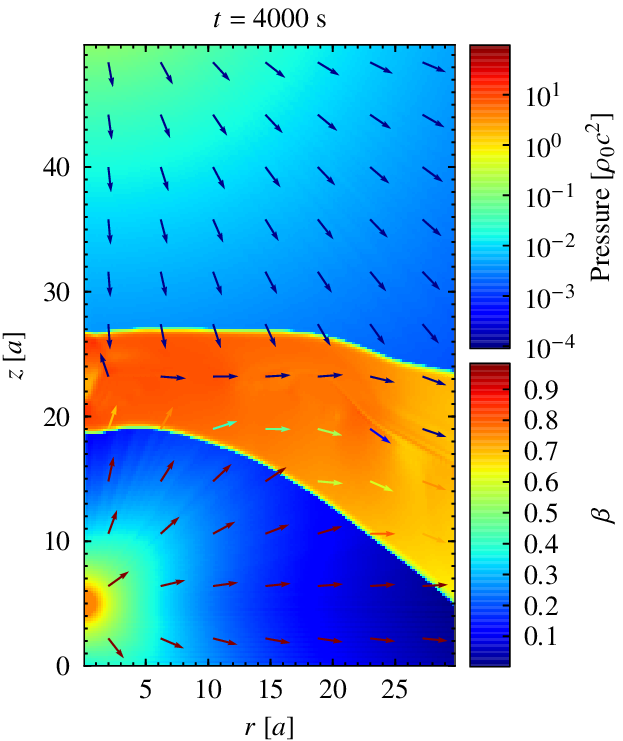}
\includegraphics[width=6.5cm]{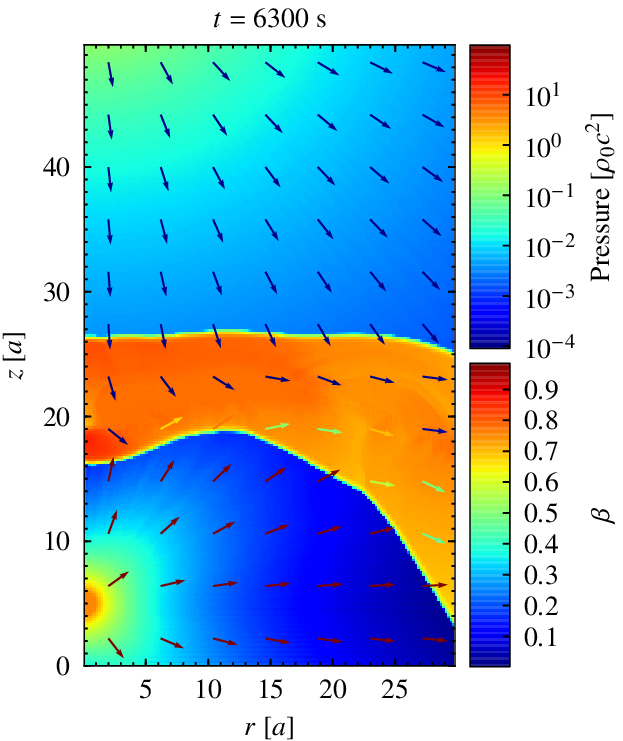}\\\vspace{0.2cm}
\includegraphics[width=6.5cm]{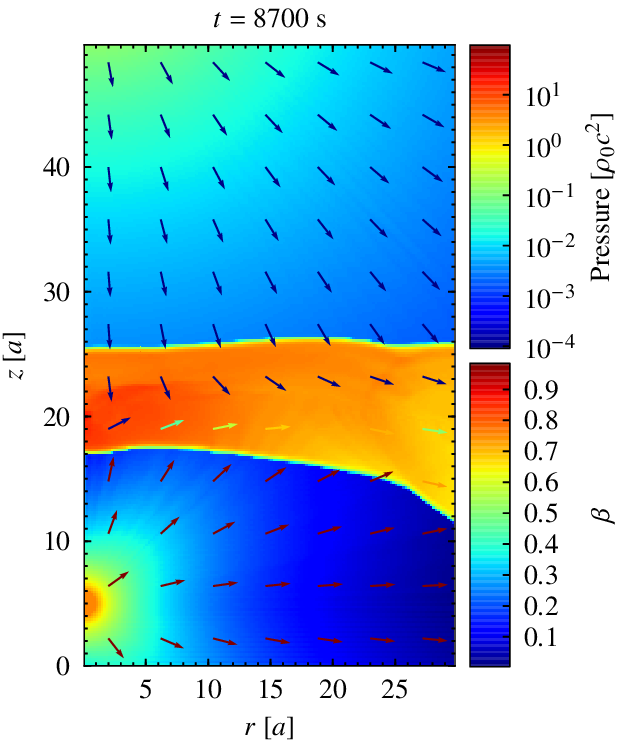}
\includegraphics[width=6.5cm]{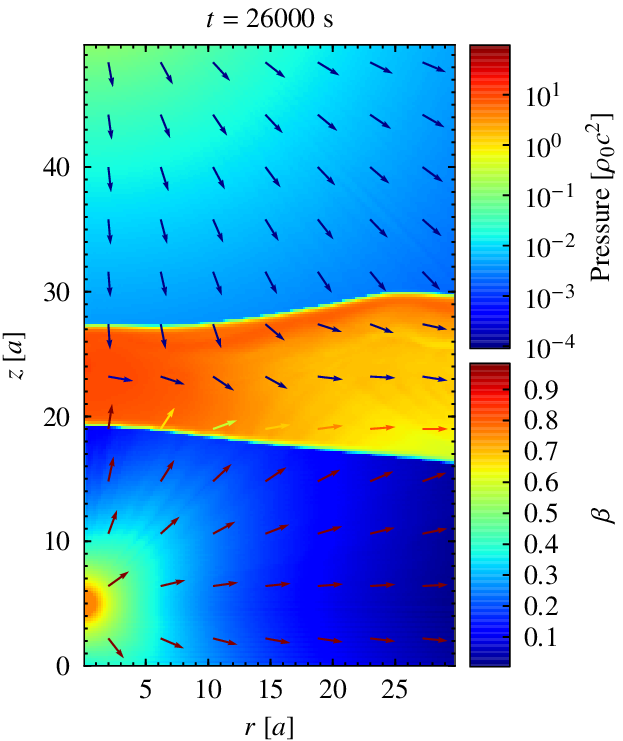}
\caption{Pressure distribution in units of $\rho_0{c^2}$ by colour for the case with clump parameters
$\chi = 10$ and $R_{\rm c} =1~a$
for the times shown at the top of each plot.
The remaining plot properties are the same as those of Fig.~\ref{f10r1_zoom}.}
\label{f10r1_pressure}
\end{figure*}

\begin{figure*}
\centering
\includegraphics[width=6.4cm]{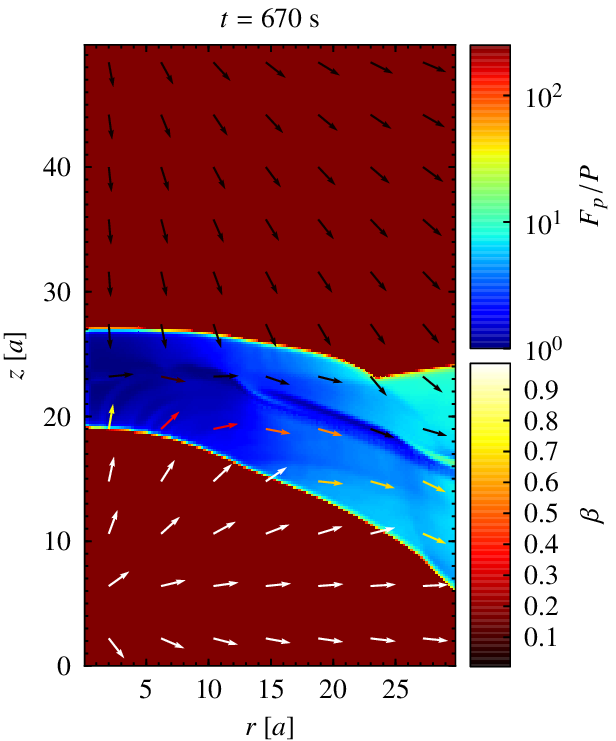}
\includegraphics[width=6.4cm]{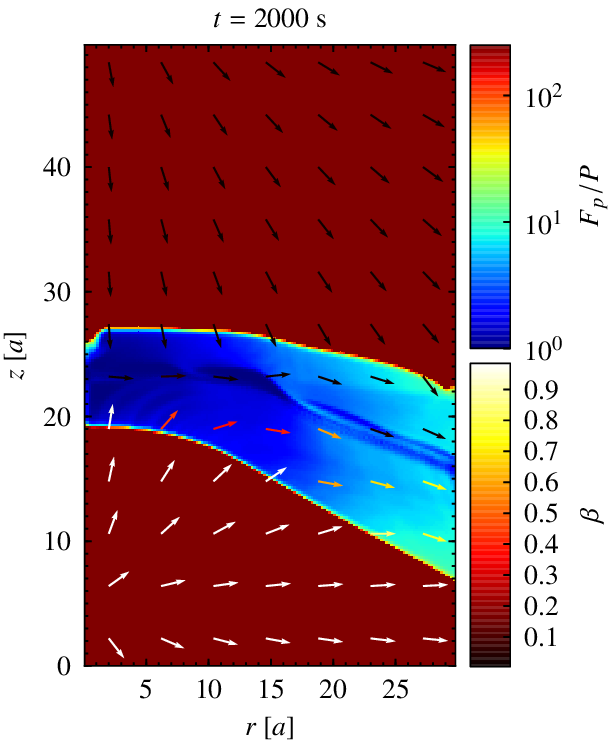}\\\vspace{0.2cm}
\includegraphics[width=6.4cm]{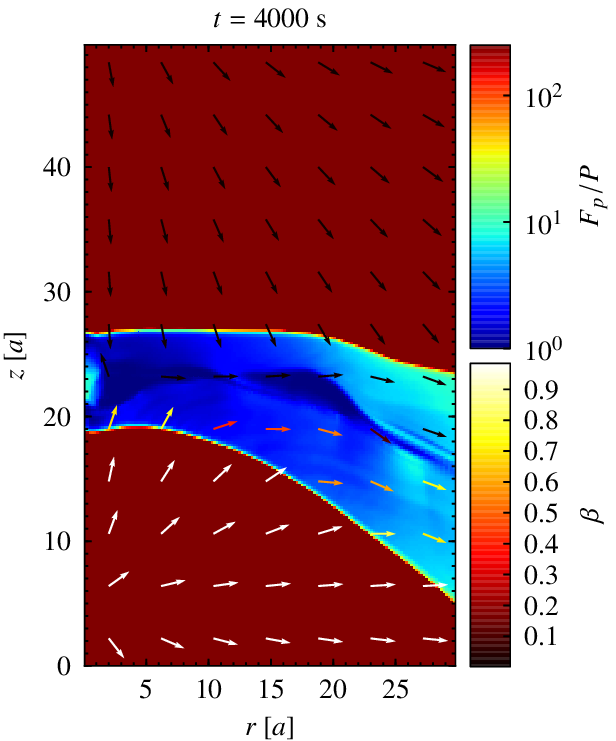}
\includegraphics[width=6.4cm]{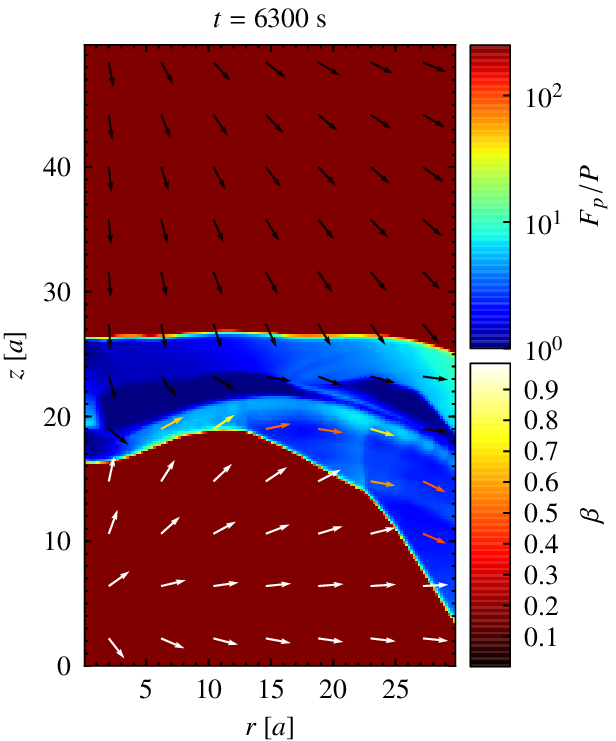}\\\vspace{0.2cm}
\includegraphics[width=6.4cm]{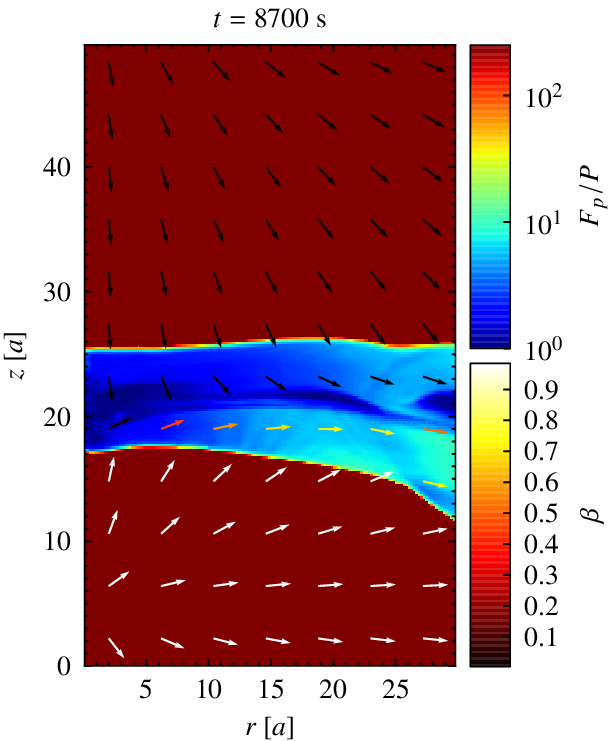}
\includegraphics[width=6.4cm]{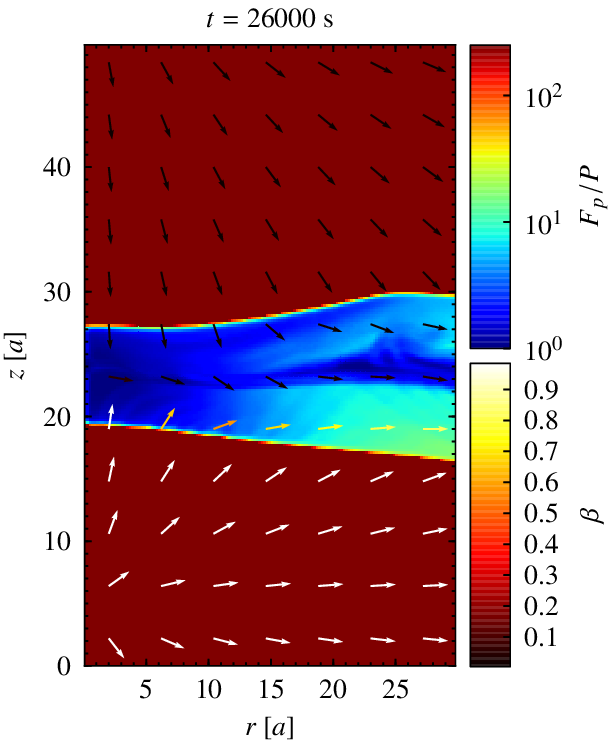}
\caption{Momentum flux over pressure distribution by colour for the case with clump parameters
$\chi = 10$ and $R_{\rm c} =1~a$
for the times shown at the top of each plot.
The momentum flux is given by $F_p = \rho~\Gamma^2 v^2 (1+\epsilon+P/{\rho})+P$.
The remaining plot properties are the same as those of Fig.~\ref{f10r1_zoom}.}
\label{f10r1_sonic}
\end{figure*}

\begin{figure*}
\centering
\includegraphics[width=6.4cm]{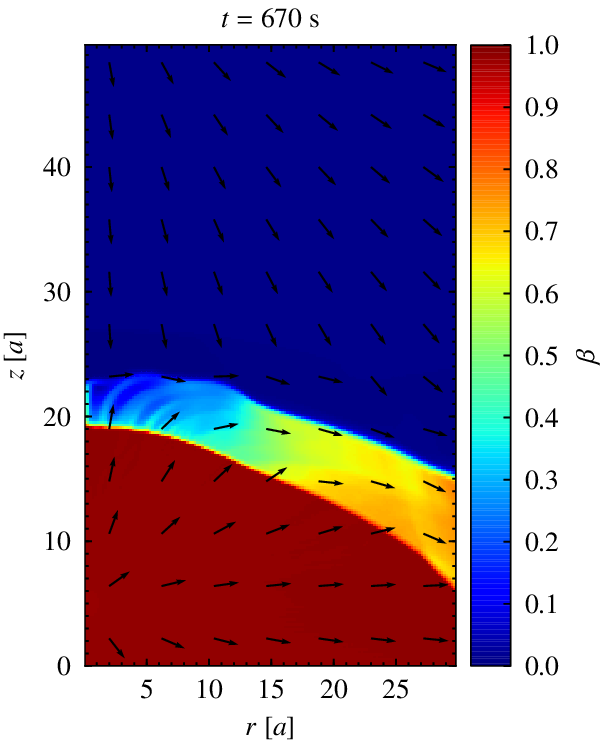}
\includegraphics[width=6.4cm]{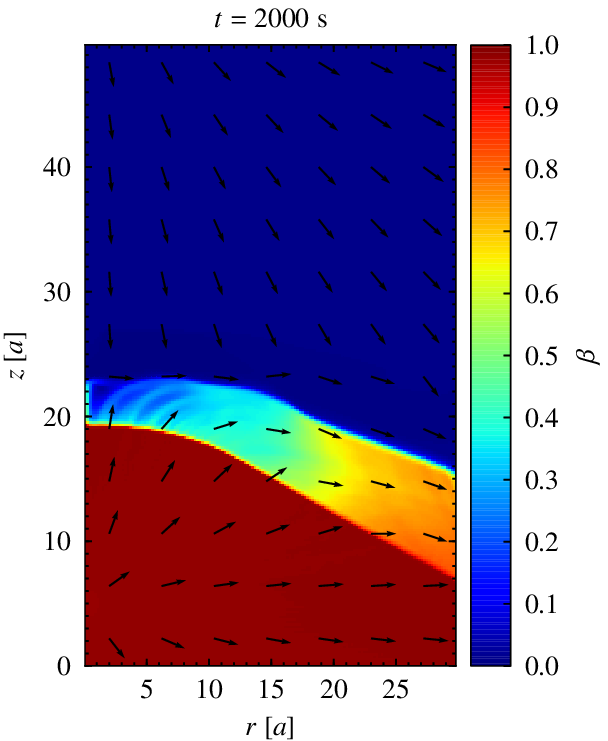}\\\vspace{0.2cm}
\includegraphics[width=6.4cm]{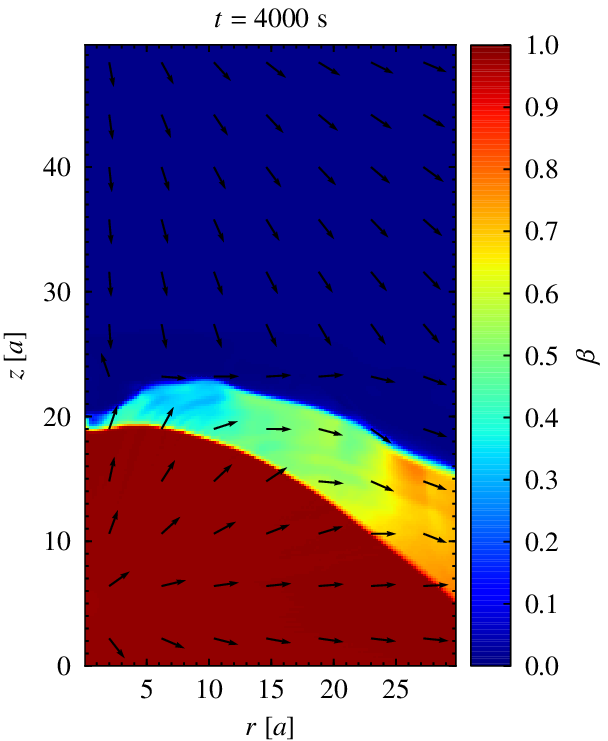}
\includegraphics[width=6.4cm]{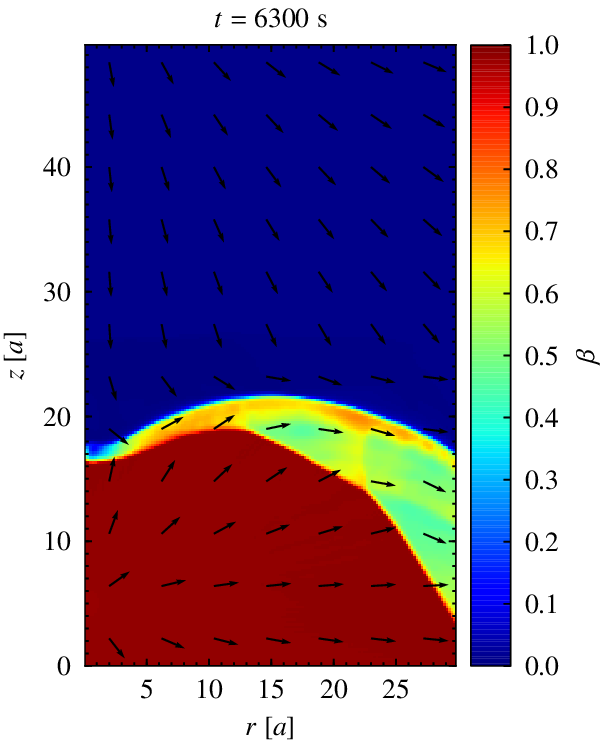}\\\vspace{0.2cm}
\includegraphics[width=6.4cm]{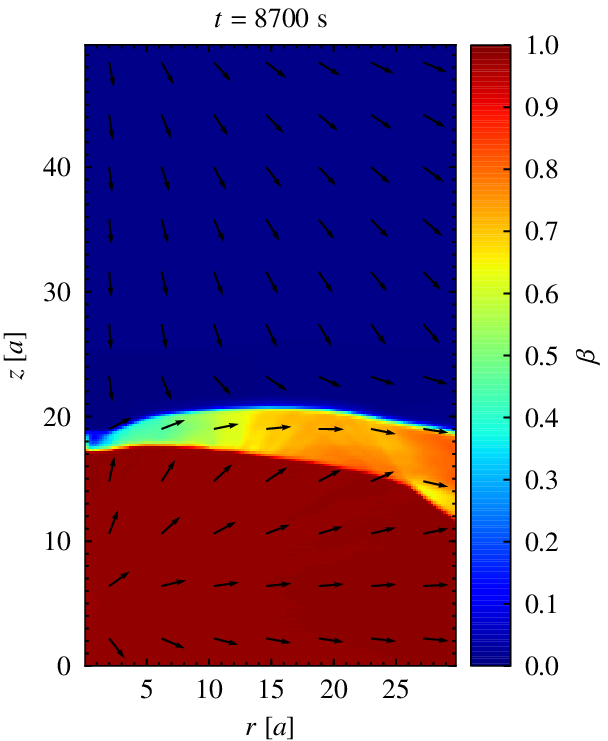}
\includegraphics[width=6.4cm]{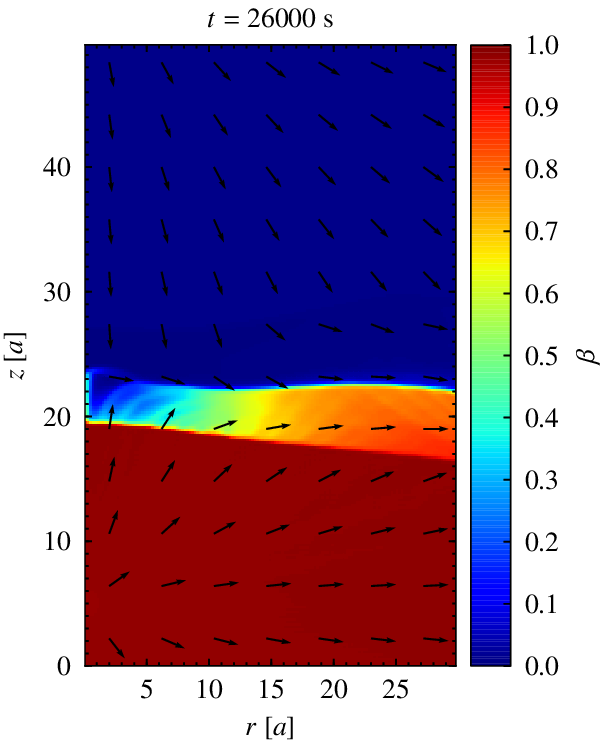}
\caption{$\beta$ distribution by colour for the case with clump parameters
$\chi = 10$ and $R_{\rm c} =1~a$
for the times shown at the top of each plot.
The remaining plot properties are the same as those of Fig.~\ref{f10r1_zoom}.}
\label{f10r1_W}
\end{figure*}

\begin{figure*}
\centering
\includegraphics[height=4.4cm,  trim= 0.0cm 0.9cm 0.0cm 0.0cm,clip]{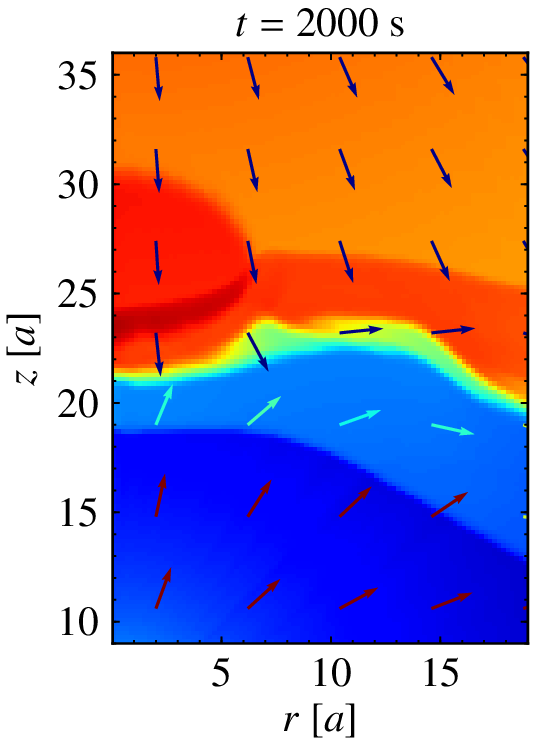}
\includegraphics[height=4.4cm,  trim= 1.0cm 0.9cm 0.0cm 0.0cm,clip]{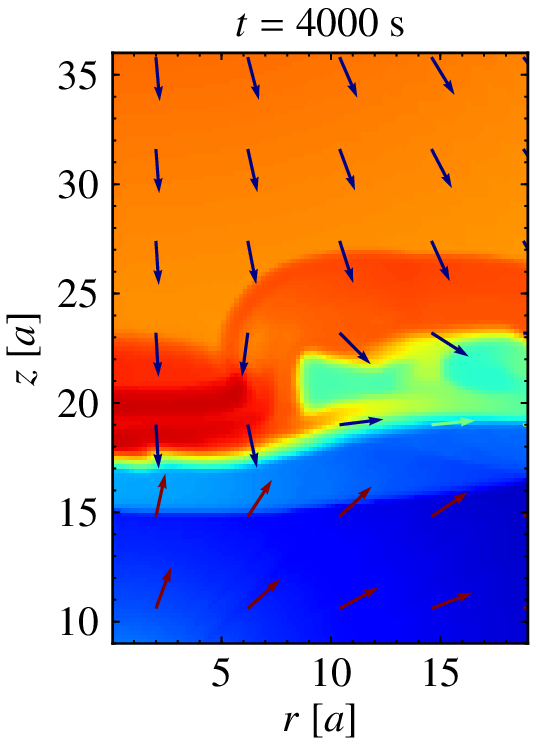}
\includegraphics[height=4.4cm,  trim= 1.0cm 0.9cm 0.0cm 0.0cm,clip]{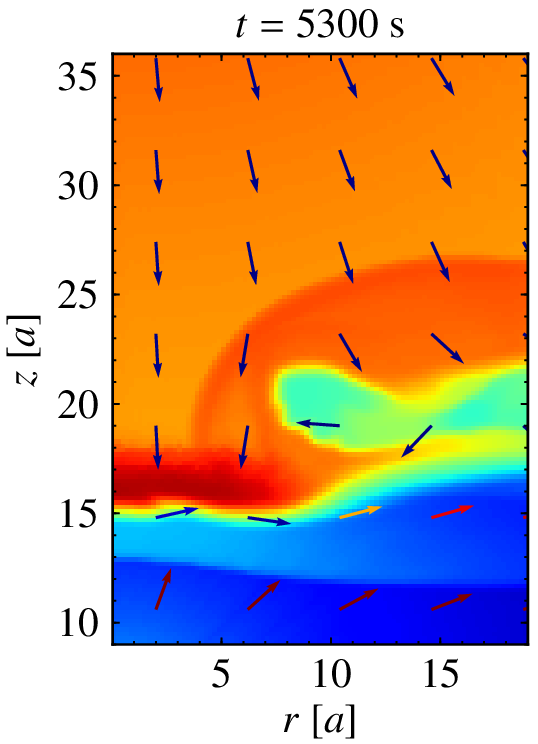}
\includegraphics[height=4.4cm,  trim= 1.0cm 0.9cm 0.0cm 0.0cm,clip]{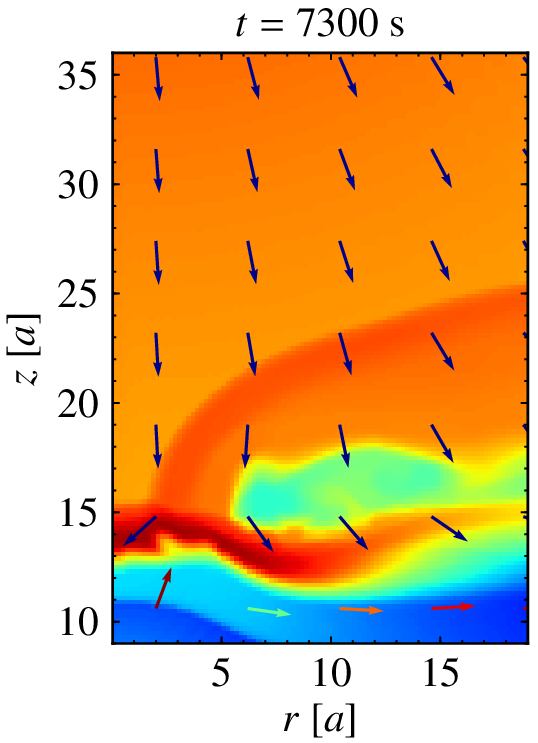}
\includegraphics[height=4.4cm,  trim= 1.0cm 0.9cm 0.0cm 0.0cm,clip]{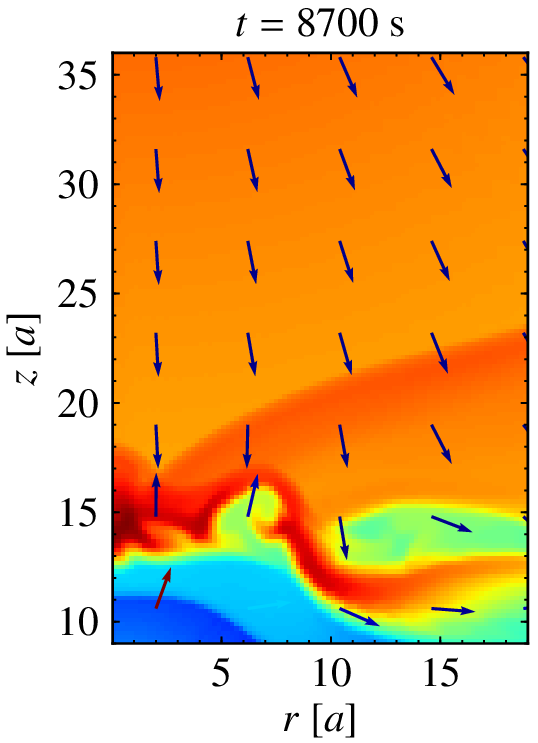}\\
\includegraphics[height=5.00cm, trim= 0.0cm 0.0cm 0.0cm 0.0cm,clip]{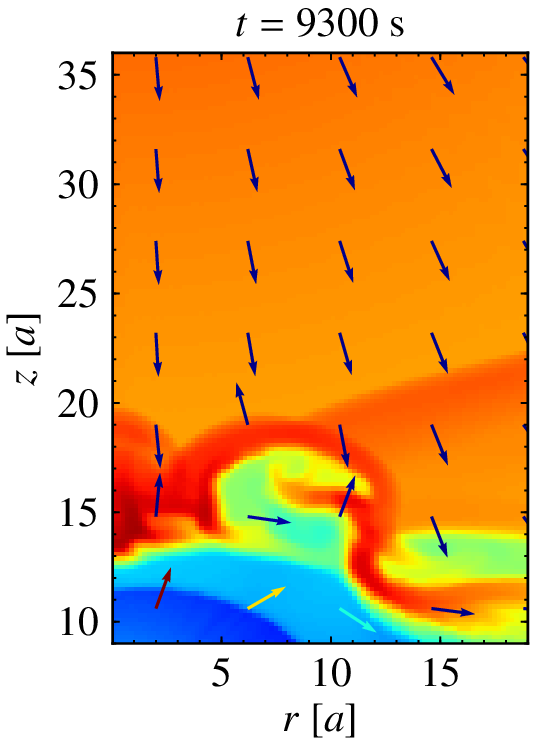}
\includegraphics[height=5.00cm, trim= 1.0cm 0.0cm 0.0cm 0.0cm,clip]{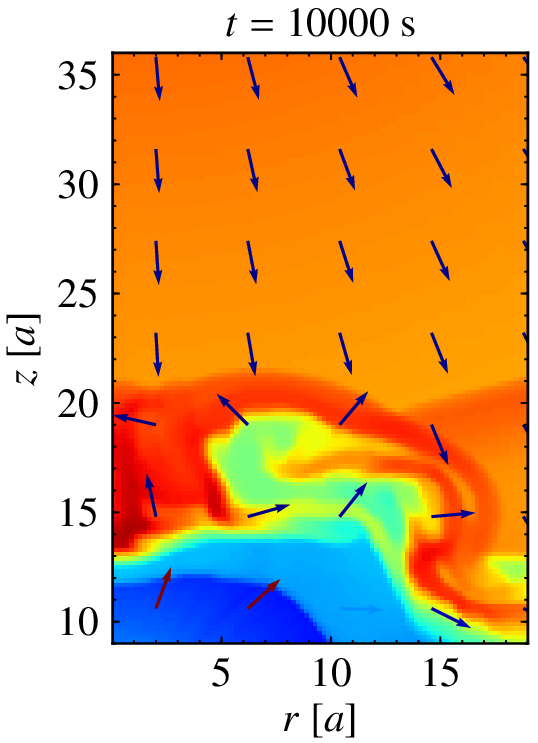}
\includegraphics[height=5.00cm, trim= 1.0cm 0.0cm 0.0cm 0.0cm,clip]{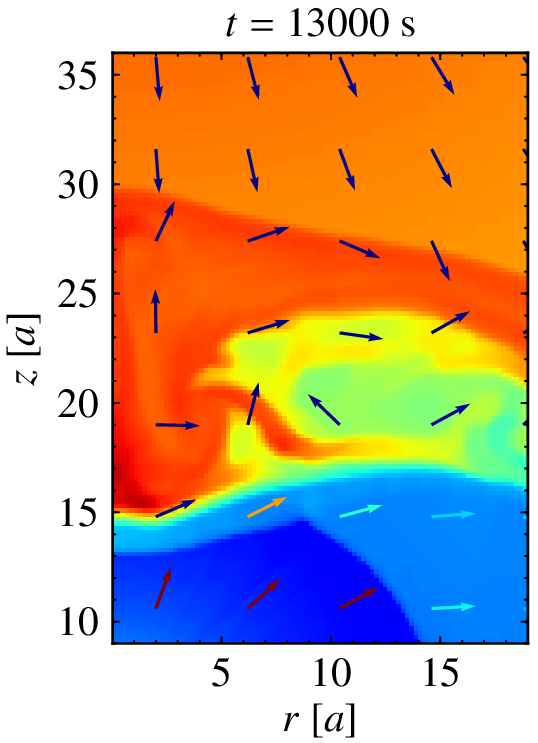}
\includegraphics[height=5.00cm, trim= 1.0cm 0.0cm 0.0cm 0.0cm,clip]{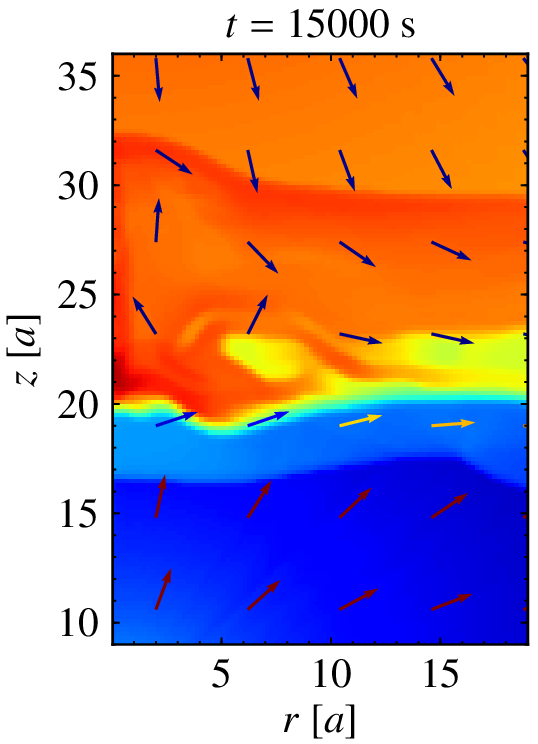}
\includegraphics[height=5.00cm, trim= 1.0cm 0.0cm 0.0cm 0.0cm,clip]{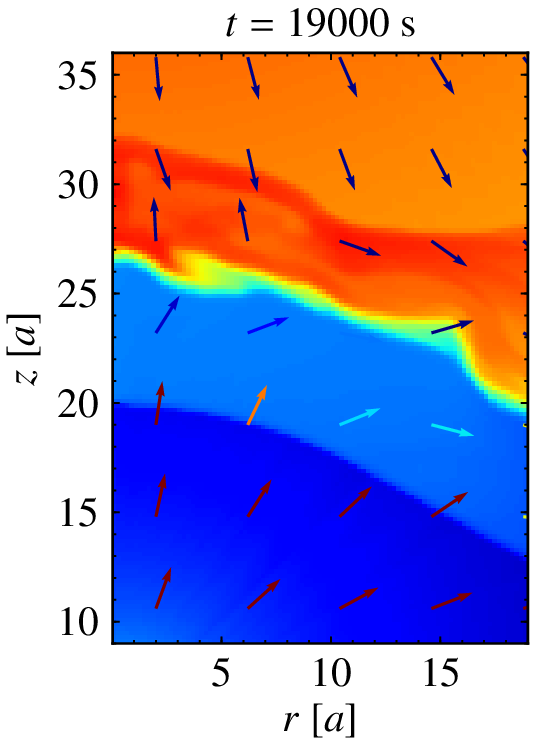}
\caption{
Zoom-in of the density distribution by colour for the case with clump parameters
$\chi = 10$ and $R_{\rm c} =5~a$
for the times shown at the top of each plot.
The remaining plot properties are the same as those of Fig.~\ref{f10r1_zoom}.}
\label{f10r5_zoom}
\end{figure*}

\begin{figure*}
\centering
\includegraphics[width=6.4cm]{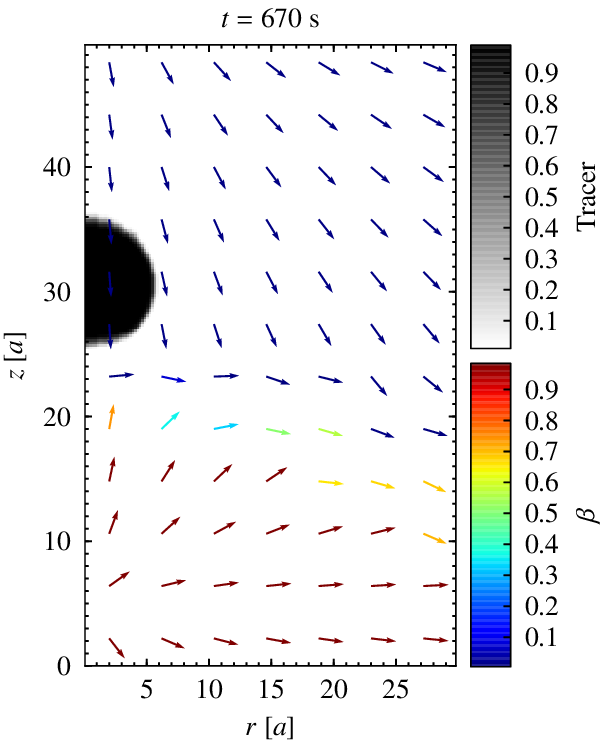}
\includegraphics[width=6.4cm]{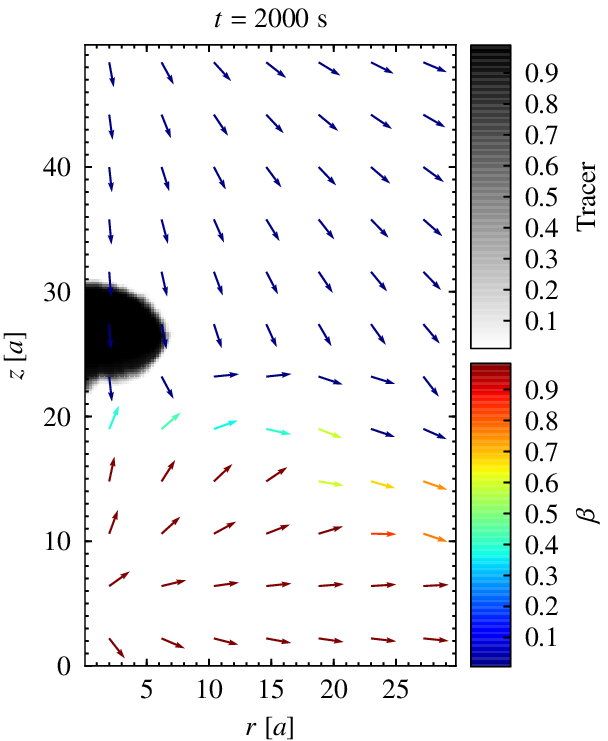}\\\vspace{0.2cm}
\includegraphics[width=6.4cm]{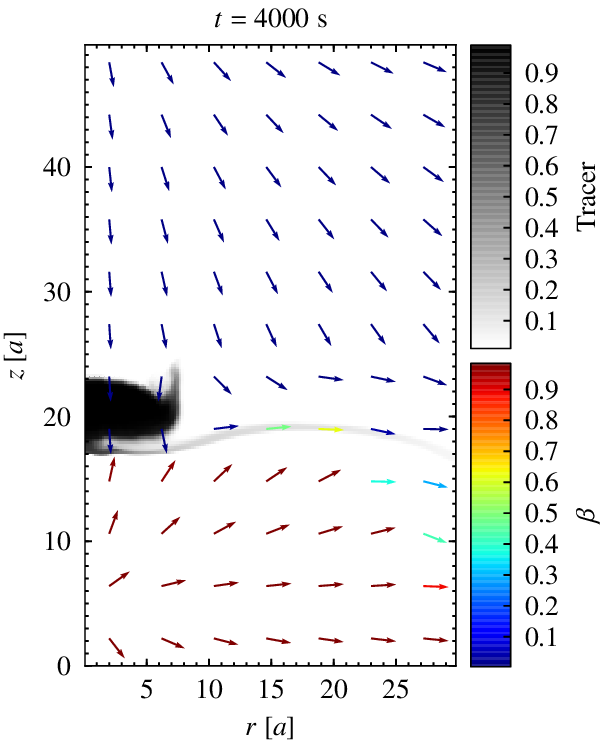}
\includegraphics[width=6.4cm]{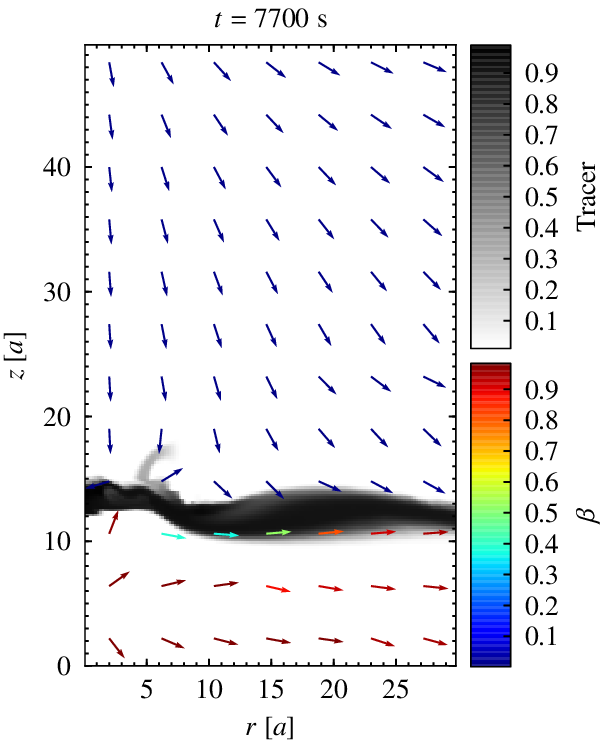}\\\vspace{0.2cm}
\includegraphics[width=6.4cm]{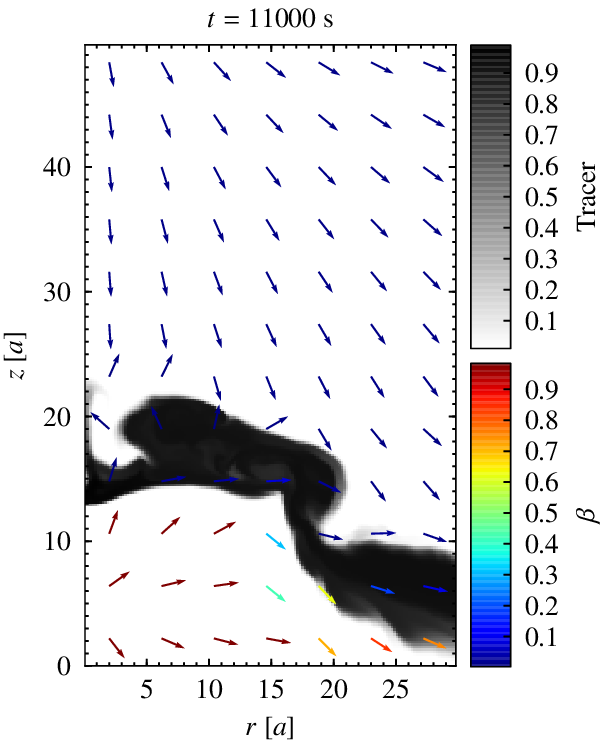}
\includegraphics[width=6.4cm]{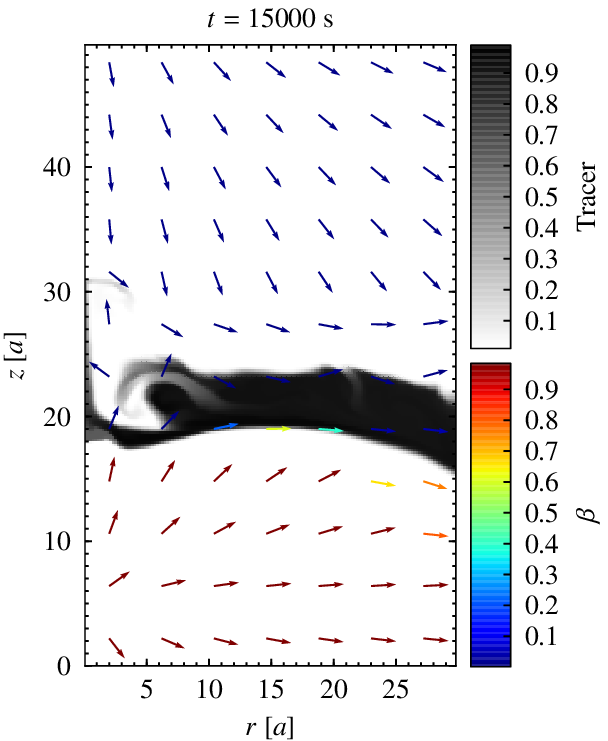}
\caption{Tracer distribution by colour for the case with clump parameters
$\chi = 10$ and $R_{\rm c} =5~a$
for the times shown at the top of each plot.
The tracer value ranges from 0 (pulsar and stellar wind) to 1 (clump).
The remaining plot properties are the same as those of Fig.~\ref{f10r1_zoom}.}
\label{f10r5_tracer}
\end{figure*}

\begin{figure*}
\centering
\includegraphics[width=6.5cm]{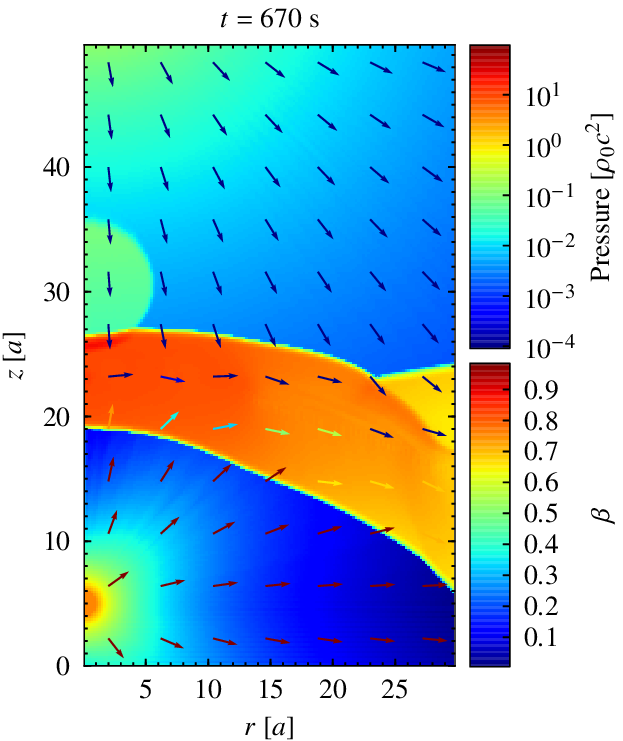}
\includegraphics[width=6.5cm]{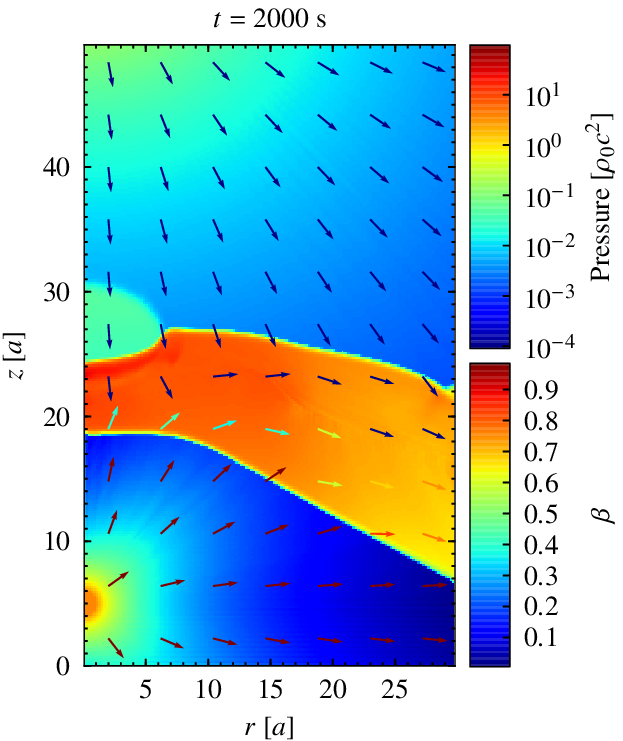}\\\vspace{0.2cm}
\includegraphics[width=6.5cm]{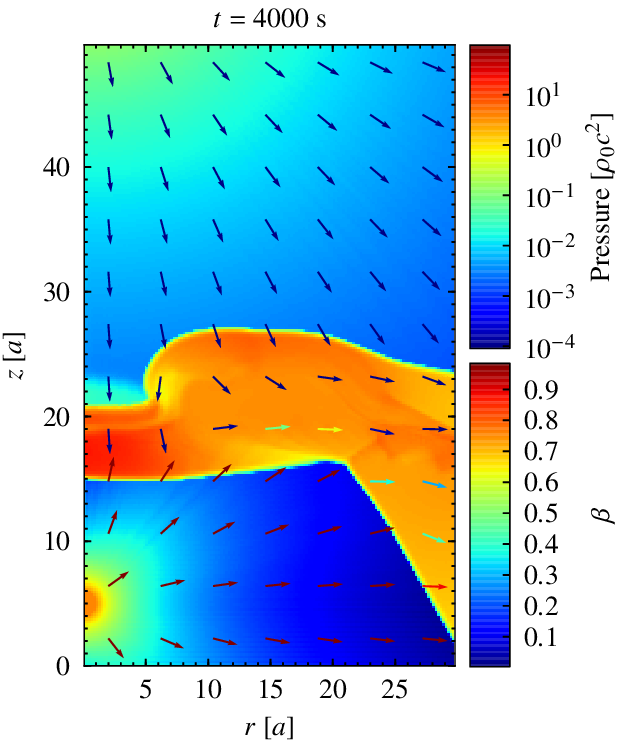}
\includegraphics[width=6.5cm]{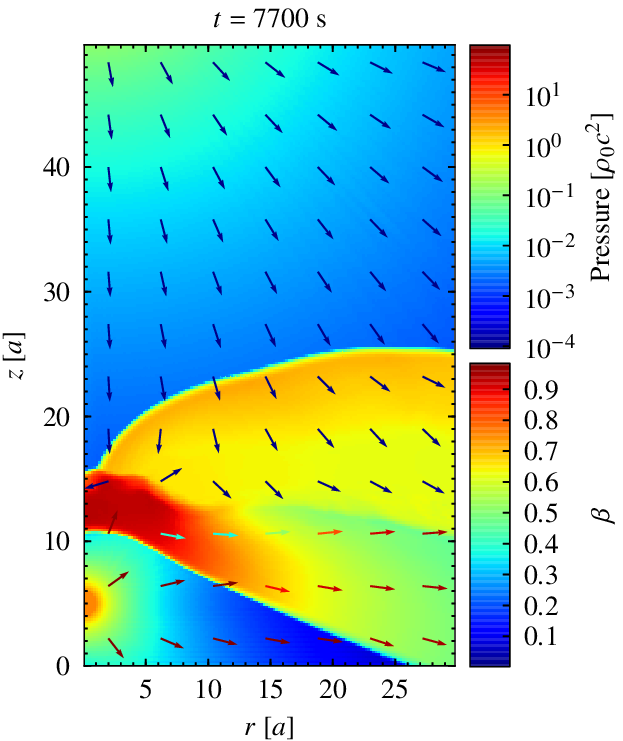}\\\vspace{0.2cm}
\includegraphics[width=6.5cm]{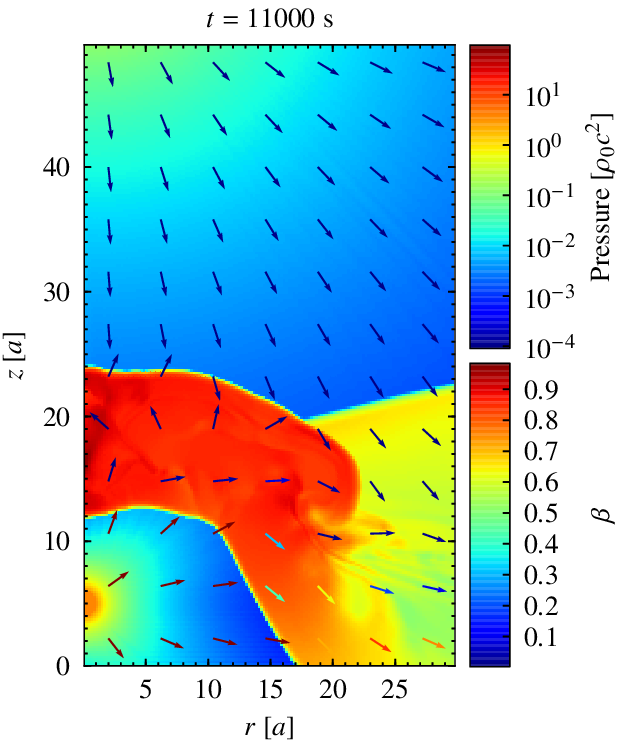}
\includegraphics[width=6.5cm]{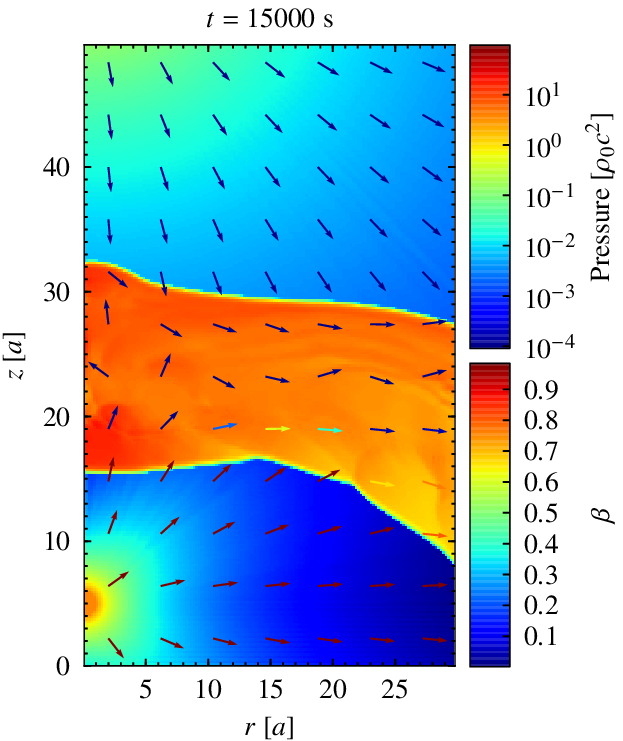}
\caption{Pressure distribution in units of $\rho_0{c^2}$ by colour for the case with clump parameters
$\chi = 10$ and $R_{\rm c} =5~a$
for the times shown at the top of each plot.
The remaining plot properties are the same as those of Fig.~\ref{f10r1_zoom}.}
\label{f10r5_pressure}
\end{figure*}

\begin{figure*}
\centering
\includegraphics[width=6.4cm]{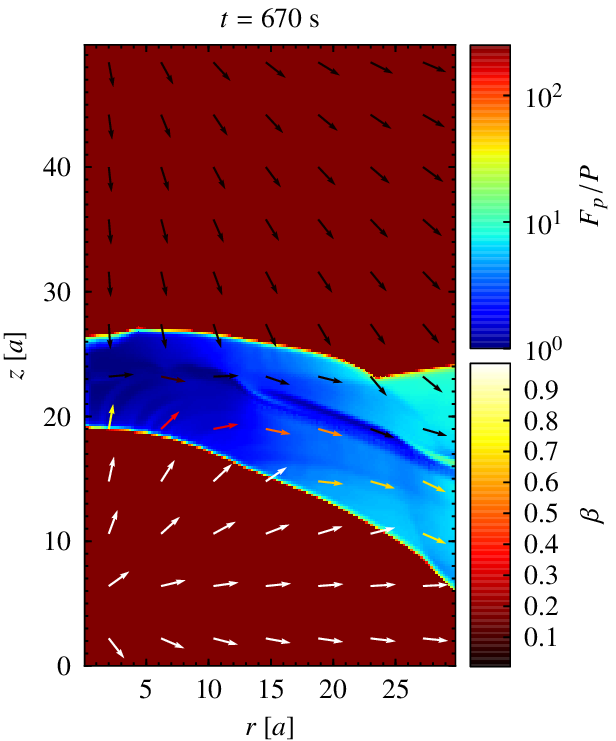}
\includegraphics[width=6.4cm]{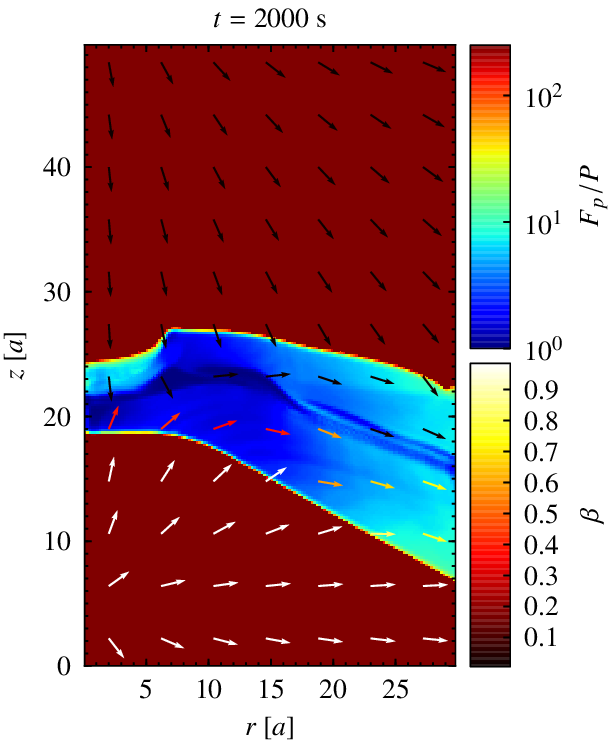}\\\vspace{0.2cm}
\includegraphics[width=6.4cm]{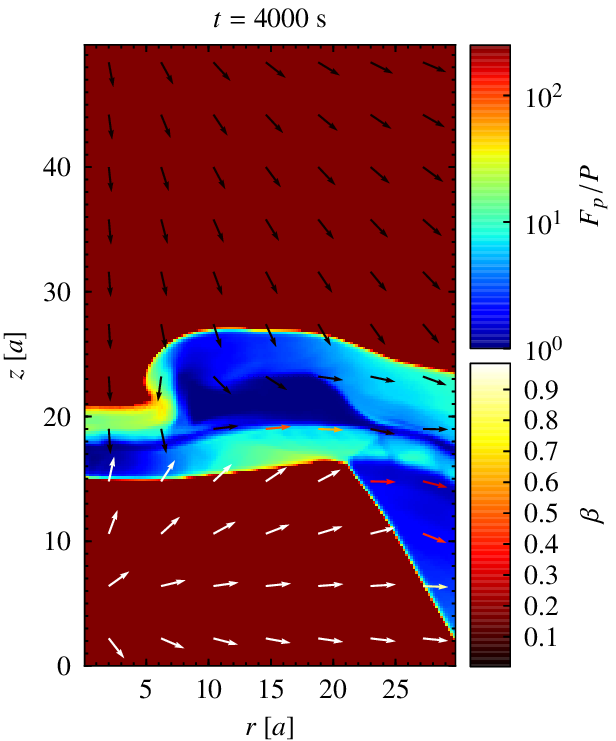}
\includegraphics[width=6.4cm]{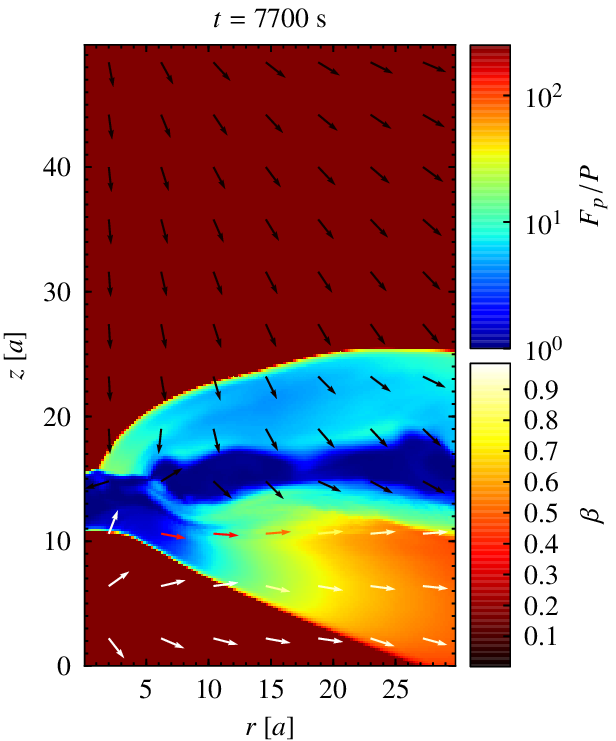}\\\vspace{0.2cm}
\includegraphics[width=6.4cm]{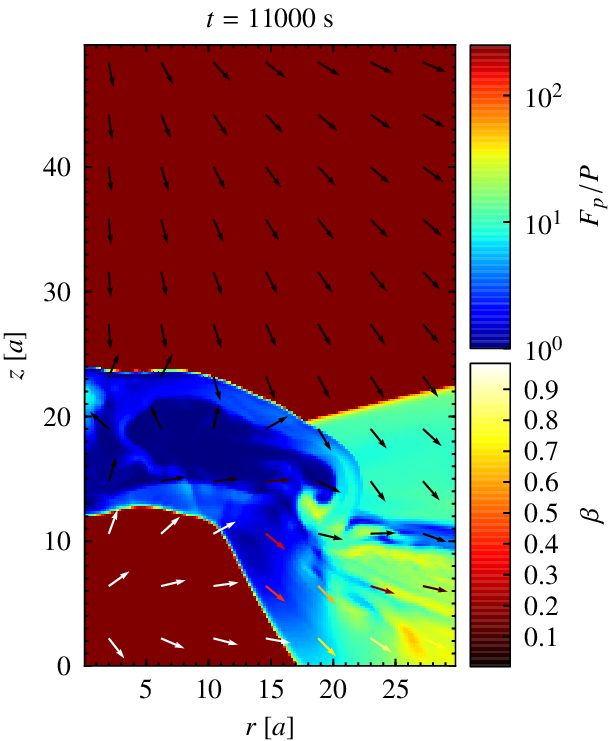}
\includegraphics[width=6.4cm]{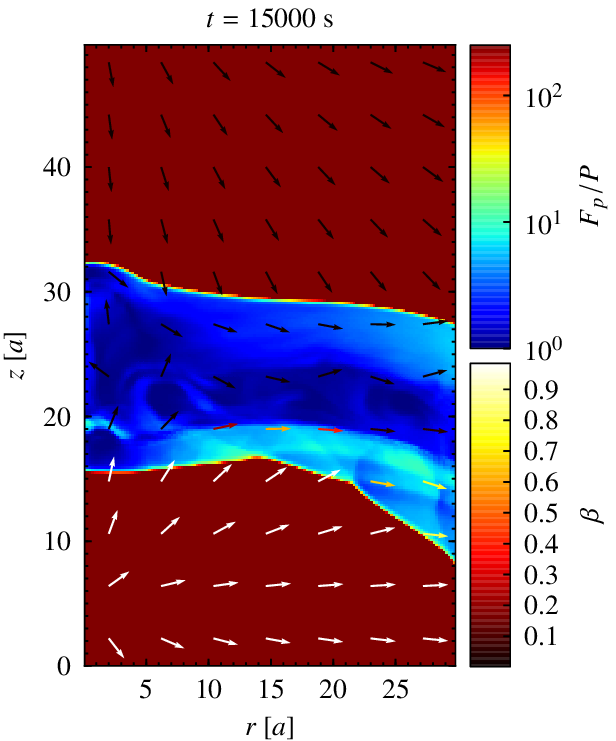}
\caption{Momentum flux over pressure distribution by colour for the case with clump parameters
$\chi = 10$ and $R_{\rm c} = 5~a$
for the times shown at the top of each plot.
The momentum flux is given by $F_p = \rho~\Gamma^2 v^2 (1+\epsilon+P/{\rho})+P$.
The remaining plot properties are the same as those of Fig.~\ref{f10r1_zoom}.}
\label{f10r5_sonic}
\end{figure*}

\begin{figure*}
\centering
\includegraphics[width=6.4cm]{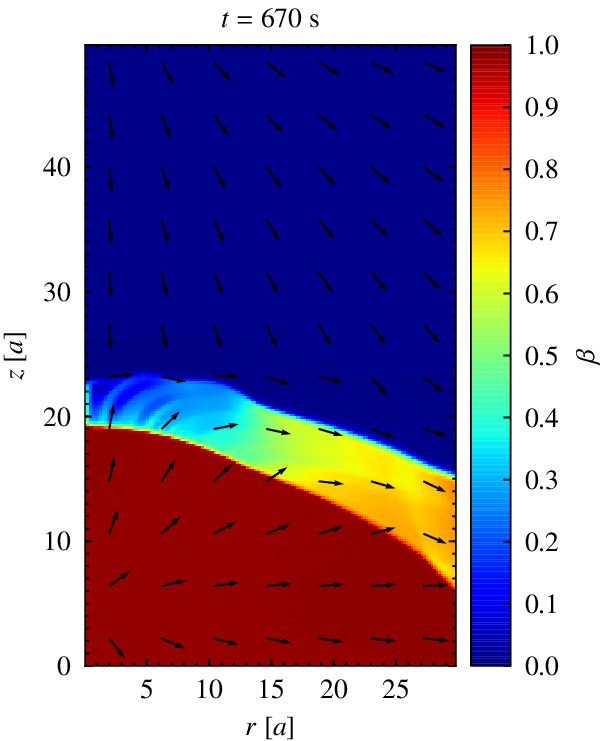}
\includegraphics[width=6.4cm]{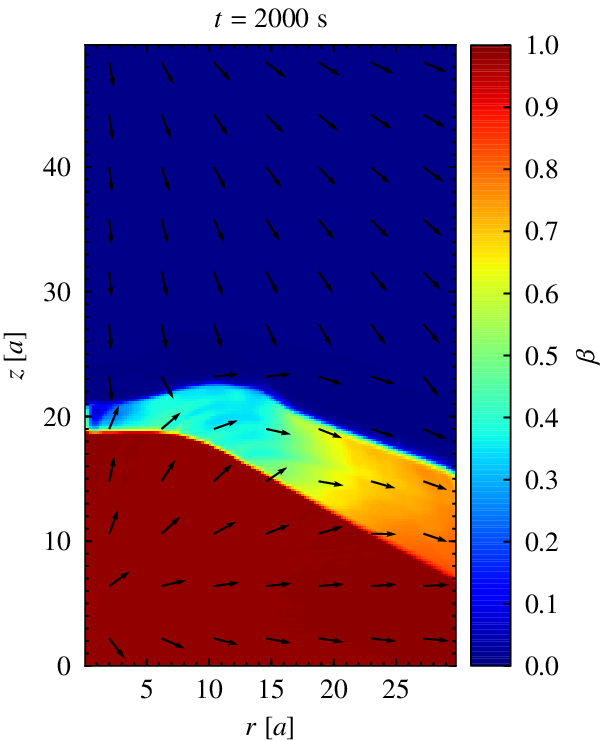}\\\vspace{0.2cm}
\includegraphics[width=6.4cm]{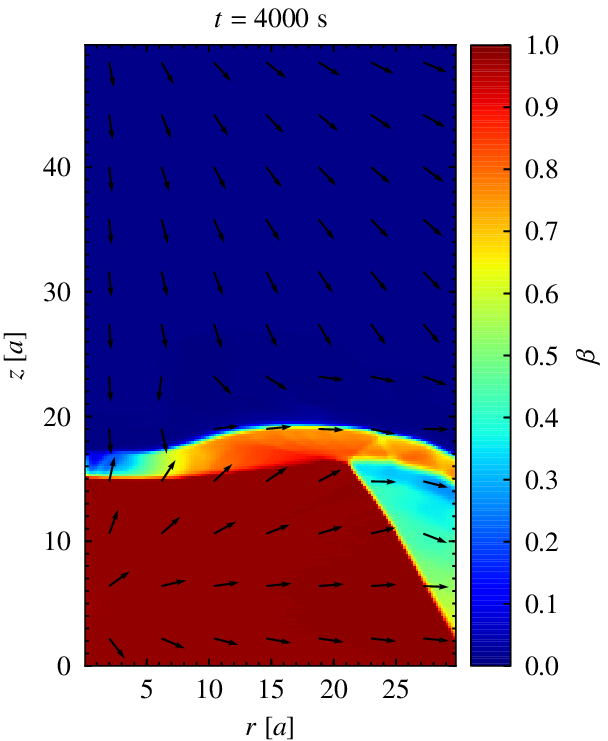}
\includegraphics[width=6.4cm]{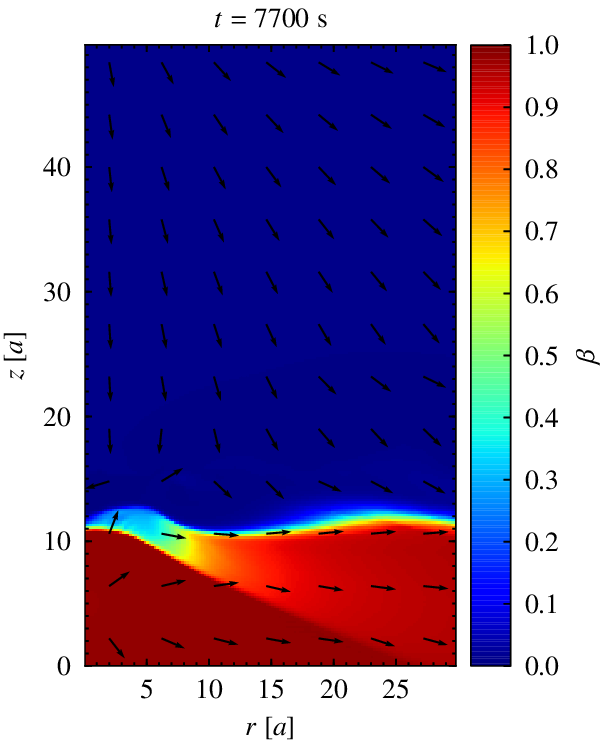}\\\vspace{0.2cm}
\includegraphics[width=6.4cm]{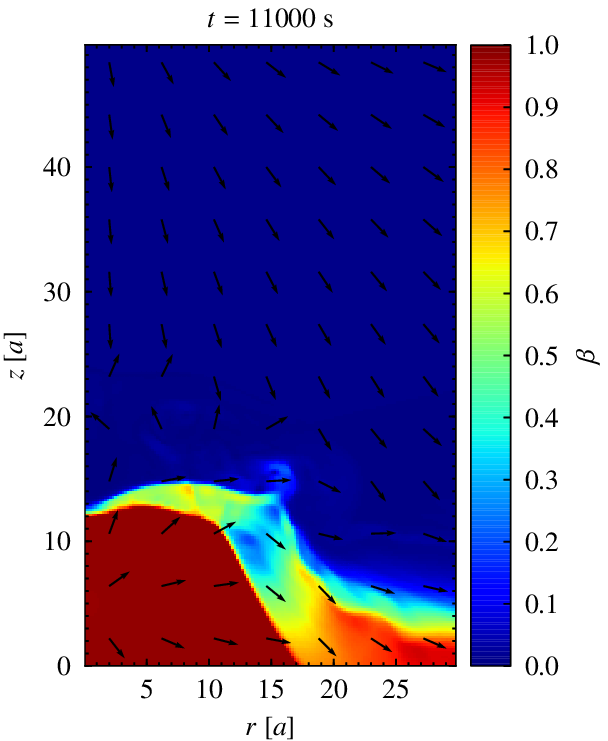}
\includegraphics[width=6.4cm]{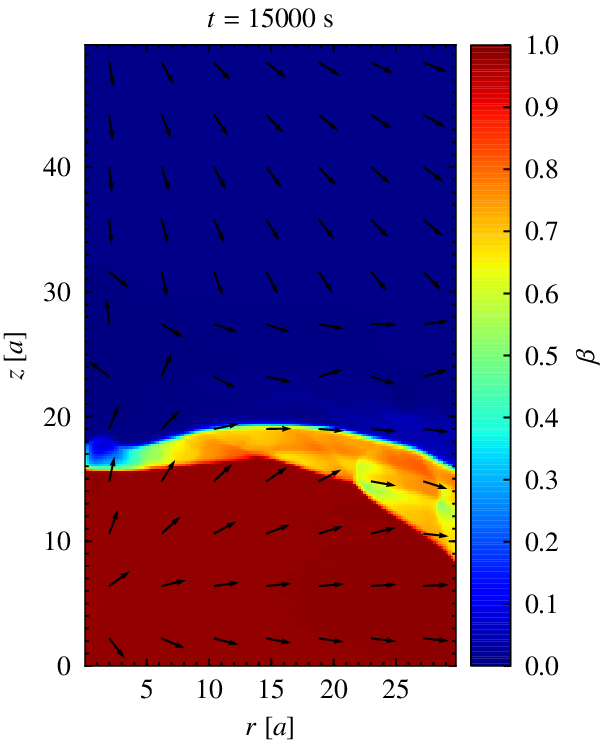}
\caption{$\beta$ distribution by colour for the case with clump parameters
$\chi = 10$ and $R_{\rm c} =5~a$
for the times shown at the top of each plot.
The remaining plot properties are the same as those of Fig.~\ref{f10r1_zoom}.}
\label{f10r5_W}
\end{figure*}

\begin{figure*}
\centering
\includegraphics[height=4.4cm,  trim= 0.0cm 0.9cm 0.0cm 0.0cm,clip]{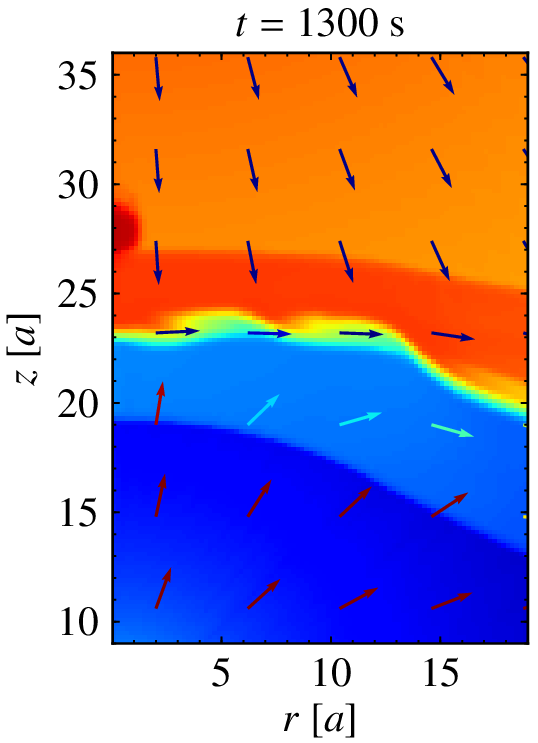}
\includegraphics[height=4.4cm,  trim= 1.0cm 0.9cm 0.0cm 0.0cm,clip]{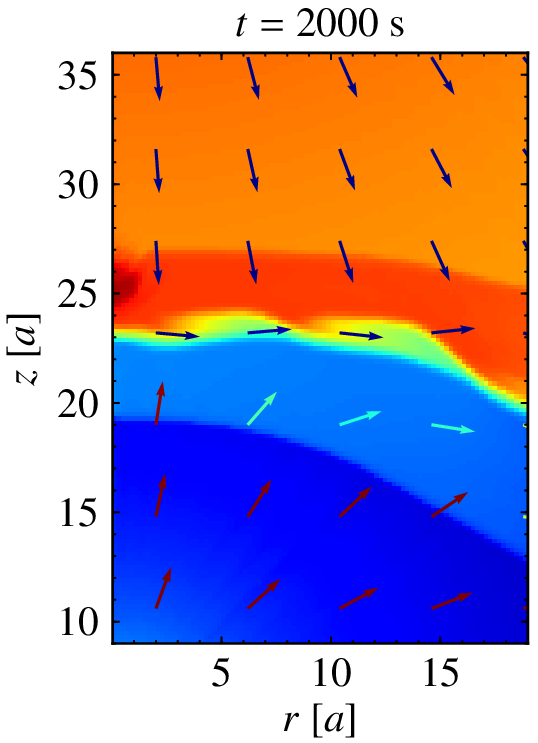}
\includegraphics[height=4.4cm,  trim= 1.0cm 0.9cm 0.0cm 0.0cm,clip]{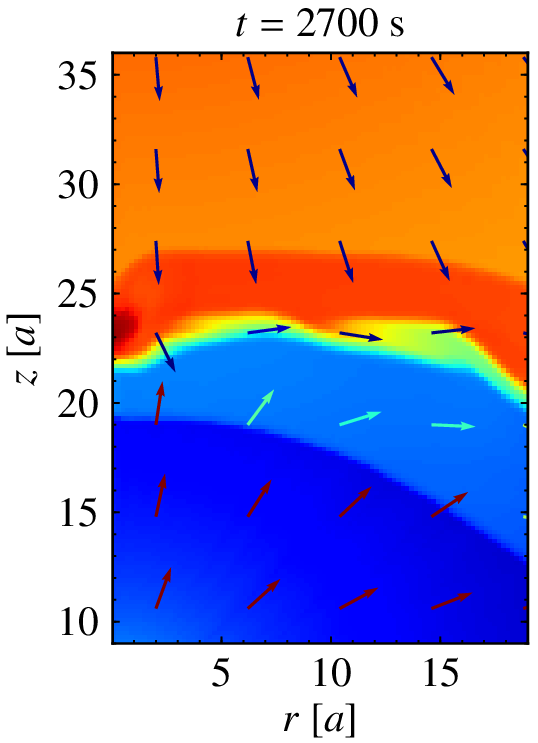}
\includegraphics[height=4.4cm,  trim= 1.0cm 0.9cm 0.0cm 0.0cm,clip]{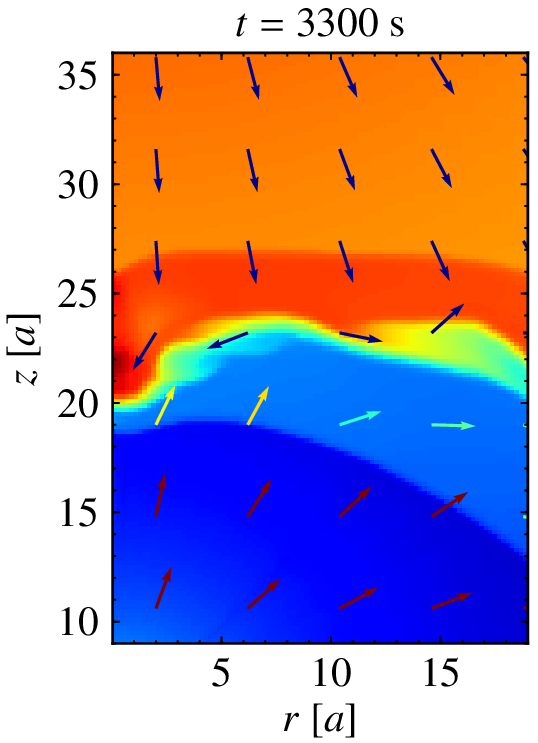}
\includegraphics[height=4.4cm,  trim= 1.0cm 0.9cm 0.0cm 0.0cm,clip]{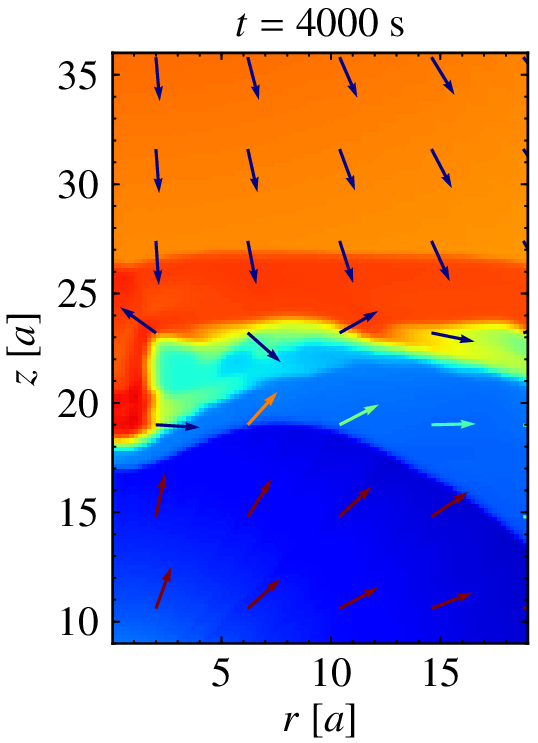}
\includegraphics[height=5.00cm, trim= 0.0cm 0.0cm 0.0cm 0.0cm,clip]{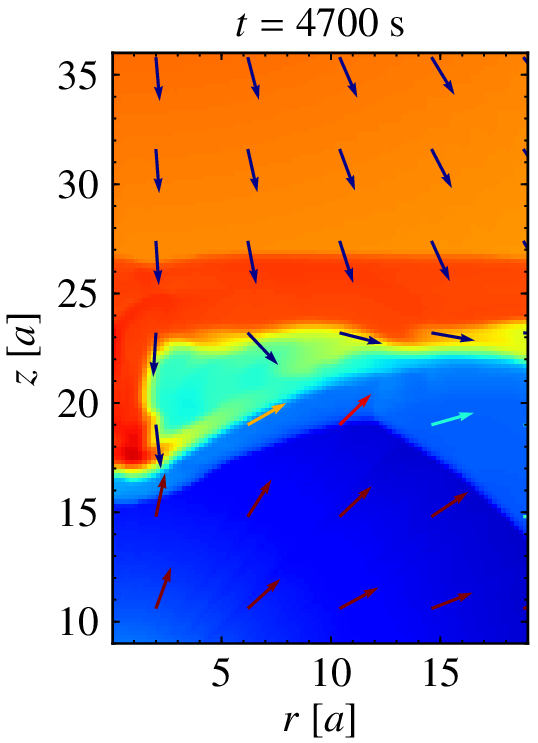}
\includegraphics[height=5.00cm, trim= 1.0cm 0.0cm 0.0cm 0.0cm,clip]{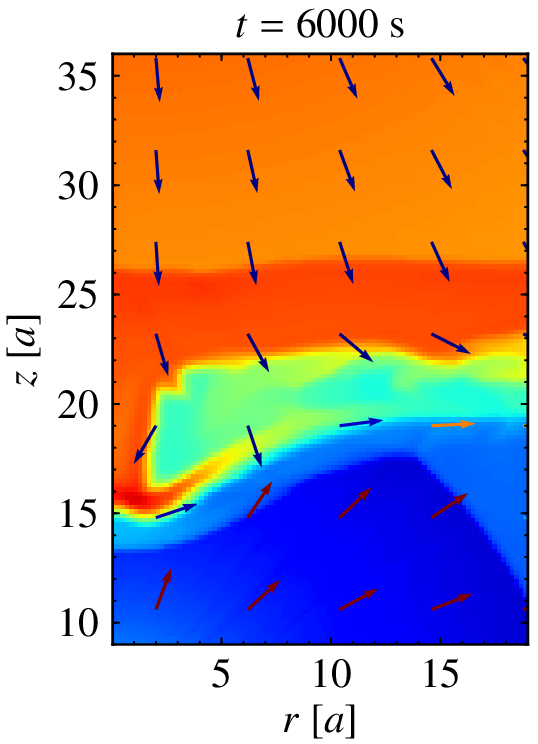}
\includegraphics[height=5.00cm, trim= 1.0cm 0.0cm 0.0cm 0.0cm,clip]{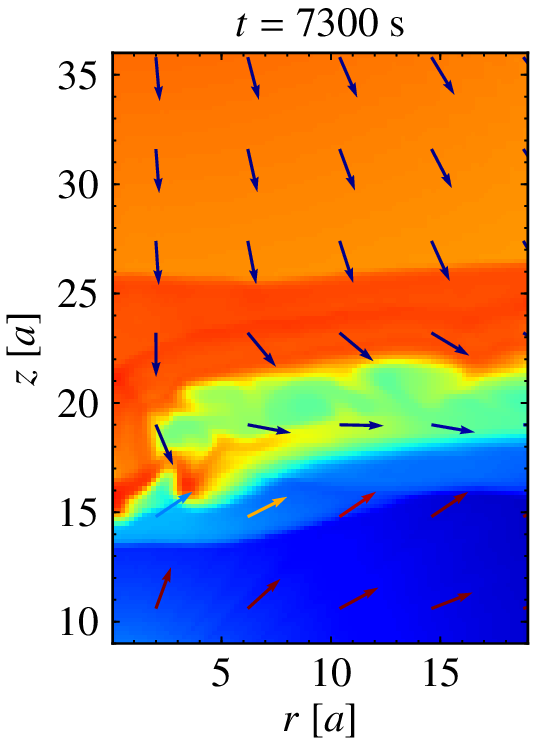}
\includegraphics[height=5.00cm, trim= 1.0cm 0.0cm 0.0cm 0.0cm,clip]{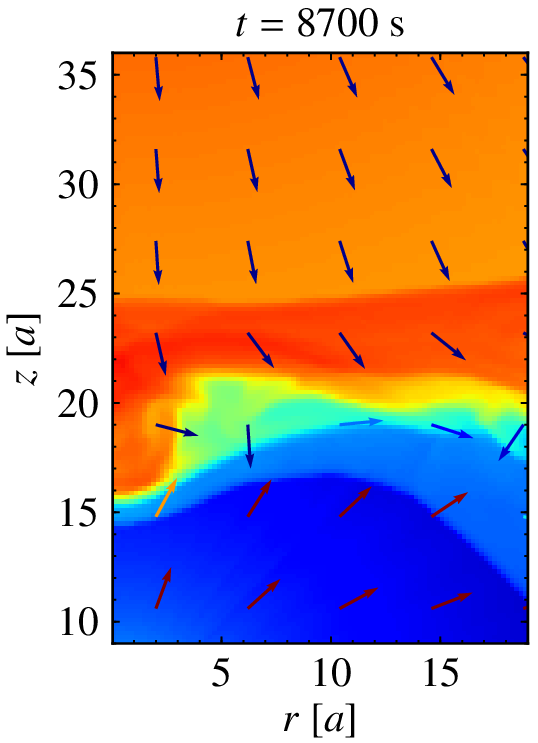}
\includegraphics[height=5.00cm, trim= 1.0cm 0.0cm 0.0cm 0.0cm,clip]{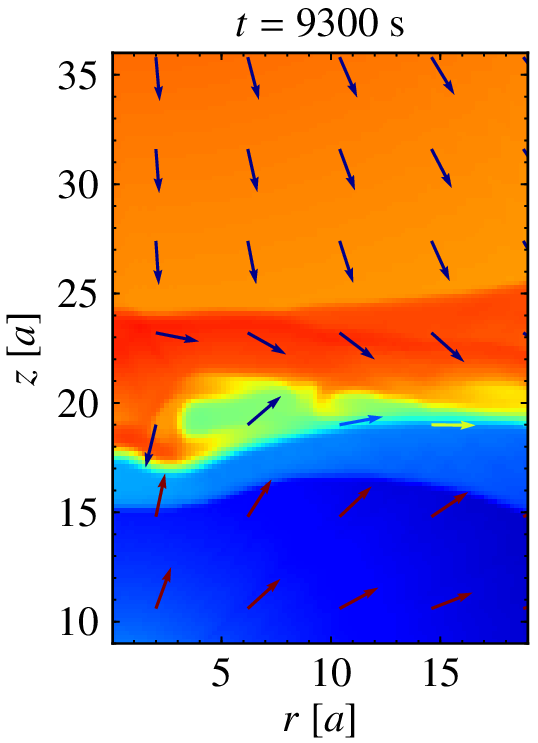}
\caption{
Zoom-in of the density distribution by colour for the case with clump parameters
$\chi = 30$ and $R_{\rm c} =1~a$
for the times shown at the top of each plot.
The remaining plot properties are the same as those of Fig.~\ref{f10r1_zoom}.}
\label{f30r1_zoom}
\end{figure*}

\begin{figure*}
\centering
\includegraphics[width=6.4cm]{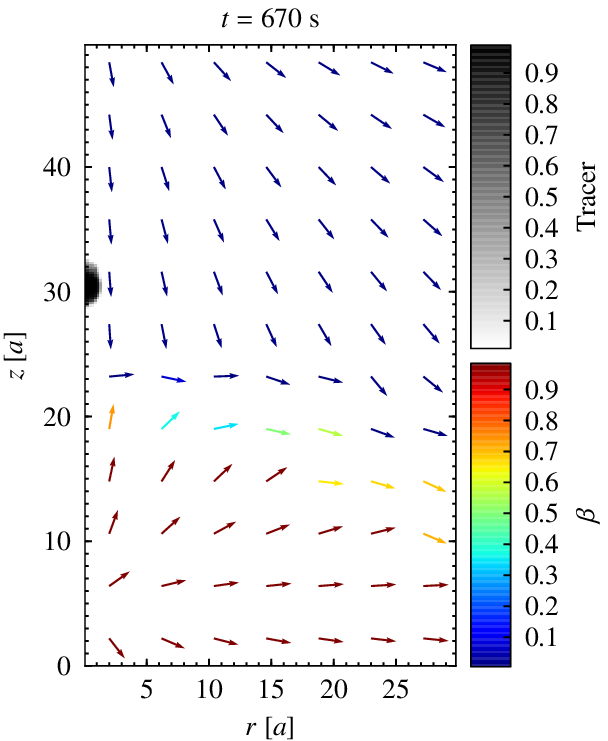}
\includegraphics[width=6.4cm]{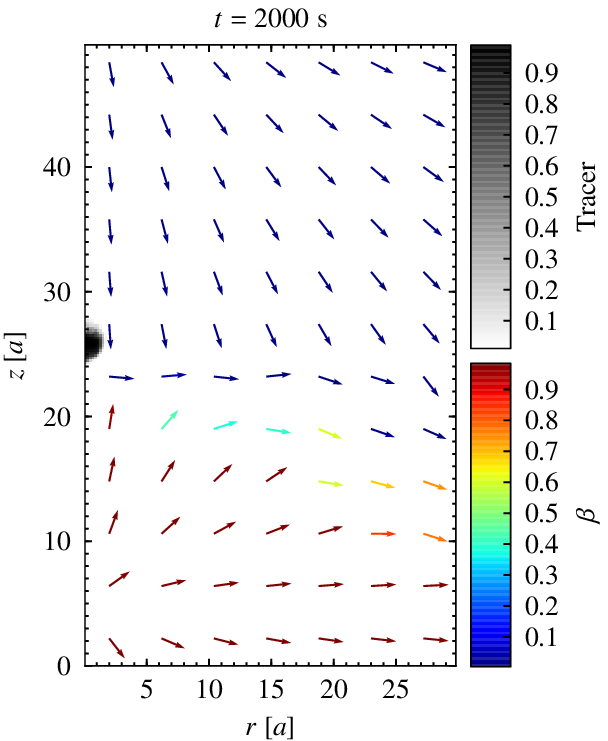}\\\vspace{0.2cm}
\includegraphics[width=6.4cm]{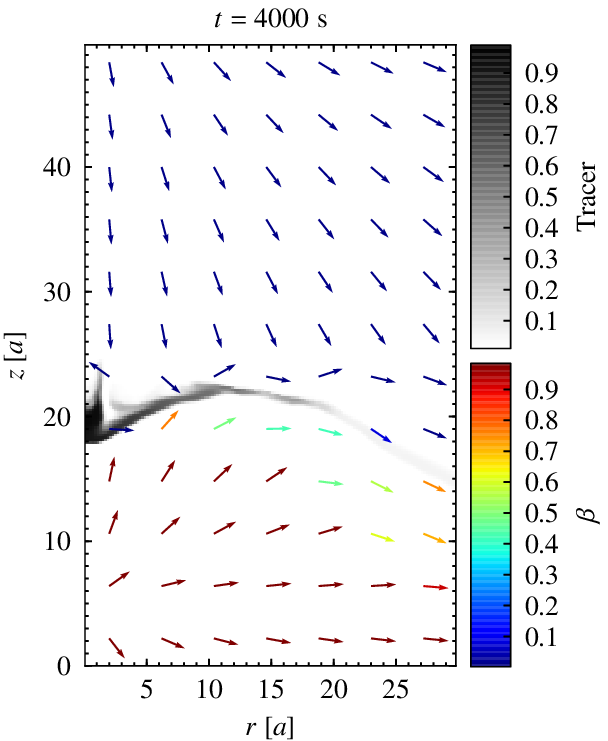}
\includegraphics[width=6.4cm]{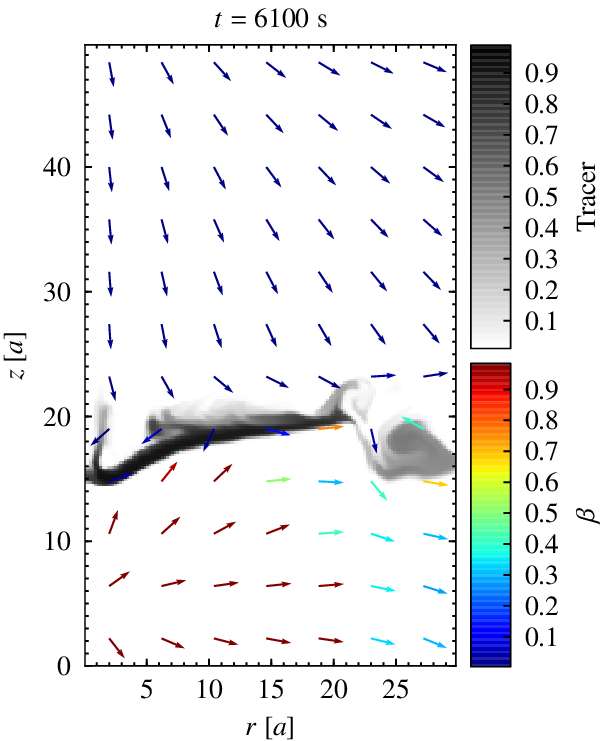}\\\vspace{0.2cm}
\includegraphics[width=6.4cm]{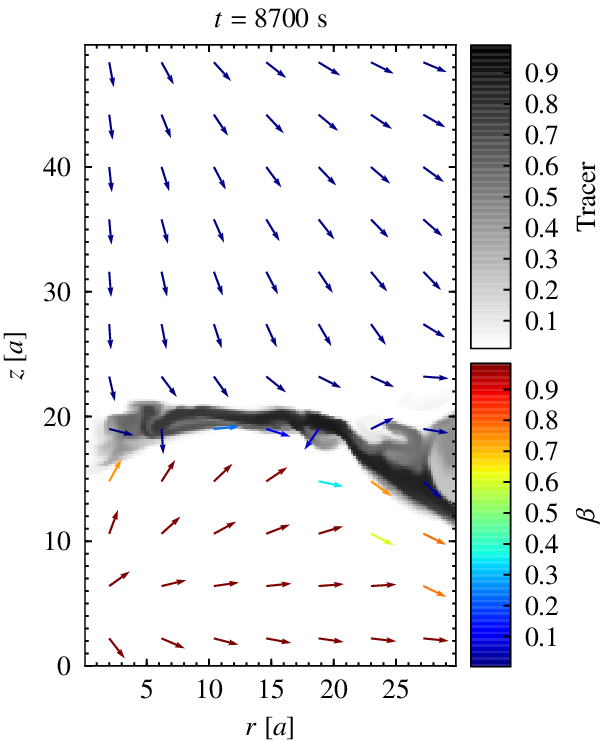}
\includegraphics[width=6.4cm]{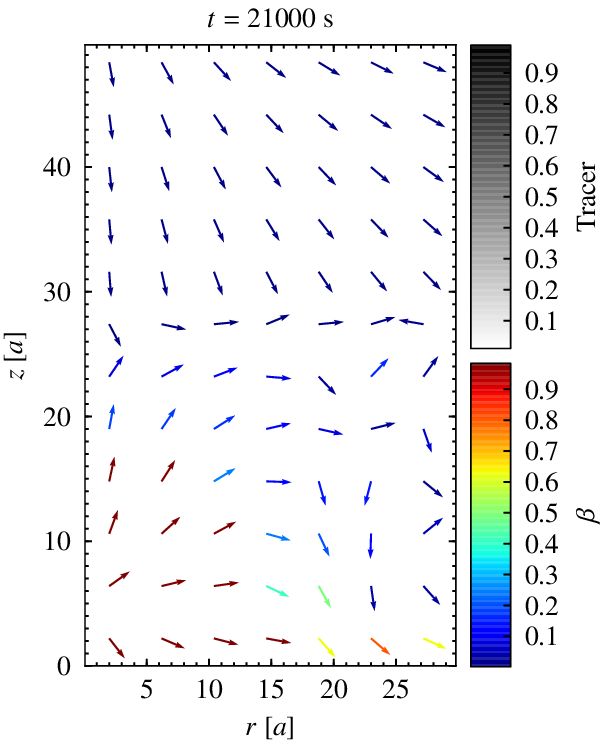}
\caption{Tracer distribution by colour for the case with clump parameters
$\chi = 30$ and $R_{\rm c} =1~a$
for the times shown at the top of each plot.
The tracer value ranges from 0 (pulsar and stellar wind) to 1 (clump).
The remaining plot properties are the same as those of Fig.~\ref{f10r1_zoom}.}
\label{f30r1_tracer}
\end{figure*}

\begin{figure*}
\centering
\includegraphics[width=6.5cm]{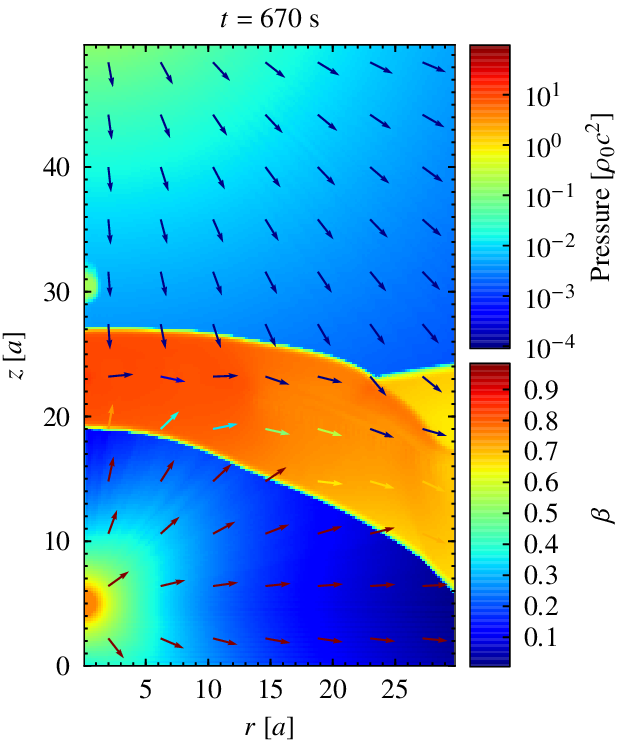}
\includegraphics[width=6.5cm]{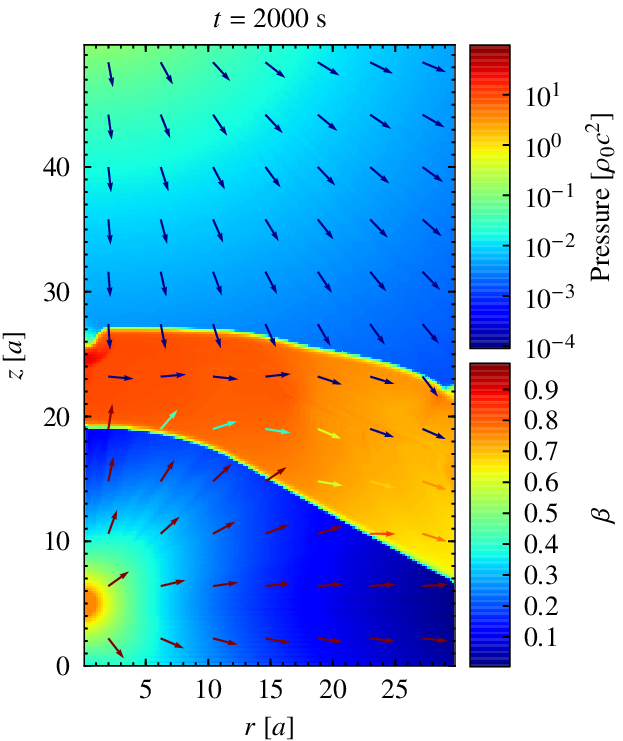}\\\vspace{0.2cm}
\includegraphics[width=6.5cm]{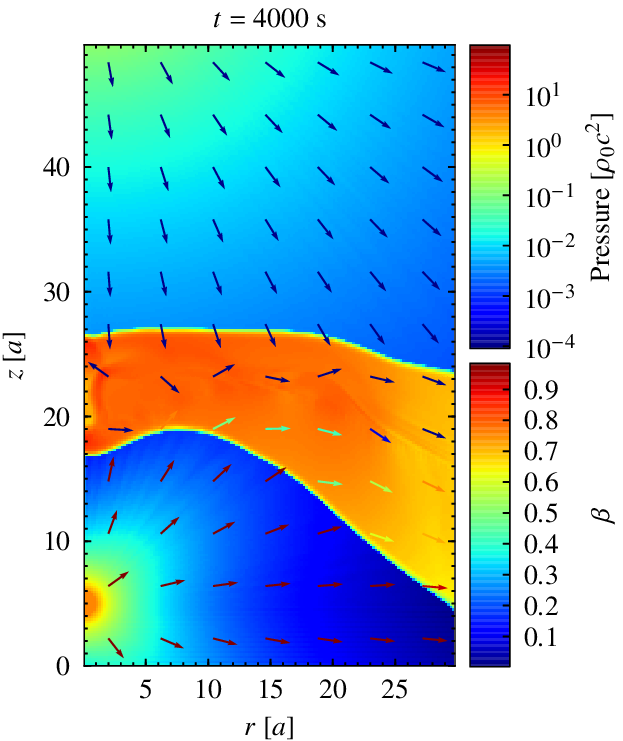}
\includegraphics[width=6.5cm]{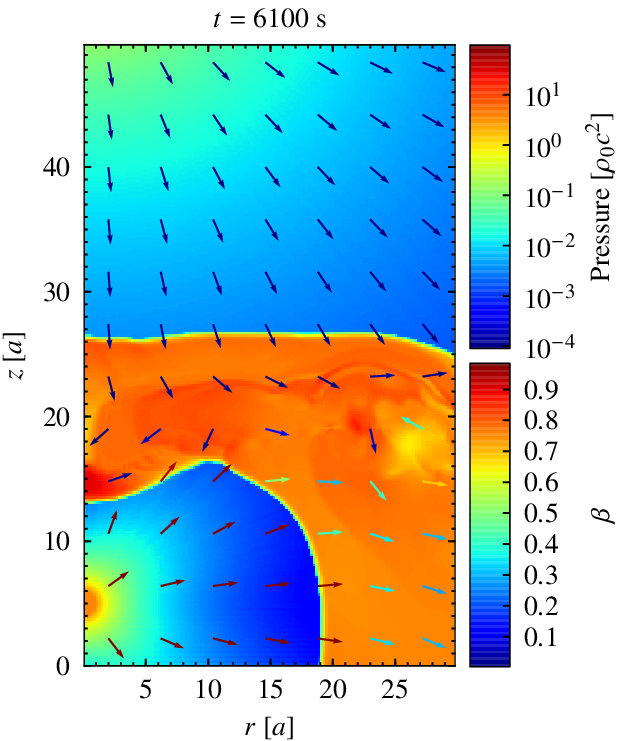}\\\vspace{0.2cm}
\includegraphics[width=6.5cm]{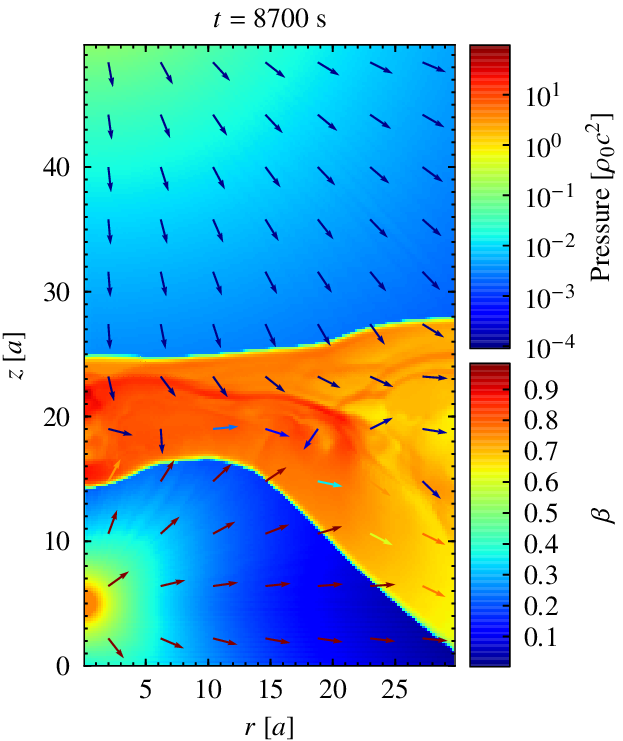}
\includegraphics[width=6.5cm]{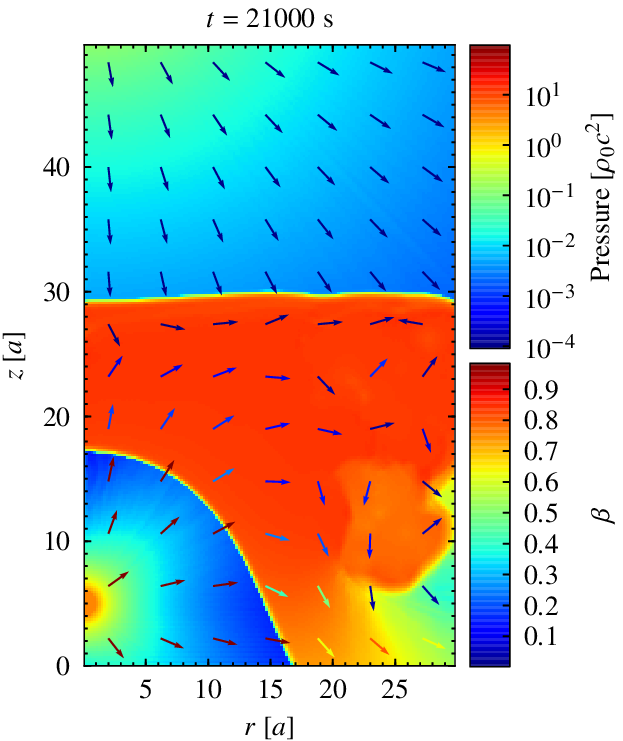}
\caption{Pressure distribution in units of $\rho_0{c^2}$ by colour for the case with clump parameters
$\chi = 30$ and $R_{\rm c} =1~a$
for the times shown at the top of each plot.
The remaining plot properties are the same as those of Fig.~\ref{f10r1_zoom}.}
\label{f30r1_pressure}
\end{figure*}

\begin{figure*}
\centering
\includegraphics[width=6.4cm]{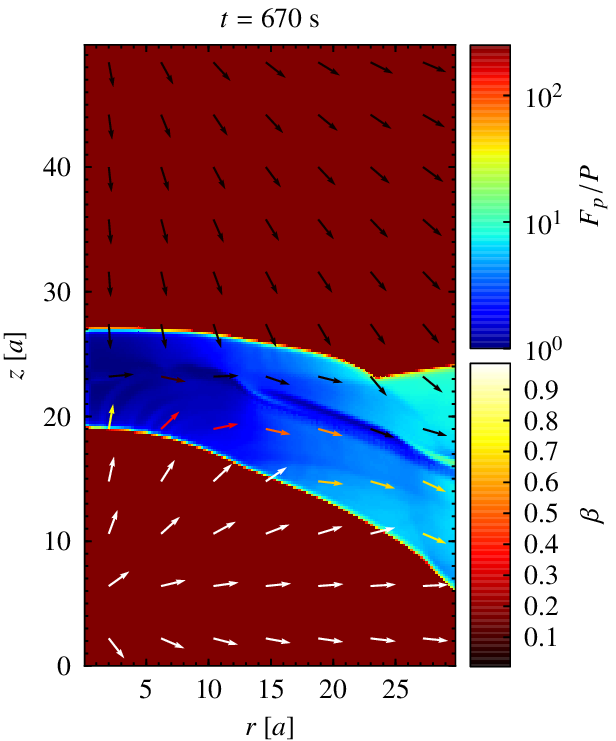}
\includegraphics[width=6.4cm]{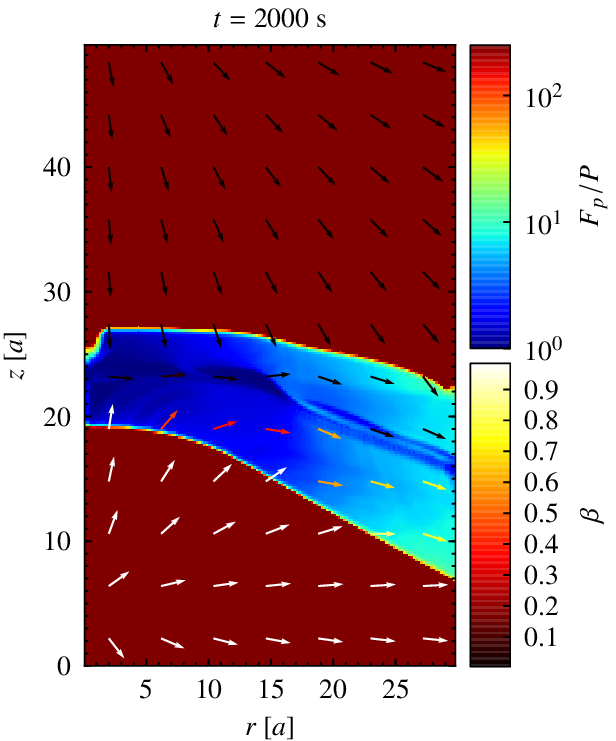}\\\vspace{0.2cm}
\includegraphics[width=6.4cm]{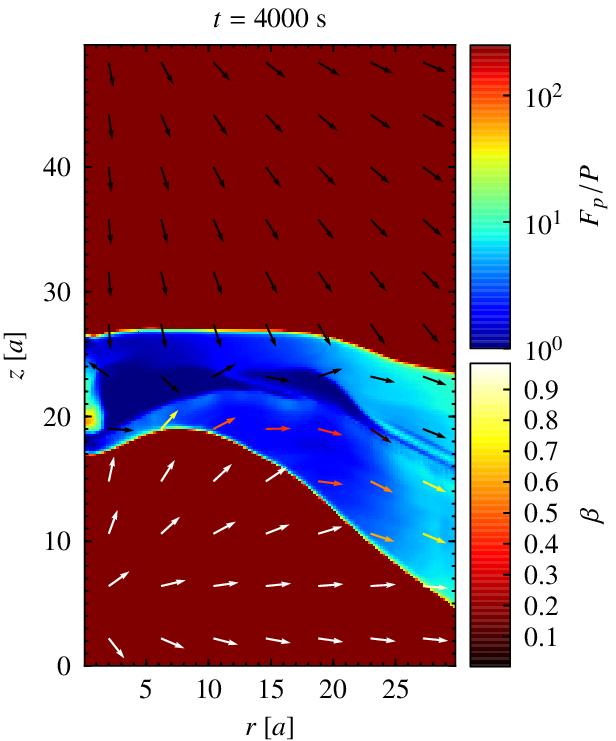}
\includegraphics[width=6.4cm]{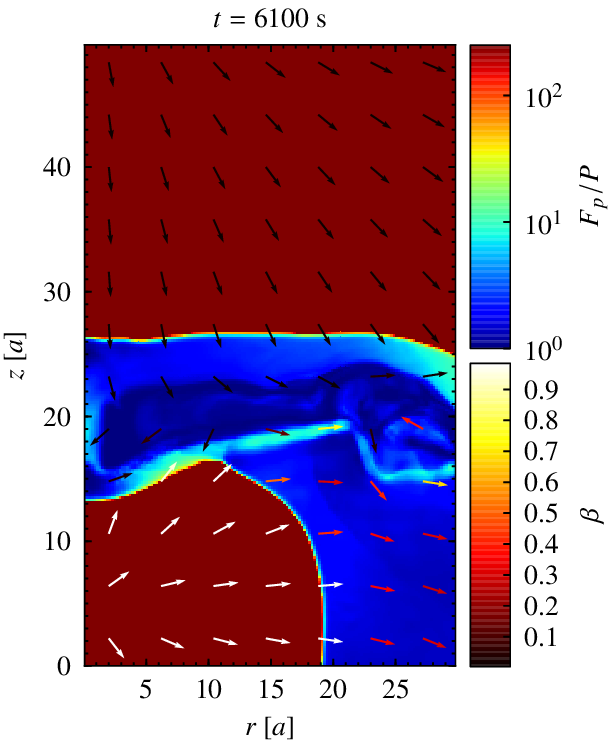}\\\vspace{0.2cm}
\includegraphics[width=6.4cm]{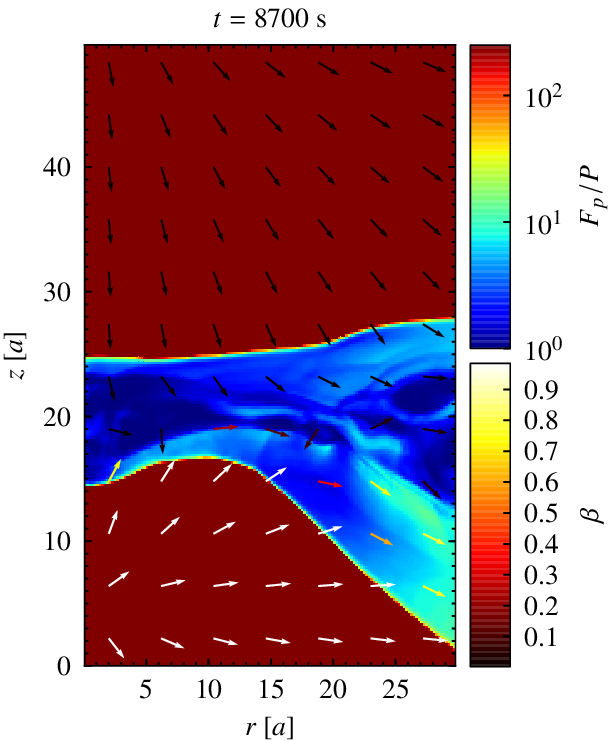}
\includegraphics[width=6.4cm]{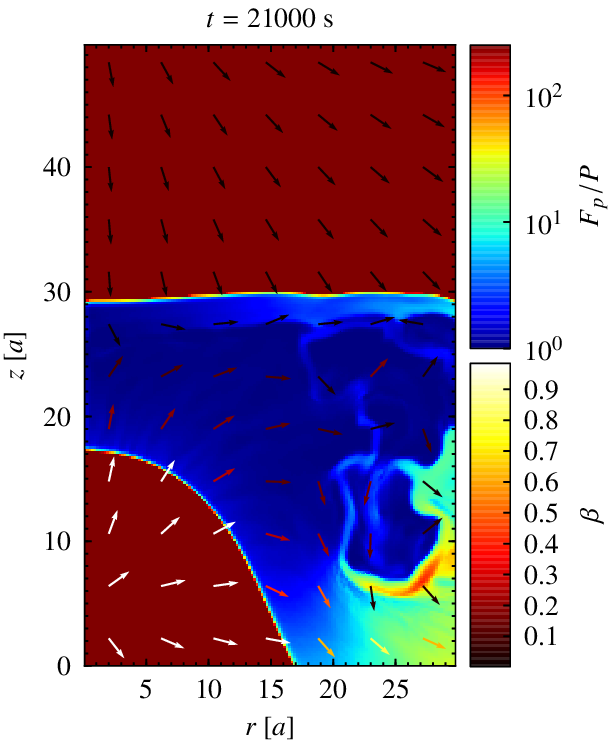}
\caption{Momentum flux over pressure distribution by colour for the case with clump parameters
$\chi = 30$ and $R_{\rm c} =1~a$
for the times shown at the top of each plot.
The momentum flux is given by $F_p = \rho~\Gamma^2 v^2 (1+\epsilon+P/{\rho})+P$.
The remaining plot properties are the same as those of Fig.~\ref{f10r1_zoom}.}
\label{f30r1_sonic}
\end{figure*}

\begin{figure*}
\centering
\includegraphics[width=6.4cm]{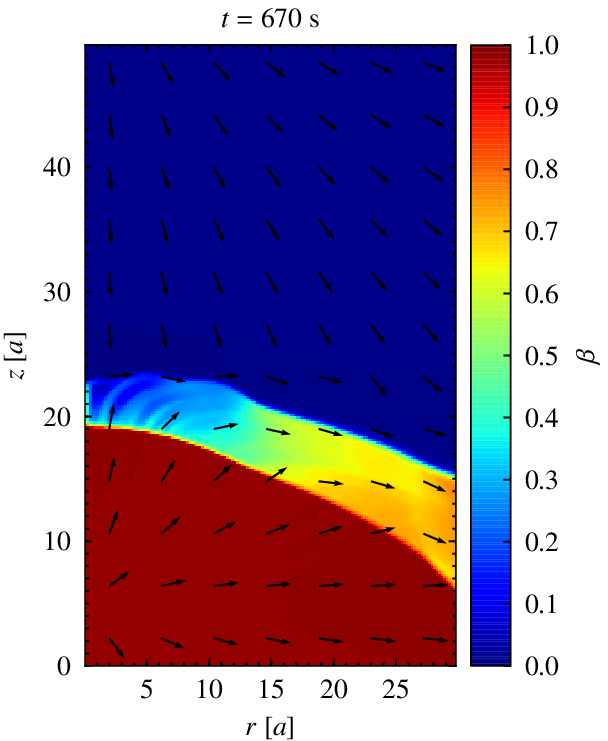}
\includegraphics[width=6.4cm]{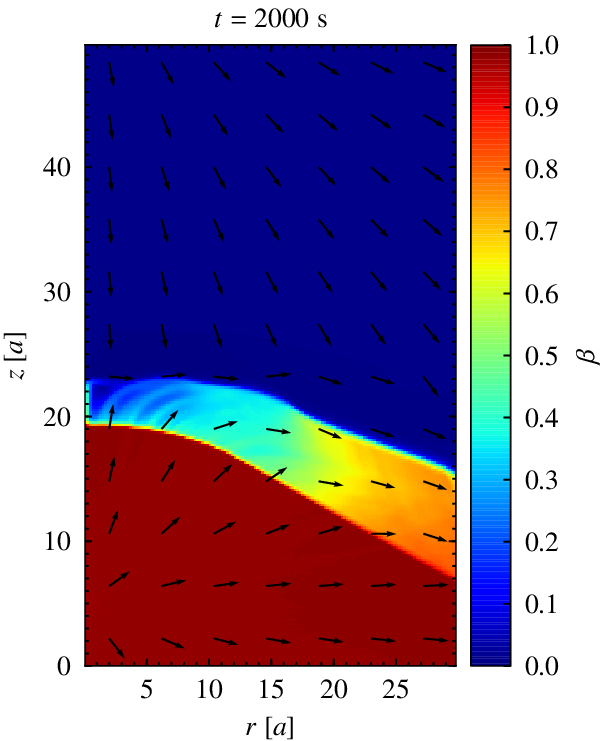}\\\vspace{0.2cm}
\includegraphics[width=6.4cm]{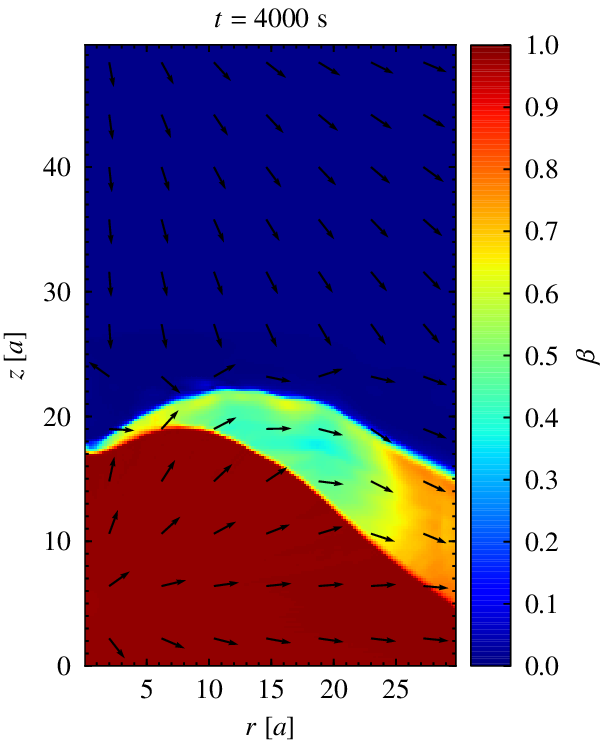}
\includegraphics[width=6.4cm]{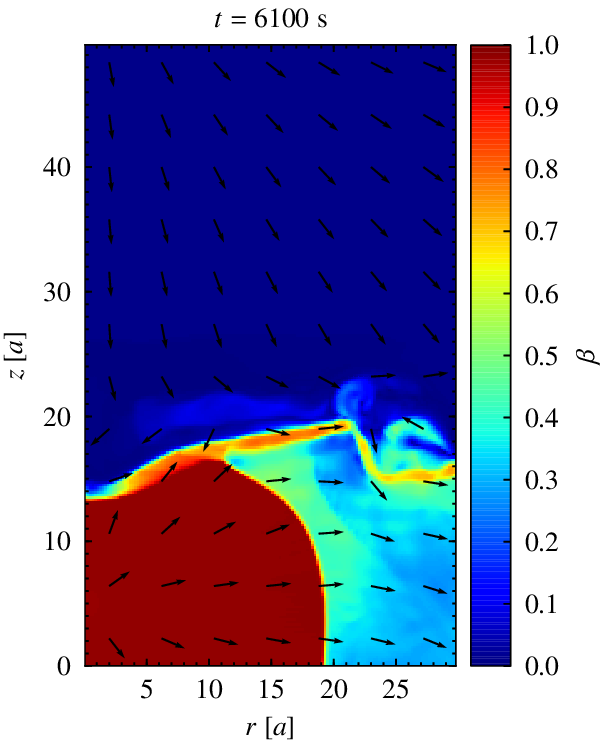}\\\vspace{0.2cm}
\includegraphics[width=6.4cm]{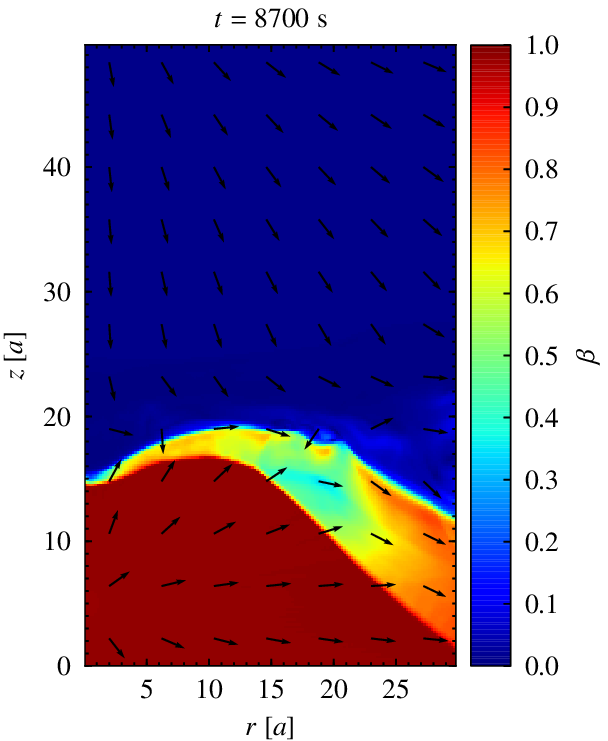}
\includegraphics[width=6.4cm]{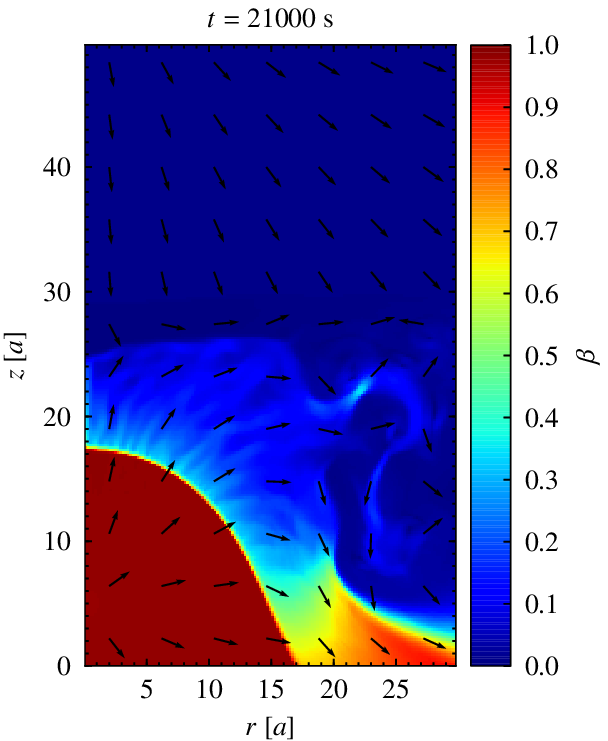}
\caption{$\beta$ distribution by colour for the case with clump parameters
$\chi = 30$ and $R_{\rm c} =1~a$
for the times shown at the top of each plot.
The remaining plot properties are the same as those of Fig.~\ref{f10r1_zoom}.}
\label{f30r1_W}
\end{figure*}

The density zoom, tracer, pressure, ratio of momentum-flux to pressure, and velocity maps are presented in Figs.~\ref{f10r1_zoom}--\ref{f30r1_W} for different clump parameters: $\chi=10$ and $R_{\rm c}=1~a$; $\chi=10$ and $R_{\rm c}=5~a$; and $\chi=30$ and $R_{\rm c}=1~a$.

%
%
%
%
%

%
%
%
%
%

%
%
%
%
%

\bibliographystyle{aa}
\bibliography{bibliography.bib}
\end{document}